%% file: article_arxiv.tex
\documentclass[hidelinks,onefignum,onetabnum]{siamart251216}

\usepackage{cite}
\usepackage{amsmath,amssymb,amsfonts}
\usepackage{algorithmic}
\usepackage{graphicx}
\usepackage{xr}
\usepackage{algorithm,algorithmic}
\usepackage{hyperref}
\hypersetup{hidelinks}
\usepackage{textcomp}
\usepackage{kotex}
\usepackage{comment}
\usepackage{csquotes}
\usepackage{upgreek}

\newcommand{\R}{\mathbb R}
\newcommand{\Z}{\mathbb Z}

\newcommand{\C}{\mathbb C}

\newcommand{\w}{\omega}

\theoremsymbol{\ensuremath{\triangle}}
\newtheorem{example}{Example}
\theoremsymbol{}

\input{ex_shared}

\theoremstyle{plain}\theoremheaderfont{\normalfont\itshape}\theorembodyfont{\normalfont}\theoremseparator{.}\theoremsymbol{\ensuremath{\triangle}}
\renewtheorem{remark}[theorem]{Remark}
\theoremsymbol{}

\ifpdf
\hypersetup{
  pdftitle={Nonlinear Moments as Equivalence Classes:
A Lyapunov--Perron Approach to the Invariance Equation},
  pdfauthor={H. Park, H. Jang, A. Astolfi, H. Shim, and J. G. Lee}
}
\fi

\begin{document}

\maketitle

\begin{abstract}
The linear moment is well defined whenever the system and the exosystem have disjoint spectra. Its nonlinear counterpart is not: the existence and uniqueness of solutions to the invariance equation that defines it are in general not guaranteed, so additional assumptions such as Poisson stability of the exosystem and the exponential stability of the system are required to define the nonlinear moment. This paper reduces this asymmetry. By using the Lyapunov--Perron fixed-point argument, we show that under a condition called the strip condition, there exist local $\mathcal C^N$ solutions, with $N$ determined by the system and the exosystem, and every pair of $\mathcal C^1$ solutions has contact of order $k^-$, where $k$ is a spectral ratio determined by the system and the exosystem; the solutions are flat-equivalent when $k=\infty$. Based on this observation, we extend the concept of nonlinear moment as an equivalence class of functions rather than a function. The polynomial of degree $N$ or formal power series ansatz of the invariance equation are the distinguished representatives of the equivalence class. We also show that, when the system is exponentially stable, the moment describes the output shared by all trajectories up to the order $k^-$, which gives the moment a time-domain interpretation when the steady-state interpretation is not available. Finally, we show that the concept of moment matching also naturally extends as the definition of nonlinear moment is extended to an equivalence class. The ansatz also plays a crucial role in the extended notion of moment matching. The results are illustrated by a numerical example.
\end{abstract}

\begin{keywords}
moment, contact of finite order, flat, invariance equation, invariant manifold, existence, uniqueness, Lyapunov--Perron fixed-point argument, moment matching, model-order reduction
\end{keywords}

\begin{MSCcodes}
34C45, 34C20, 37D10, 93B11, 93C10
\end{MSCcodes}


\section{Introduction}\label{sec:introduction}

The transfer function $G(s)=C(sI-A)^{-1}B$ and its moments $D^jG(s)$ of the linear system $\dot{x} = Ax +Bu,~y=Cx$ are fundamental objects in linear systems theory \cite[Chapter 11]{antoulas2005approximation}. Classically, these objects are interpreted in the time domain through the sinusoidal asymptotic steady-state response, characterized by the gain $|G(iw)|$ and the phase $\angle G(iw)$ for $w\in\R$ \cite[Chapter 6]{franklin1994feedback}. This interpretation is tied to imaginary-axis frequencies and, more fundamentally, to the existence of an asymptotic steady-state response. Linear moment theory \cite{astolfi2010model} removes both restrictions by introducing the exosystem $\dot\omega=S\omega$, $u=L\omega$ and the invariant hyperplane $\{(\omega,x):x=\Pi\omega\}$ defined by the unique solution $\Pi$ to the \emph{Sylvester equation} $A\Pi+BL=\Pi S$, provided that $\sigma(A)\cap\sigma(S)=\varnothing$. There, the linear moment is defined as $C\Pi$, which encodes the moments of the linear system at the eigenvalues of $S$. In this perspective, the linear moment $C\Pi$ is the static map from the exosystem state $\omega$ to the system output $y$ when the system state $x$ is restricted to the invariant hyperplane $x = \Pi \omega$. This alternative interpretation is independent of the asymptotic steady-state response and provides the physical meaning of the classical notion of moments---$G(s)$ and its derivatives---outside the imaginary axis.

The same geometric idea extends formally to nonlinear systems $\dot x=f(x,u)$, $y=h(x)$, driven by a nonlinear exosystem $\dot\omega=s(\omega)$, $u=l(\omega)$. The nonlinear analogue of the invariant hyperplane is an invariant manifold defined by the graph $\{(\omega,x):x=\pi(\omega)\}$ passing through the origin, and the \emph{invariance equation} $D\pi(\omega)s(\omega)=f(\pi(\omega),l(\omega))$ is required for its invariance. If the map $\pi$ is unique, or if it is possible to define a unique canonical solution to the invariance equation, the nonlinear moment is defined analogously as $h\circ\pi$. This viewpoint underlies nonlinear moment matching, data-driven moment estimation, least-squares formulations, and related applications \cite{scarciotti2017data, padoan2023model, scarciotti2016moment, faedo2018energy, faedo2020nonlinear}; see \cite{astolfi2020model} for a review.

The problem is that, unlike the Sylvester equation, which has a unique solution whenever $\sigma(A)\cap\sigma(S)=\varnothing$, the existence and uniqueness of solutions to the invariance equation are not guaranteed in the general nonlinear setting---even when $\sigma(S)\subset i\R$ and $\sigma(A)\cap i \R = \varnothing$, we can guarantee existence by using the center manifold theorem \cite{carr1981applications}, but not uniqueness; here, $A:=D_xf(0,0)$ and $S:=Ds(0)$. For example, consider the scalar system $\dot x=-x+u$ connected to the exosystem
\begin{equation}
\label{eq:sig-locally-flat}
    \dot{\omega} = s(\omega) = \begin{bmatrix}
        -\omega_{2}-\omega_{1}(\omega_{1}^{2}+\omega_{2}^{2}) \\[3pt]
        \phantom{-}\omega_{1}-\omega_{2}(\omega_{1}^{2}+\omega_{2}^{2})
    \end{bmatrix},~ u = \omega_1.
\end{equation}
The linearization gives $A=-1$ and $S=[\begin{smallmatrix} 0 & -1 \\ 1 & \phantom{-}0\end{smallmatrix}]$.
The invariance equation is given as $D \pi (\omega) s(\omega) = -\pi(\omega) + \omega_1,~\pi(0) = 0,$ and the function
\begin{equation}\label{eq:locally-flat}
\phi(\omega) := \begin{cases}
    0,& \text{if}~\omega =0 \\
    e^{-\frac{1}{2\|\omega\|^{2}}},&\text{if}~\omega\neq 0
\end{cases}
\end{equation}
satisfies $D\phi(\omega) s(\omega)=-\phi(\omega)$. Therefore, if $\pi$ is a $\mathcal{C}^1$ solution to the invariance equation, so is $\pi+c\phi$ for every $c\in\R$. Since the solution to the invariance equation exists by the center manifold theorem, the invariance equation has infinitely many solutions, and the nonlinear moment is not well-defined as a single function.


This discussion leads to the following fundamental questions.
\begin{itemize}
    \item \emph{When does a solution to the invariance equation exist, in general, e.g., when it is impossible to apply the center manifold theorem?}
    \item \emph{If a solution to the invariance equation exists but is not unique, how should we define the nonlinear moment?}
\end{itemize}

Similar questions have been investigated in the field of output regulation (OR), in which the invariance equation is a part of the Francis--Byrnes--Isidori (FBI) equation. In OR, the invariance equation is central because it characterizes the zero-error invariant manifold $\{(\omega,x):x=\pi(\omega)\}$ induced by the exosystem \cite{isidori1990output}. Focusing on this structural similarity, recent work \cite{Jang2026steady} shows that, for linear systems, the steady-state assignment aspect of OR can be viewed as a moment assignment for each subsystem, with the regulator equations being reformulated as a set of Sylvester equations associated with each subsystem. In OR, existence of the invariant manifold of the form $\{(\omega,x): x = \pi(\omega)\}$ is typically guaranteed by the center manifold theorem, which yields such a manifold only when $\sigma (S) \subset i\R$. Moreover, since center manifolds are not unique in general \cite[Section 2.12]{perko2013differential}, the canonical solution is used instead of establishing uniqueness. It is typically assumed that $\sigma(A) \subset \mathbb{C}_{<0}$ and also that the exosystem is Poisson stable \cite[Section 3.2]{huang2004nonlinear}, which also implies $\sigma(S) \subset i\mathbb{R}$: otherwise, the local stable or unstable manifold of the origin would contain points $\omega\neq0$ converging to the origin in forward or backward time, and such points are not Poisson stable \cite[Section~2.7]{perko2013differential}. These additional assumptions guarantee that there exists a solution to the invariance equation that corresponds to the asymptotic steady-state response \cite[Chapter 8]{isidori1995nonlinear}, which is considered the canonical solution. Most of the nonlinear moment literature bypasses the uniqueness problem by also defining the nonlinear moment through this canonical solution \cite{ionescu2016nonlinear, scarciotti2017nonlinear, faedo2021approximation, moreschini2025closed}. In \cite[Definition~2]{shakib2023model}, the moment is defined through the graph of a unique invariant function, without appealing to the center manifold theorem; however, the existence and uniqueness of this function are guaranteed by the input-to-state convergence of the system and by the boundedness of the exosystem trajectories in forward and backward time \cite[Assumptions~1 and~2, Lemma~5]{shakib2023model}, under which the graph describes the steady-state solution; see Remark~\ref{rem:shakib} for a further comparison. In all these cases, the nonlinear moment is once again tied to the asymptotic steady-state response, as in the classical transfer function interpretation.

The existence and smoothness of invariant manifolds is a classical topic in dynamical systems theory. Integral manifolds of perturbed systems over a compact or slowly varying base are constructed by the Lyapunov--Perron (L--P) fixed-point argument in \cite{hale1961integral,sakamoto1990invariant,yi1993generalized}, but these results are perturbative in a small parameter, whereas the invariance equation involves neither a small parameter nor such a base. Closer to the present setting, local invariant manifolds of an equilibrium tangent to a spectral subspace exist under a spectral gap condition, and their order of smoothness is governed by the ratio of the gap \cite{kelley1967stable,hirsch1977invariant,irwin1980pseudostable,vanderbauwhede1989centre}. The cascade of the system and the exosystem falls into this framework: in this paper, the gap is formulated as the strip condition, which distinguishes the growth of the exosystem in forward and backward time, and the existence result (Theorem~\ref{thm:existence}) is a version of these results adapted to the cascade, proved by the standard L--P argument with cutoff functions in the spirit of \cite{vanderbauwhede1989centre}. The uniqueness statements in this literature, however, concern a single manifold selected by the cutoff function \cite{vanderbauwhede1989centre} or by global growth conditions on orbits \cite{irwin1980pseudostable,delallave1995irwin}, or manifolds that are sufficiently smooth \cite{cabre2003parameterization1}; when the exosystem has purely imaginary spectrum, distinct center manifolds are known to differ only by flat terms \cite{sijbrand1985properties}. From the viewpoint of moment theory, these statements are not sufficient: every $\mathcal C^1$ solution to the invariance equation defines a candidate for the nonlinear moment, so what is needed is a statement about all $\mathcal C^1$ solutions, without selecting one of them and without assuming additional smoothness. We show that all $\mathcal C^1$ solutions have contact of order $k^-$ with each other, where $k$ is determined by the spectra of the system and the exosystem, which recovers the flatness result of \cite{sijbrand1985properties} when $k=\infty$ and extends it beyond the case of purely imaginary spectrum, and we interpret the resulting contact class as the nonlinear moment.

Motivated by these observations, we study the local existence, smoothness, and uniqueness of solutions to the invariance equation by using the L--P fixed-point argument. The strip condition, which will be defined in Definition~\ref{def:strip-condition}, is required for the L--P fixed-point argument, and also makes it possible to show that all $\mathcal C^1$ solutions to the invariance equation have contact of order $k^-$ with each other, where $k$ is a spectral ratio determined by the system and the exosystem. This is possible because the contact is derived from the variation-of-constants formula applied to the difference of two arbitrary $\mathcal C^1$ solutions (Lemma~\ref{lem:order-raising}), not from the L--P fixed-point argument. This result suggests defining the nonlinear moment as an equivalence class, whose interpretation does not rely on the stability of the steady-state response, thus reducing the asymmetry that existed between the linear and nonlinear moment theories. We also show that, first, this equivalence class has a distinguished polynomial or formal power series representative; second, the moment retains a time-domain interpretation beyond the classical setting: the classical interpretation of the nonlinear moment relies on the steady-state response, which becomes trivial when the trajectories of the exosystem converge to the origin, whereas we show that, when the system is exponentially stable, the moment describes the output shared by all trajectories up to the order $k^-$, the same order that limits the contact between solutions; and third, the definition of nonlinear moment-matching model reduction can be extended as we define the nonlinear moment as an equivalence class rather than a single function. The polynomial or the formal power series representative plays a central role in the extended notion of nonlinear moment matching.

In summary, the contributions of this paper are twofold. From the viewpoint of invariant manifold theory, all $\mathcal C^1$ solutions to the invariance equation have contact of order $k^-$, without any smoothness assumption and beyond the case of purely imaginary spectrum; to the best of our knowledge, this result is not available in the classical theory recalled above. From the viewpoint of moment theory, this contact class defines the nonlinear moment, which admits a time-domain interpretation that does not rely on the steady-state response and leads to an extended notion of moment matching.

The remainder of the paper is organized as follows. Section~\ref{sec:prelim} reviews the necessary background on moment theory and formulates the problem precisely. The existence of solutions to the invariance equation is derived in Section~\ref{sec:existence}, and the contact of solutions to the invariance equation is established in Section~\ref{sec:uniqueness}. The polynomial or formal power series representative of the equivalence class, the time-domain interpretation of the moment, and the extended definition of nonlinear moment matching are described in Section~\ref{sec:consequences}. Section~\ref{sec:example} presents an illustrative example with a decaying exosystem, for which the steady-state interpretation of the nonlinear moment is not available, and finally, Section~\ref{sec:conclusion} concludes the paper.

\textbf{Notation.} The sets of integers, real, and complex numbers are denoted by $\mathbb Z$, $\mathbb R$, and $\mathbb C$, respectively. The set of positive integers, the set of nonnegative integers, the set of positive real numbers, open left and right half-planes, and the imaginary axis are denoted by $\mathbb{Z}_{>0}$, $\mathbb{Z}_{\ge 0}$, $\R_{>0}$, $\mathbb C_{<0}$, $\mathbb C_{>0}$, and $i \R$, respectively. Since $\pi$ is reserved for the solution to the invariance equation, we denote the real number 3.1415... as $\uppi$ rather than $\pi$.

Given $A \in \R^{n \times n}$, the spectrum of $A$ is denoted by $\sigma(A)$. If $\sigma(A)\cap i\mathbb R=\varnothing$, we denote the unstable and stable subspaces of $A$ by $E_+$ and $E_-$, respectively. The projection from $\R^n$ onto $E_+$ is denoted by $P_+$, and the restriction of $A$ to $E_+$ is denoted by $A_+$. We similarly define the projection $P_-$ and the restriction $A_-$. The identity matrix of appropriate dimension is denoted by $I$. The direct sum of two vector spaces $V$ and $W$ is denoted as $V \oplus W$.

Let $N,n_1,n_2 \in \Z_{>0}$. The space of $N$-times continuously differentiable functions from $\R^{n_1}$ to $\R^{n_2}$ is denoted by $\mathcal C^N(\R^{n_1}, \R^{n_2})$; smoothness means $\mathcal C^\infty$. The space of symmetric $N$-linear functions from $\R^{n_1}$ to $\R^{n_2}$ is denoted by $\mathrm{Sym}^N(\R^{n_1}, \R^{n_2})$. For a $\mathcal{C}^N$ function $f$ defined around the origin, the Taylor polynomial approximation of $f$ of order $N$ at the origin is denoted as $T_{N} f$. For a differentiable function $(x,u) \mapsto f(x,u)$, the partial derivative with respect to $x$ at $(x_0,u_0)$ is written as $D_xf(x_0,u_0)$. When $f$ depends on a single variable, we write $Df$. Open and closed balls of radius $\rho$ centered at the origin with various dimensions are uniformly denoted by $B_\rho$ and $\overline{B}_\rho$, respectively. All norms are Euclidean, and the norm of a multilinear map is the induced operator norm. 

For a function $g$ defined around the origin and $\phi:\R^\nu\to[0,\infty)$, we write $g(\omega)=O(\phi(\omega))$ if there exist $c>0$ and a neighborhood $U$ of the origin such that $\|g(\omega)\|\le c\,\phi(\omega)$ for all $\omega\in U$, and $g(\omega)=o(\phi(\omega))$ if, for every $c>0$, such a neighborhood exists; see \cite[Chapter 1]{debruijn1981asymptotic}, \cite[Section 1, Chapter V]{bourbaki2004functions}. In particular, $g(\omega)=O(\|\omega\|^a)$ with $a>0$ implies $g(0)=0$.

\section{Preliminaries and problem formulation}\label{sec:prelim}
We begin by recalling the essential elements of the moment theory developed in~\cite{astolfi2010model}, and then formulate the problem addressed in this paper.

Consider the linear time-invariant system $\dot x = Ax + Bu, y = Cx,$ and the exosystem $\dot\omega = S\omega, u = L\omega,$ with $x\in\mathbb{R}^{n}$, $u\in\mathbb{R}^{n_u}$, $y\in\mathbb{R}^{n_y}$, $\omega\in\mathbb{R}^{\nu}$, $S\in\mathbb{R}^{\nu\times\nu}$, and $L\in\mathbb{R}^{n_u\times\nu}$. If $\sigma(S)\cap\sigma(A)=\varnothing$, the Sylvester equation
\begin{equation}\label{eq:sylvester}
  A\Pi + BL = \Pi S
\end{equation}
has a unique solution $\Pi\in\mathbb{R}^{n\times\nu}$, and the hyperplane $\{(\omega,x) : x = \Pi\omega\}$ is an invariant hyperplane of the interconnection of the system and the exosystem. The matrix $C\Pi \in \R^{n_y\times\nu}$ describes the map $\omega\mapsto y$ on this invariant hyperplane and is referred to as the \emph{moment} of the linear system at $(L, S)$. The terminology is justified by a connection to the frequency domain: {if $S$ is non-derogatory with characteristic polynomial $p_{S}(s) = \prod_{i=1}^{\eta}(s-s_{i}^{\star})^{\nu_{i}}$ and the pair $(L,S)$ is observable, then there exists a one-to-one correspondence between the linear moment $C\Pi$ and the values $D^{j} G(s_{i}^{\star})$, $j = 0,\ldots,\nu_{i}-1$, $i=1,\ldots,\eta$, of the transfer function $G$ of the system \cite{astolfi2010model, shakib2023time}.}

This geometric construction extends to nonlinear systems as follows. Consider the nonlinear system
\begin{equation}\label{eq:nl}
  \dot x = f(x,u),~ y = h(x);~f(0,0)=0,~h(0)=0,
\end{equation}
and the nonlinear exosystem
\begin{equation}\label{eq:nl-exo}
    \dot\omega = s(\omega),~ u = l(\omega);~s(0)=0,~l(0)=0,
\end{equation}
with $x \in \R^n$, $u \in \R^{n_u}$, $y \in \R^{n_y}$, and $\omega \in \R^\nu$. Throughout the paper, the functions $h$, $f$, $l$, and $s$ are assumed to be smooth. Let $A := D_{x}f(0,0), B := D_{u}f(0,0), C := Dh(0), S := Ds(0), L := Dl(0),$ so that $(C,A,B)$ and $(L,S)$ are the linearizations at the origin of the system \eqref{eq:nl} and of the exosystem \eqref{eq:nl-exo}, respectively. The interconnection of the system \eqref{eq:nl} and the exosystem \eqref{eq:nl-exo} is the cascade $\dot\omega = s(\omega),\dot x = f(x,l(\omega)),$ and the central object is the \textit{invariance equation}
\begin{equation}\label{eq:inv}
  D\pi(\omega) s(\omega) = f(\pi(\omega), l(\omega)),
  ~\pi(0) = 0,
\end{equation}
whose solution, if it exists, defines a local invariant manifold $\{(\omega,x) : x = \pi(\omega)\}$ of the cascade. The boundary condition $\pi(0)=0$ is imposed because the invariant manifold is a nonlinear generalization of the invariant hyperplane and is therefore required to pass through the origin. In parallel with the linear moment $C\Pi$, the nonlinear moment is defined in \cite[Definition~2]{astolfi2010model} as $h\circ\pi$. However, as illustrated in the Introduction, the existence and uniqueness of solutions to the invariance equation \eqref{eq:inv} are far from trivial, and this causes an ambiguity in defining the nonlinear moment in terms of functions. The ambiguity persists even when the system and the exosystem are both linear, if the invariance equation is considered in place of the Sylvester equation.

\begin{example}\label{ex:non-unique}
    Consider a linear one-dimensional system and exosystem $\dot x = 2x + u,~ y = x;~\dot\omega = \omega,~ u = \omega$. The linear moment is uniquely defined by the invariant hyperplane $\{(\omega,x): x = - \omega\}$. From the viewpoint of nonlinear moment theory, however, the invariance equation becomes $\omega D\pi(\omega) = 2\pi(\omega) + \omega$ and admits infinitely many solutions $\pi_{c}(\omega) = c\omega^{2} - \omega$, $c \in \R$. The invariant hyperplane corresponds to $\pi_0$, but without further understanding, there is no clear reason for the solution $\pi_0$ to be the distinguished representative among solutions. Why $\pi_{0}$ is nevertheless the distinguished representative in this example will be explained in Remark~\ref{rem:boundary-condition} and also at the end of Section~\ref{subsec:representatives}. 
\end{example}

The discussion in the Introduction and Example~\ref{ex:non-unique} illustrate that the ambiguity in defining nonlinear moments stems from the lack of uniqueness for the solutions to the invariance equation. Thus, the goal of this paper is to discover additional properties of these possibly many solutions using the L--P approach and to define the nonlinear moment as a single mathematical object based on this understanding.

Before we close this section, we show that the polynomial ansatz solution of the invariance equation is a good approximation of the solution to the invariance equation around the origin. To obtain the ansatz, we introduce several definitions. For $j\in\Z_{>0}$, let $\mathcal P_j$ be the space of homogeneous polynomials of degree $j$ from $\mathbb R^\nu$ to $\mathbb R^n$, and define the \emph{homological operator} $\mathcal L_j:\mathcal P_j\to\mathcal P_j$ by
\begin{equation}\label{eq:homological-operator}
    (\mathcal L_jp)(\omega):=Dp(\omega)S\omega-Ap(\omega).
\end{equation}
For each $j \in \Z_{>0}$, $\mathcal L_j$ is a linear operator on the finite-dimensional space $\mathcal P_j$, and it holds that \cite[Chapter 5]{Arnold1988}, \cite{broer2022normal} 
\begin{equation}\label{eq:homological-operator-spectrum}
\sigma(\mathcal L_j)=\{\lambda_1+\cdots+\lambda_j - \mu: \lambda_1,\cdots, \lambda_j \in \sigma(S),~\mu \in \sigma(A) \}.
\end{equation}
Additionally, for a neighborhood $U \subset \R^\nu$ around the origin, we define the residual operator $\mathcal R: \mathcal C^1(U, \R^n) \rightarrow \mathcal C^0 (U, \R^n)$, which calculates the discrepancy between both sides of the invariance equation \eqref{eq:inv},
\begin{equation}\label{eq:residual-operator}
    \mathcal R(q)(\omega):=Dq(\omega)s(\omega)-f(q(\omega),l(\omega)),
\end{equation}
so that $\pi$ solves the invariance equation precisely when $\mathcal R(\pi)\equiv0$ and $\pi(0)=0$.

\begin{lemma}\label{lem:lower-order-property-for-E}
Let $j\in\Z_{>0}$, let $q$ be a smooth function defined around the origin with $q(0)=0$, and let $p\in\mathcal P_j$. Then
\begin{equation}\label{eq:homological-identity}
    \mathcal R(q+p)(\omega) - \mathcal R(q)(\omega)=\mathcal L_jp(\omega)+O(\|\omega\|^{j+1}).
\end{equation}
In particular,
\begin{equation}\label{eq:residual-taylor}
T_{j-1} \mathcal R(q+p) = T_{j-1} \mathcal R(q),~T_{j} \mathcal R(q+p) = T_{j} \mathcal R(q)+\mathcal L_jp.
\end{equation}
\end{lemma}
\begin{proof}
From the definition of $\mathcal R$, we have
\begin{equation}\label{eq:lem1-0}
\mathcal R(q+p)(\omega) - \mathcal R(q)(\omega) = D p(\omega)s(\omega) -f(q(\omega)+p(\omega),l(\omega)) + f(q(\omega), l(\omega))
\end{equation}
Then, from $s(\omega)=S\omega+O(\|\omega\|^{2})$ and $Dp(\omega)=O(\|\omega\|^{j-1})$ we obtain
\begin{equation}\label{eq:lem1-1}
Dp(\omega)s(\omega) =Dp(\omega)S\omega+O(\|\omega\|^{j+1}).
\end{equation}
Since $q(\omega) = O(\|\omega\|)$, $l(\omega) = O(\|\omega\|)$---because both are $\mathcal C^1$ and equal zero at the origin---and $p(\omega)=O(\|\omega\|^{j})$, we obtain
\begin{equation}\label{eq:bounding-arguments}
\|(q(\omega)+\vartheta p(\omega), l(\omega))\|=O(\|\omega\|)
\end{equation}
uniformly in $\vartheta\in[0,1]$. Thus,
\begin{align}
\begin{split}\label{eq:lem1-2}
f(q(\omega)+p(\omega),l(\omega))-f(q(\omega),l(\omega)) &= \int_0^1D_xf(q(\omega)+\vartheta p(\omega),l(\omega))p(\omega) d\vartheta \\ &=Ap(\omega)+O(\|\omega\|^{j+1}),
\end{split}
\end{align}
in which we have used $D_x f(x,u) = A + O(\|(x,u)\|)$ together with the estimate \eqref{eq:bounding-arguments} in the second equality. Substituting the identities \eqref{eq:lem1-1} and \eqref{eq:lem1-2} into the equation \eqref{eq:lem1-0} {gives the identity \eqref{eq:homological-identity}. Since $\mathcal R(q+p)-\mathcal R(q)-\mathcal L_jp$ is smooth and $O(\|\omega\|^{j+1})$, its Taylor polynomial of order $j$ vanishes, and the identities \eqref{eq:residual-taylor} follow from $\mathcal L_jp\in\mathcal P_j$.}
\end{proof}

We now construct the polynomial ansatz. Let $N\in\Z_{>0}$ and assume that the homological operators $\mathcal L_1,\ldots,\mathcal L_N$ are invertible. Set $q_0:=0$. For $j=1,\ldots,N$, we wish to define $p_j\in\mathcal P_j$ and $q_j$ recursively by
\begin{equation}\label{eq:qN-recursion}
    \mathcal L_jp_j=-T_j\mathcal R(q_{j-1}),~q_j:=q_{j-1}+p_j.
\end{equation}
Since $\mathcal L_j:\mathcal P_j\to\mathcal P_j$, this is possible only if $T_j\mathcal R(q_{j-1})\in\mathcal P_j$ for $j=1,\ldots,N$, which is shown in Lemma~\ref{lem:recursion}. Since $p_j$ does not depend on $N$, the polynomials $q_1,\ldots,q_N$ are the same for every $N$ for which $\mathcal L_1,\ldots,\mathcal L_N$ are invertible. Thus, if $\mathcal L_N$ is invertible for all $N \in \Z_{>0}$, we can construct the formal power series $\hat q$.

\begin{lemma}\label{lem:recursion}
Let $N\in\Z_{>0}$ and assume that $\mathcal L_1,\ldots,\mathcal L_N$ are invertible. Then, for $j=1,\ldots,N$, $T_{j-1}\mathcal R(q_{j-1})=0$, so that $T_j\mathcal R(q_{j-1})\in\mathcal P_j$ and $p_j$ in the recursion \eqref{eq:qN-recursion} is well defined, and $T_j\mathcal R(q_j)=0$.
\end{lemma}

\begin{proof}
We use induction on $j$. For $j=1$, $T_0\mathcal R(q_0)=\mathcal R(0)(0)=-f(0,0)=0$. Let $j\in\{1,\ldots,N\}$ and assume $T_{j-1}\mathcal R(q_{j-1})=0$. Then $T_j\mathcal R(q_{j-1})$ has no terms of degree less than $j$, that is, $T_j\mathcal R(q_{j-1})\in\mathcal P_j$, and $p_j$ is well defined since $\mathcal L_j$ is invertible. By the identities \eqref{eq:residual-taylor} with $q=q_{j-1}$ and $p=p_j$, we obtain $T_j\mathcal R(q_j)=T_j\mathcal R(q_{j-1})+\mathcal L_jp_j=0$, which also gives the induction hypothesis for $j+1$.
\end{proof}

\begin{theorem}\label{thm:approximate-jet}
Let $N \in \Z_{>0}$, assume that the homological operators $\mathcal L_1, \ldots, \mathcal L_N$ are invertible, and let $q_N$ be defined by the recursion \eqref{eq:qN-recursion}. Then $q_N$ is the unique element of $\mathcal P_1 \oplus \cdots \oplus \mathcal P_N$ that satisfies
\begin{equation}\label{eq:qN-residual}
    \mathcal R(q_N)(\omega)=O(\|\omega\|^{N+1}).
\end{equation}
\end{theorem}

\begin{proof}
Since $\mathcal R(q_N)$ is smooth and $T_N\mathcal R(q_N)=0$ by Lemma~\ref{lem:recursion}, $q_N$ satisfies the estimate \eqref{eq:qN-residual}. Conversely, let $q'=p_1'+\cdots+p_N'$ with $p_j'\in\mathcal P_j$ satisfy $\mathcal R(q')(\omega)=O(\|\omega\|^{N+1})$, so that $T_N\mathcal R(q')=0$, and let $q_j':=p_1'+\cdots+p_j'$ for $j=0,\ldots,N$. We show $p_j'=p_j$ by induction on $j$. Let $j\in\{1,\ldots,N\}$ and assume $q_{j-1}'=q_{j-1}$. Applying the first of the identities \eqref{eq:residual-taylor} with $p=p_i'$ for $i=N,N-1,\ldots,j+1$ and using $T_j=T_j\circ T_{i-1}$ for $i\ge j+1$, we obtain $T_j\mathcal R(q_j')=T_j\mathcal R(q')=0$. The second of the identities \eqref{eq:residual-taylor} with $q=q_{j-1}$ and $p=p_j'$ then gives $0=T_j\mathcal R(q_{j-1})+\mathcal L_jp_j'$. Comparing with the recursion \eqref{eq:qN-recursion} yields $\mathcal L_jp_j'=\mathcal L_jp_j$, and therefore $p_j'=p_j$ since $\mathcal L_j$ is invertible. Thus $q'=q_N$.
\end{proof}

\begin{proposition}\label{prop:unique-jets}
    Let $N \in \Z_{>0}$, let the homological operators $\mathcal L_1, \ldots, \mathcal L_N$ be invertible, let $q_N$ be as in Theorem~\ref{thm:approximate-jet}, and let $\pi$ be a $\mathcal C^N$ solution to the invariance equation. Then, $T_N \pi = q_N$.
\end{proposition}
\begin{proof}
Let $r_N:= \pi - T_N \pi$. Then $r_N(\omega) = o(\| \omega\|^N)$ and $D r_N(\omega) = o(\| \omega\|^{N-1})$ by Taylor's theorem. On the other hand, we have $\mathcal R(T_N \pi)(\omega) =\mathcal R(T_N \pi)(\omega)-\mathcal R(\pi)(\omega) =-Dr_N(\omega)s(\omega) + \int_0^1D_xf(T_N \pi(\omega)+\vartheta r_N(\omega), l(\omega))r_N(\omega) d\vartheta,$ since $\mathcal R(\pi)\equiv0$. The first term is $o(\|\omega\|^{N})$ because $s(\omega)=O(\|\omega\|)$, and the second is $o(\|\omega\|^{N})$ because $\|D_xf\|$ is bounded near the origin. Since $\mathcal R(T_N \pi)$ is smooth, $\mathcal R(T_N \pi)=O(\|\omega\|^{N+1})$, and the proof is complete by Theorem~\ref{thm:approximate-jet}.
\end{proof}

Proposition~\ref{prop:unique-jets} implies that although there might be multiple $\mathcal C^N$ solutions, their Taylor polynomials of degree $N$ coincide, provided that $\mathcal L_1,\ldots,\mathcal L_N$ are invertible. The polynomial $q_N$ will play a central role throughout the paper. 

\section{Lyapunov--Perron construction and the order of smoothness}\label{sec:existence}
To apply the L--P fixed-point argument, we assume that the system and the exosystem satisfy the strip condition.

\begin{definition}
    \label{def:strip-condition}
    Consider the system \eqref{eq:nl} and the exosystem \eqref{eq:nl-exo}, and let
    \begin{align}
    \begin{split}\label{eq:spectral-params}
        &\alpha_+ := \max_{\lambda\in\sigma(S)\cup\{0\}}\mathrm{Re}\lambda,~\alpha_- := \max_{\lambda\in\sigma(S)\cup\{0\}}\left(-\mathrm{Re}\lambda\right), \\ 
        &\kappa_- := \begin{cases}
            \min_{\mu\in\sigma(A), \mathrm{Re}\mu\le 0}\left(-\mathrm{Re}\mu \right), & \text{if}~\{\mu\in\sigma(A): \mathrm{Re}\mu\le 0\} \neq \varnothing, \\ \infty, &\text{if}~ \{\mu\in\sigma(A): \mathrm{Re}\mu\le 0\} = \varnothing,
        \end{cases} \\ & \kappa_+ :=\begin{cases}
            \min_{\mu\in\sigma(A),\mathrm{Re}\mu\ge 0}\mathrm{Re}\mu,&\text{if}~\{\mu\in\sigma(A):\mathrm{Re}\mu\ge 0\} \neq \varnothing, \\
            \infty,&\text{if}~\{\mu\in\sigma(A):\mathrm{Re}\mu\ge 0\} = \varnothing.
        \end{cases}
    \end{split}
    \end{align}
    The system and the exosystem satisfy the \emph{strip condition} if 
    \begin{equation}\label{eq:strip}
    \alpha_- < \kappa_-,~\alpha_+ < \kappa_+.
    \end{equation}
\end{definition}
The condition is named after the following geometric interpretation: the strip condition holds if and only if $\sigma(S)\cup i\R$ is contained in the open vertical strip $\{z\in\C:-\kappa_-<\mathrm{Re}z<\kappa_+\}$, while $\sigma(A)$ lies outside this strip by the definitions of $\kappa_-$ and $\kappa_+$. Thus the strip separates $\sigma(S)$ from $\sigma(A)$. Since $\kappa_->0$ and $\kappa_+>0$ under the strip condition, $A$ is hyperbolic and $\R^n=E_-\oplus E_+$, where $E_-=\{0\}$ if $\kappa_-=\infty$ and $E_+=\{0\}$ if $\kappa_+=\infty$. For every $\kappa_-' \in (0,\kappa_-)$ and $\kappa_+' \in (0,\kappa_+)$,
there exists $K = K(\kappa_-',\kappa_+') > 0$ such that
\begin{equation}\label{eq:dichotomy}
  \|e^{A_- t}P_-\| \le K e^{-\kappa_-' t},~
  \|e^{-A_+ t}P_+\| \le K e^{-\kappa_+' t},~ \forall t \ge 0.
\end{equation}

The main objective of this section is to prove Theorem~\ref{thm:existence}.
\begin{theorem}\label{thm:existence}
    Consider the system \eqref{eq:nl} and the exosystem \eqref{eq:nl-exo} that satisfy the strip condition. Then, for every $N \in \Z_{>0}$ that satisfies
    \begin{equation}\label{eq:N-cond}
    N\alpha_- < \kappa_-,~N\alpha_+ < \kappa_+,
    \end{equation}
    there exist a neighborhood $U_N$ of the origin and a map $\pi_N\in\mathcal C^N(U_N,\mathbb R^n)$ that satisfies the invariance equation \eqref{eq:inv} for all $\omega \in U_N$. 
\end{theorem}

Theorem~\ref{thm:existence} can be regarded as a version, adapted to the cascade, of the classical existence results for invariant manifolds under spectral gap conditions \cite{irwin1980pseudostable,cabre2003parameterization1}. We include its proof because the construction and the estimates in Appendix~\ref{app:existence} are used in Remark~\ref{rem:boundary-condition} and estimates in Lemma~\ref{lem:cutoff} are used in Lemma~\ref{lem:order-raising} and Proposition~\ref{prop:transient}.

If $N\in \Z_{>0}$ satisfies the condition \eqref{eq:N-cond}, then $\mathcal L_1,\ldots,\mathcal L_N$ are all invertible, and we can construct $q_N$ as in Theorem~\ref{thm:approximate-jet}. Specifically, for $j=1,\ldots,N$, $\lambda_1,\ldots,\lambda_j \in \sigma(S)$, $ - N \alpha_- \le -j\alpha_- \le \text{Re}(\lambda_1 + \cdots + \lambda_j) \le j \alpha_+\le N\alpha_+$. Since $A$ is hyperbolic, every $\mu\in\sigma(A)$ belongs to $\sigma(A_-)$ or to $\sigma(A_+)$. If $\mu \in \sigma(A_-)$, then $E_- \neq \{0\}$, $\mathrm{Re}\mu\le-\kappa_-$, and from the first inequality in the condition \eqref{eq:N-cond} we get $\text{Re}(\lambda_1 +\cdots + \lambda_j - \mu) \ge \kappa_- - N \alpha_- >0$; a similar argument shows that $\text{Re}(\lambda_1 + \cdots + \lambda_j - \mu) \le N \alpha_+ - \kappa_+ < 0$ if $\mu \in \sigma(A_+)$. Therefore $\mathcal L_j$ is invertible for $j=1,\ldots,N$. As a consequence of Proposition~\ref{prop:unique-jets}, $T_N \pi_N = q_N$. Since $\sigma(S) \cap \sigma(A) = \varnothing$ is equivalent to the invertibility of $\mathcal L_1$, the Sylvester equation has the unique solution $\Pi$. Note that $N=1$ always satisfies the condition \eqref{eq:N-cond} under the strip condition, so the existence of a local $\mathcal C^1$ solution is guaranteed. Since $\sigma(S)$ is not assumed to lie on the imaginary axis, the standard center-manifold argument is not directly applicable. The proof proceeds in two steps: we first remove a polynomial approximation of order $N$ to the solution and then construct the resulting higher-order remainder by the L--P fixed-point argument. The $\mathcal C^N$ smoothness is obtained by proving that the remainder is $\mathcal C^N$.

Let $N \in \Z_{>0}$ satisfy the condition \eqref{eq:N-cond} and let $q_N$ be given by Theorem~\ref{thm:approximate-jet}. Writing $\pi(\omega)=q_N(\omega)+r(\omega),$ the invariance equation is equivalent to
\begin{equation}\label{eq:remainder-invariance}
    Dr(\omega)s(\omega)=Ar(\omega)+\mathcal F_N(\omega,r(\omega)),~r(0)=0,
\end{equation}
where
\begin{align}
\mathcal F_N(\omega,z):= f(q_N(\omega)+z,l(\omega))-f(q_N(\omega),l(\omega))-Az-\mathcal R(q_N)(\omega).
\label{eq:FN-def}
\end{align}

The main idea for the construction is described as follows. Denote the flow generated by the vector field $s$ that starts at $\omega$ at $t=0$ as $\varphi_t(\omega)$ and abbreviate it as $\omega_t$. Suppose that there exists a $\mathcal C^N$ solution $r$ to the equation \eqref{eq:remainder-invariance} and the flow $\varphi$ is complete. Then $\frac{d}{dt}r(\omega_t) = Ar(\omega_t) + \mathcal{F}_{N}(\omega_t , r(\omega_t))$. If $e^{A_-t} P_-r(\omega_{-t})$ and $e^{-A_+ t}P_+r(\omega_{t})$ converge to zero as $t$ goes to infinity, the variation-of-constants formula applied to $P_+ r(\omega_t)$ and $P_- r(\omega_t)$ gives
\begin{equation}\label{eq:fixed-point}
r(\omega) = \int_0^\infty e^{A_-t}P_- \mathcal F_{N}(\omega_{-t},r(\omega_{-t})) dt -\int_0^\infty e^{-A_+t}P_+ \mathcal F_{N}(\omega_t,r(\omega_t)) dt.
\end{equation}
Thus, if we define the L--P operator $\mathcal I_{N}$ as the operator that maps $r$ to the right-hand side of the formula \eqref{eq:fixed-point} 
and find the fixed point of the operator $\mathcal I_N$, then it will be a solution to the equation \eqref{eq:remainder-invariance}. However, this is just a sketch because in general the flow $\omega_t$ is not complete. Additionally, we have to check whether any fixed-point theorems are applicable and check the order of smoothness of this fixed point. To resolve these obstacles, we truncate the vector field $s$ and the function $\mathcal F_N$ as follows.

Let $\chi$ be a smooth \emph{cutoff function} satisfying $\chi(z)=1$ for $\|z\|\le1$ and $\chi(z)=0$ for $\|z\|\ge2$~\cite[Section 13]{tu2011manifolds}; the same symbol is used for cutoff functions on spaces of different dimensions. For a \emph{truncation radius} $\rho>0$, define
\begin{equation}\label{eq:truncated}
\begin{aligned}
    s_\rho(\omega)
        &:=S\omega+\chi\!\left(\frac{\omega}{\rho}\right)(s(\omega)-S\omega),\\
    \mathcal F_{N,\rho}(\omega,z)
        &:=\chi\!\left(\frac{\omega}{\rho}\right)
          \chi\!\left(\frac{z}{\rho}\right)\mathcal F_N(\omega,z).
\end{aligned}
\end{equation}

\begin{lemma}\label{lem:cutoff}
    Let $N\in\Z_{>0}$ satisfy the condition \eqref{eq:N-cond}, and let $s_\rho$ and $\mathcal F_{N,\rho}$ be defined as in the definition \eqref{eq:truncated}. There exists $\rho_{\mathrm{loc}}>0$ with the following properties:

    \noindent\textbf{(a)} For every $\epsilon>0$, there exist $\rho_\epsilon\in(0,\rho_{\mathrm{loc}}]$ and $C_\epsilon>0$ such that for every $\rho \in (0,\rho_\epsilon)$, the vector field $s_\rho$ has a complete flow $\varphi_t^\rho$, and for all $t \ge 0$ and for all $\omega \in \R^\nu$,
    
    \begin{align}
        \|\varphi_t^\rho(\omega)\| & \le C_\epsilon\|\omega\|e^{(\alpha_++\epsilon)t},~\|\varphi_{-t}^\rho(\omega)\| \le C_\epsilon\|\omega\|e^{(\alpha_-+\epsilon)t}, \label{eq:flow-bound} \\
        \|D\varphi_t^\rho(\omega)\| & \le C_\epsilon e^{(\alpha_++\epsilon)t},~\|D\varphi_{-t}^\rho(\omega)\| \le C_\epsilon e^{(\alpha_-+\epsilon)t}. \label{eq:flow-deriv-bound}
    \end{align}
    Furthermore, for each $N \ge 2$ and $\rho \in (0,\rho_{\epsilon})$, there also exist $C_{2,\rho,\epsilon},\ldots,C_{N,\rho,\epsilon}>0$ that depend on $\rho$ such that for $j=2,\ldots,N$, for all $t \ge 0$ and for all $\omega \in \R^\nu$,
    
    \begin{equation}\label{eq:higher-flow-deriv-bound}
        \|D^j\varphi_t^\rho(\omega)\|
        \le C_{j,\rho,\epsilon}e^{j(\alpha_++\epsilon)t},~~\|D^j\varphi_{-t}^\rho(\omega)\|
        \le C_{j,\rho,\epsilon}e^{j(\alpha_-+\epsilon)t}.
    \end{equation}

    \noindent\textbf{(b)} There are constants $C_z>0$ and $C_j>0$, $j=0,\ldots,N$, independent of $\rho\in(0,\rho_{\mathrm{loc}})$, such that for all $(\omega,z)\in\mathbb R^\nu\times\mathbb R^n$ and $j = 0,\ldots, N$,
    \begin{align}
        \|D_z\mathcal F_{N,\rho}(\omega,z)\| &\le C_z\rho, \label{eq:FN-rho-lipschitz} \\
        \|D_\omega^j\mathcal F_{N,\rho}(\omega,0)\| &\le C_j\|\omega\|^{N+1-j}. \label{eq:FN-rho-jet-bound}
    \end{align}
    Furthermore, for each fixed $\rho\in(0,\rho_{\mathrm{loc}})$, $\mathcal F_{N,\rho}$ and its derivatives $D_z^a D_\omega^b \mathcal F_{N,\rho}$ for $a,b \in \Z_{\ge 0}$, $a+b \le N+1$ are bounded. Finally, there is a constant $C_{\mathrm{loc}}>0$ such that
    \begin{equation}\label{eq:FN-z-local-main}
        \|D_z\mathcal F_N(\omega,z)\|\le C_{\mathrm{loc}}(\|\omega\|+\|z\|)~\text{whenever}~\|\omega\|,\|z\|\le 2\rho_{\mathrm{loc}}.
    \end{equation}
\end{lemma}

Here, $\rho_{\mathrm{loc}}$ is a localization radius, which is fixed once and for all and on which the local bound \eqref{eq:FN-z-local-main} holds, whereas the truncation radius $\rho\in(0,\rho_{\mathrm{loc}})$ is at our disposal and is eventually taken small. The proof is given in Appendix~\ref{app:cutoff}. The bounds of Lemma~\ref{lem:cutoff} are used to show that the truncated counterpart of $\mathcal I_N$ is a contraction and that its fixed point is $\mathcal C^N$, and also in Section~\ref{sec:uniqueness} to derive the contact between two solutions to the invariance equation.

Replacing $s$ and $\mathcal F_N$ in the formula \eqref{eq:fixed-point} by their truncated counterparts $s_\rho$ and $\mathcal F_{N,\rho}$, we define the \emph{truncated L--P operator}
\begin{align}\label{eq:T-def}
\begin{split}
(\mathcal I_{\rho,N}r)(\omega)
    :=&\int_0^\infty e^{A_-t}P_-
        \mathcal F_{N,\rho} (\omega_{-t}^\rho,r(\omega_{-t}^\rho) ) dt -\int_0^\infty e^{-A_+t}P_+
        \mathcal F_{N,\rho} (\omega_t^\rho,r(\omega_t^\rho) ) dt,
\end{split}
\end{align}
where $\omega_t^\rho:=\varphi_t^\rho(\omega)$ is defined for all $t\in\R$ by Lemma~\ref{lem:cutoff}\textbf{(a)}. Lemma~\ref{lem:weighted-LP} in Appendix~\ref{app:existence} shows that, for a suitable $\epsilon>0$ and a sufficiently small $\rho \in (0,\rho_\epsilon)$, $\mathcal I_{\rho,N}$ has a fixed point $r_\rho\in\mathcal C^N(\R^\nu,\R^n)$ with $r_\rho(0)=0$.

\begin{proof}[Proof of Theorem~\ref{thm:existence}]
Let $r_\rho$ be the fixed point of $\mathcal I_{\rho,N}$ given by Lemma~\ref{lem:weighted-LP}. Evaluating the identity $r_\rho=\mathcal I_{\rho,N}r_\rho$ at $\omega_t^\rho$ and using the group property of $\varphi_t^\rho$ give
\begin{align*}
P_-r_\rho(\omega_t^\rho)&=\int_{-\infty}^t e^{A_-(t-\tau)}P_-\mathcal F_{N,\rho}(\omega_\tau^\rho,r_\rho(\omega_\tau^\rho))d\tau,\\
P_+r_\rho(\omega_t^\rho)&=-\int_t^{\infty} e^{A_+(t-\tau)}P_+\mathcal F_{N,\rho}(\omega_\tau^\rho,r_\rho(\omega_\tau^\rho))d\tau.
\end{align*}
Differentiating these identities with respect to $t$ at $t=0$, which is possible because $r_\rho\in\mathcal C^1$, gives
\begin{equation}\label{eq:trunc-inv}
    Dr_\rho(\omega)s_\rho(\omega) =Ar_\rho(\omega)+\mathcal F_{N,\rho}(\omega,r_\rho(\omega)),~\forall \omega \in \R^\nu.
\end{equation}
Since $r_\rho$ is continuous and $r_\rho(0)=0$, there exists $\delta\in(0,\rho)$ such that $\|r_\rho(\omega)\|<\rho$ for $\omega\in B_\delta$. Then $\chi(\omega/\rho)=\chi(r_\rho(\omega)/\rho)=1$ on $B_\delta$, so that $s_\rho=s$ and $\mathcal F_{N,\rho}(\omega,r_\rho(\omega))=\mathcal F_N(\omega,r_\rho(\omega))$ there, and the equation \eqref{eq:trunc-inv} reduces to the equation \eqref{eq:remainder-invariance} on $B_\delta$. Therefore $\pi_N:=q_N+r_\rho$ is $\mathcal C^N$ on $U_N:=B_\delta$ and satisfies the invariance equation \eqref{eq:inv}.
\end{proof}

Motivated by Theorem~\ref{thm:existence}, if $A$ is hyperbolic, that is, if $\kappa_->0$ and $\kappa_+>0$, define the \emph{spectral ratio} $k$ between the system and the exosystem, and the associated order $N_{\max}$, by
\begin{equation}\label{eq:Nmax}
    k:= \min \left\{ \frac{\kappa_+}{\alpha_+}, \frac{\kappa_-}{\alpha_-} \right\},~N_{\max}:=
    \begin{cases}
        \infty, & \text{if}~k=\infty,\\
        \max\{N\in\Z_{>0}:N < k\}, & \text{if}~k<\infty.
    \end{cases}
\end{equation}
Since $\kappa_->0$ and $\kappa_+>0$, $\frac{0}{0}$ does not occur, and it is understood that $\frac{c}{0} = \infty$ for $c \in \R_{>0} \cup \{\infty\}$ and that $\frac{\infty}{c}=\infty$ for $c\in\R_{>0}$ when defining $k$. For every $b>0$, $b\alpha_-<\kappa_-$ and $b\alpha_+<\kappa_+$ hold if and only if $b<k$. In particular, the strip condition holds if and only if $A$ is hyperbolic and $k>1$, and the condition \eqref{eq:N-cond} is equivalent to $N<k$. Under the strip condition, $N_{\max}$ is the largest positive integer satisfying the condition \eqref{eq:N-cond}; it therefore determines the maximal order of smoothness supplied by Theorem~\ref{thm:existence}. In Section~\ref{sec:uniqueness}, we will show that the spectral ratio $k$ also determines the order of contact of solutions to the invariance equation.

\begin{corollary}\label{thm:main-existence}
    Consider the system \eqref{eq:nl} and the exosystem \eqref{eq:nl-exo} satisfying the strip condition, and define $k$ and $N_{\max}$ by the definition \eqref{eq:Nmax}.

    \noindent\textbf{(a)} If $k<\infty$, the invariance equation admits a local $\mathcal C^{N_{\max}}$ solution.

    \noindent\textbf{(b)} If $k=\infty$, then for every $N\in\Z_{>0}$ the invariance equation admits a local $\mathcal C^N$ solution. 
\end{corollary}

Part \textbf{(b)} does not imply the existence of a single $\mathcal C^\infty$ solution. For instance, \cite{van1979center} shows that the interconnection of the exosystem $\dot \omega_1 = \omega_1^2 -\omega_2^2,\dot{\omega}_2=0,u=\omega_1^2 -\omega_2^2$ and the scalar system $\dot x=x+u$, for which $\alpha_+=\alpha_-=0$, $\kappa_+=1$, and $\kappa_-=\infty$, so that $k=\infty$, admits a $\mathcal C^N$ invariant manifold for every $N \in \Z_{>0}$ but none of them is $\mathcal C^\infty$.

At this point, we pose two questions.
\begin{itemize}
    \item \emph{Under the hypotheses of Corollary~\ref{thm:main-existence}, is it possible to derive higher-order smoothness than $\mathcal C^{N_{\max}}$ when $N_{\max}<\infty$?}
    \item \emph{If $N_{\max}>1$, can every $\mathcal C^1$ solution be asserted to be $\mathcal C^{N_{\max}}$?}
\end{itemize}
The answer to both questions is negative without additional assumptions, as Examples~\ref{ex:counterexample-1} and~\ref{ex:counterexample-2} show. The order of smoothness supplied by Theorem~\ref{thm:existence} is therefore intrinsic to the problem rather than an artifact of the construction.

\begin{example}\label{ex:counterexample-1}
    Let $N\in\Z_{>0}$ and consider the scalar system and exosystem $\dot x=(N+1)x+u^{N+1};~\dot\omega=\omega,~u=\omega.$ Here $\alpha_+=1$, $\alpha_-=0$, $\kappa_+=N+1$, and $\kappa_-=\infty$, so that $k=N+1$ and $N_{\max}=N$, and the invariance equation is $\omega D\pi(\omega)=(N+1)\pi(\omega)+\omega^{N+1}.$ Its general solution is $\pi_c(\omega)=c\omega^{N+1}+\omega^{N+1}\log|\omega|$, with $\pi_c(0)=0$. Here, $\pi_c$ is $\mathcal C^N$ but not $\mathcal C^{N+1}$; the order $N_{\max}$ cannot be improved in general. 
\end{example}

\begin{example}\label{ex:counterexample-2}
    Let $N\ge2$ and $\mu\in(N,N+1)$. Consider the scalar system $\dot x=\mu x+u$ and the two-dimensional exosystem $\dot\omega_1=\omega_1,~\dot\omega_2=\tfrac12\omega_2,~u=\omega_1+\omega_2$, so that $\alpha_+=1$, $\alpha_-=0$, $\kappa_+=\mu$, $\kappa_-=\infty$, $k=\mu$, and $N_{\max}=N$. The invariance equation is $\omega_1D_1\pi(\omega)+\tfrac12\omega_2D_2\pi(\omega) =\mu\pi(\omega)+\omega_1+\omega_2.$ The function $\omega \mapsto \Pi\omega$ with $\Pi=\left(\frac{1}{1-\mu},\frac{2}{1-2\mu}\right)$ is a particular solution, and $\delta\pi(\omega)=|\omega_1|^{\mu-3/4}|\omega_2|^{3/2}$
    solves the homogeneous equation. Thus, $\Pi\omega+\delta\pi(\omega)$ is a $\mathcal C^1$ solution that is not $\mathcal C^2$, although $N_{\max}=N\ge2$. 
\end{example}

The classical case when the exosystem is Poisson stable and the system is exponentially stable, and more generally, when the exosystem has purely imaginary spectrum and the system is hyperbolic so that the existence of the invariant manifold of the form $x = \pi(\w)$ can be guaranteed by the center manifold theorem, corresponds to the case $k = \infty$. However, $k = \infty$ is possible even when $\sigma(S) \not\subset i\R$, as Example~\ref{ex:hyperbolic-S-but-infinite-k} shows.
\begin{example}\label{ex:hyperbolic-S-but-infinite-k}
For the interconnection of $\dot{x} = -x+u,~y=x$ and $\dot \w=3\w, u=\w$, we have $\alpha_+ = 3$, $\alpha_- = 0$, $\kappa_- = 1$, and $\kappa_+ = \infty$, so that $k=\infty$. In this example, the invariance equation $3\w D\pi(\w)=-\pi(\w)+\w$ has the general solution $\pi(\w)=\w/4+c|\w|^{-1/3}$ on each half-line, so that $\pi(\w)=\w/4$, which defines the unstable manifold of the cascade, is the unique continuous solution. 
\end{example}

\begin{remark}\label{rem:boundary-condition}
From the L--P viewpoint, the linear moment is special among the solutions to the invariance equation: it is selected by a \emph{global} fixed point of the L--P operator, without any truncation. To make this precise, suppose that the flow $\omega_t$ of $s$ is complete, let $N\in\Z_{>0}$ satisfy the condition \eqref{eq:N-cond}, let $q_N$ be as in Theorem~\ref{thm:approximate-jet}, let $\gamma$ satisfy $N<\gamma<\min\{N+1,k\}$, and let $\mathcal I_N$ map a function $r$ to the right-hand side of the formula \eqref{eq:fixed-point}. By the variation-of-constants argument leading to the formula \eqref{eq:fixed-point}, every $\mathcal C^1$ solution $r:\R^\nu\to\R^n$ of the equation \eqref{eq:remainder-invariance} satisfying
\begin{equation}\label{eq:LP-limit}
    \lim_{t\to\infty}e^{A_-t}P_-r(\omega_{-t})=0,\quad \lim_{t\to\infty}e^{-A_+t}P_+r(\omega_t)=0
\end{equation}
is a fixed point of $\mathcal I_N$. Conversely, suppose that the flow $\omega_t$ of $s$ satisfies the estimates \eqref{eq:flow-bound}--\eqref{eq:higher-flow-deriv-bound} for all $\omega\in\R^\nu$ and $t\ge0$, with $\varphi_{t}^\rho$ and $\varphi_{-t}^\rho$ replaced by $\omega_{t}$ and $\omega_{-t}$, respectively, that $\mathcal I_N$ maps $\mathcal X_\gamma$ into itself and is a contraction on $\mathcal X_\gamma$, and that, for $j=1,\ldots,N$, the operators $\mathcal A^{(j)}_r$ and $\mathcal H^{(j)}[r]$ of Appendix~\ref{app:existence}, defined with $s$ and $\mathcal F_N$ in place of $s_\rho$ and $\mathcal F_{N,\rho}$, satisfy the estimates established there. Then the arguments in the proof of Theorem~\ref{thm:existence} yield a unique fixed point $r\in\mathcal X_\gamma$ of $\mathcal I_N$, which is $\mathcal C^N$ and, since $\gamma<k$, satisfies the limit conditions \eqref{eq:LP-limit}; thus $q_N+r$ is a distinguished $\mathcal C^N$ solution to the invariance equation defined on $\R^\nu$. Any other solution $q_N+\tilde r$ with $\tilde r\in\mathcal X_\gamma$ violates the limit conditions \eqref{eq:LP-limit}, since otherwise $\tilde r$ would be a second fixed point of $\mathcal I_N$ in $\mathcal X_\gamma$; this argument does not exclude solutions satisfying \eqref{eq:LP-limit} with $\tilde r\notin\mathcal X_\gamma$.

All of these hypotheses hold for linear systems and exosystems: the flow of $S\omega$ is complete, $q_N(\omega)=\Pi\omega$ for every $N$, and $\mathcal F_N\equiv0$, so that $\mathcal I_N$ maps every $r$ to zero and $r\equiv0$ is its unique fixed point, in $\mathcal X_\gamma$ as well as in any other class of functions. Hence the invariant hyperplane $\{(\omega,x):x=\Pi\omega\}$ is the solution selected by the global L--P fixed point, whereas the other solutions violate the limit conditions \eqref{eq:LP-limit}. For instance, in Example~\ref{ex:non-unique}, $E_-=\{0\}$, $k=2$, $N=1$, and $q_1(\omega)=-\omega$, so that the fixed point $r=0$ corresponds to $\pi_0$, whereas, for $c\neq0$, $r_c:=\pi_c-q_1$ violates the limit conditions, since $
\lim_{t \to \infty}e^{-2t}r_c(\omega_t)=c\omega^2\neq0$.
\end{remark}

\begin{remark}
The L--P argument also underlies the construction of the polynomial part $q_N$. Let $N$ satisfy the condition \eqref{eq:N-cond}, choose $\epsilon>0$ and $\kappa_-'$, $\kappa_+'$ in the inequality \eqref{eq:dichotomy} so that $N(\alpha_-+\epsilon)<\kappa_-'$ and $N(\alpha_++\epsilon)<\kappa_+'$, and let $j\in\{1,\ldots,N\}$ and $g\in\mathcal P_j$. Since there exists $C>0$ such that $\|g(e^{-St}\omega)\|\le Ce^{j(\alpha_-+\epsilon)t}\|\omega\|^j$ and $\|g(e^{St}\omega)\|\le Ce^{j(\alpha_++\epsilon)t}\|\omega\|^j$ for all $t\ge0$, the inequality \eqref{eq:dichotomy} implies that
\begin{equation}\label{eq:homological-LP}
    p(\omega):=\int_0^\infty e^{A_-t}P_-g(e^{-St}\omega)dt-\int_0^\infty e^{-A_+t}P_+g(e^{St}\omega)dt
\end{equation}
is well-defined, and $p\in\mathcal P_j$. A direct computation along the flow $\tau\mapsto e^{S\tau}\omega$ shows that $Dp(\omega)S\omega=Ap(\omega)+g(\omega)$, that is, $p=\mathcal L_j^{-1}g$. In particular, for $j=1$ and $g(\omega)=BL\omega$, the formula \eqref{eq:homological-LP} gives the L--P representation of the unique solution to the Sylvester equation,
\begin{equation}\label{eq:Pi-LP}
\Pi=\int_0^\infty e^{A_-t}P_-BLe^{-St}dt-\int_0^\infty e^{-A_+t}P_+BLe^{St}dt.
\end{equation}
Together with Remark~\ref{rem:boundary-condition}, this shows that, in the linear case, both parts of the solution $q_N+r=\Pi\omega$ are given by the L--P argument: the polynomial part by the formula \eqref{eq:Pi-LP}, and the remainder by the global fixed point $r\equiv0$.
\end{remark}

\section{Equivalence class definition of nonlinear moments}\label{sec:uniqueness}
The main objective of this section is to show that every $\mathcal C^1$ solution to the invariance equation and its composition with the system output map $h$ form a contact class of order $k^-$, and to define the nonlinear moment as this contact class. In Section~\ref{subsec:on-uniqueness}, we relate this definition to the classical definition of the nonlinear moment \cite{astolfi2010model}.

\begin{definition}\label{def:finite-order-flat-equivalence}
Let $n_1,n_2 \in \Z_{>0}$, $a >0$, let $U_1,U_2 \subset \R^{n_1}$ be open neighborhoods of the origin, and let $g_1:U_1 \rightarrow \R^{n_2}$, $g_2:U_2 \rightarrow \R^{n_2}$. If there exists an open neighborhood $U \subset U_1 \cap U_2$ of the origin such that $\left( \left. g_1 \right|_U - \left. g_2 \right|_U \right)(\omega) = O(\|\omega\|^a)$, then we say that $g_1$ and $g_2$ \emph{have contact of order $a$} and write $g_{1}\sim_a g_{2}$. If $g_1$ and $g_2$ have contact of order $b$ for all $b \in (0,a)$, then we say that $g_1$ and $g_2$ \emph{have contact of order $a^-$} and write $g_{1}\sim_{a^-} g_{2}$. In particular, we say that $g_1$ and $g_2$ are \emph{flat-equivalent} if $g_1 \sim_{\infty^-} g_2$, that is, if $g_1-g_2$ is flat at the origin in the sense that it is $O(\|\omega\|^b)$ for every $b>0$. 
\end{definition}

We mention that $\sim_{a}$ for all $a \in \R_{>0}$ and $\sim_{a^-}$ for all $a \in \R_{>0} \cup \{\infty\}$ are equivalence relations, i.e., reflexive, symmetric, and transitive. Therefore, these relations induce equivalence classes. Given a function $\pi$ defined around the origin and $a \in \R_{>0} \cup \{\infty\}$, we denote the equivalence class of $\pi$ induced by the equivalence relation $\sim_{a^-}$ as $[\pi]_{a^-}$ and call it the \emph{contact class of order $a^-$} of $\pi$; if $a=\infty$, we call it the \emph{flat-equivalence class} of $\pi$. By definition, the order of contact can always be lowered, i.e., if $0<a<b,$
\begin{equation}\label{eq:basic-prop-for-equiv-1}
    g_1\sim_b g_2 \Rightarrow g_1\sim_a g_2,~g_1\sim_{b^-}g_2 \Rightarrow g_1 \sim_{a^-} g_2,
\end{equation}
but in general, cannot be raised. Lemma~\ref{lem:order-raising}, which is proved in Appendix~\ref{app:order-raising}, is used to raise the order of contact throughout this paper.

\begin{lemma}\label{lem:order-raising}
    Consider the system \eqref{eq:nl} and the exosystem \eqref{eq:nl-exo} satisfying the strip condition, and define $k$ as in the definition \eqref{eq:Nmax}. Let $0<a<b\le a+1$ with $b<k$, and let $U\subset\R^\nu$ be a neighborhood of the origin. Assume that $z\in\mathcal C^1(U,\mathbb R^n)$, $\Delta\in\mathcal C^0(U,\R^{n\times n})$, and $g\in\mathcal C^0(U,\R^n)$ satisfy
    \begin{equation}\label{eq:order-raising-equation}
        Dz(\omega)s(\omega)
        =(A+\Delta(\omega))z(\omega)+g(\omega),~\forall\omega\in U,
    \end{equation}
    and
    \begin{equation}\label{eq:order-raising-assumptions}
        z(\omega)=O(\|\omega\|^a),~\Delta(\omega)=O(\|\omega\|),~g(\omega)=O(\|\omega\|^b).
    \end{equation}
    Then, $z(\omega)=O(\|\omega\|^b)$. 
\end{lemma}

\begin{theorem}
\label{thm:finite-order}
    Consider the system \eqref{eq:nl} and the exosystem \eqref{eq:nl-exo} satisfying the strip condition, and define $k$ as in the definition \eqref{eq:Nmax}. Then any pair of local $\mathcal C^1$ solutions $\pi^1$ and $\pi^2$ to the invariance equation satisfies $\pi^1 \sim_{k^-} \pi^2$. In particular, all $\mathcal C^1$ solutions to the invariance equation belong to a single contact class of order $k^-$. 
\end{theorem}

\begin{proof}
    Define $\delta \pi:= \pi^2 - \pi^1$ and fix $k^\prime \in (1,k)$. It is enough to prove $\delta \pi(\omega) = O(\| \omega\|^{k^\prime})$, since $k^\prime$ is arbitrary in $(1,k)$ and the implications \eqref{eq:basic-prop-for-equiv-1} cover the orders below $1$. Let $N \in \Z_{>0}$ be such that $k^\prime \in (N,N+1]$. Proposition~\ref{prop:unique-jets} with $N=1$ gives $\delta \pi(0) = 0,~D\delta \pi(0) = 0,$ which implies $\delta \pi(\omega) = O(\| \omega\|)$. By the fundamental theorem of calculus, $D\delta \pi(\omega) s(\omega) = \left( A + \Delta(\omega) \right) \delta \pi(\omega),$
    where
    \begin{equation}\label{eq:B-def}
        \Delta(\omega)
        :=\int_0^1 [
        D_xf (\pi^1(\omega)+\vartheta\delta\pi(\omega),l(\omega) )-A
         ]d\vartheta.
    \end{equation}
    Since $f$ is smooth, $l(0)=0$, and $\pi^i(\omega)=O(\|\omega\|)$, we obtain $\Delta(\omega) = O(\|\omega\|)$.

    Starting from $\delta \pi(\omega)=O(\|\omega\|)$, apply Lemma~\ref{lem:order-raising} successively. Specifically, if $N \ge 2$, apply Lemma~\ref{lem:order-raising} to the pairs $(a,b) = (1,2),(2,3),\ldots,(N-1,N)$ and then $(N , k^\prime)$; if $N = 1$, apply Lemma~\ref{lem:order-raising} once with $(a,b) = (1,k^\prime)$. In every case the zero function takes the role of $g$ in Lemma~\ref{lem:order-raising}. Successive application is legitimate because each pair satisfies $0<a<b\le a+1$, because $0 = O(\|\omega\|^b)$, and because $b\le k^\prime$ and $k^\prime<k$. The successive application yields $\delta \pi(\omega) = O(\|\omega\|^{k^\prime})$, and the proof is complete.
\end{proof}

Note that Example~\ref{ex:hyperbolic-S-but-infinite-k} is consistent with Theorem~\ref{thm:finite-order} since $\pi(\w) = \w/4$ is the only local $\mathcal C^1$ solution to the invariance equation.

When $\sigma(S)\subset i\R$, the graph of a solution is a center manifold of the cascade, and Theorem~\ref{thm:finite-order} recovers the classical fact that distinct center manifolds differ only by flat terms \cite{sijbrand1985properties}. Combined with Proposition~\ref{prop:unique-jets}, Theorem~\ref{thm:finite-order} also complements the uniqueness result of the parameterization method \cite[Theorem~1.2]{cabre2003parameterization1}. That result is formulated for maps and applies to flows through the time-$t$ map \cite[Remark~6]{cabre2003parameterization1}. For the cascade with $\sigma(S)\subset\C_{<0}$, it asserts that, if $N_*>\frac{\max_{\mu\in\sigma(A_-)}|\mathrm{Re}\mu|}{\min_{\lambda\in\sigma(S)}|\mathrm{Re}\lambda|}$ and $\mathcal L_1,\ldots,\mathcal L_{N_*-1}$ are invertible, then the invariance equation has, locally, a unique solution among the Lipschitz functions of the form $q_{N_*-1}+O(\|\omega\|^{N_*})$, and in particular a unique $\mathcal C^{N_*}$ solution \cite[Theorem~1.2(b1) and (b2)]{cabre2003parameterization1}. Theorem~\ref{thm:finite-order} shows that, without any smoothness assumption or prescribed jet, all $\mathcal C^1$ solutions have contact of order $k^-$. This is further illustrated in Example~\ref{ex:transient-example} in Section~\ref{sec:example}.

\begin{remark}\label{rem:k-minus}
The definitions of $\sim_{k}$ and $\sim_{k^-}$ raise the question whether the conclusion $\sim_{k^-}$ of Theorem~\ref{thm:finite-order} can be strengthened to $\sim_{k}$. In general, this is not possible, as Example~\ref{ex:k-andk} shows. The order $k^-$ originates from two kinds of margins: the margins $\kappa_--\kappa_-'$ and $\kappa_+-\kappa_+'$ in the inequality \eqref{eq:dichotomy}, which absorb the polynomial factors produced by nontrivial Jordan blocks of $A_-$ and $A_+$, and the margin $\epsilon$ in the estimates \eqref{eq:flow-bound}--\eqref{eq:higher-flow-deriv-bound} of Lemma~\ref{lem:cutoff}, which absorbs the polynomial factors produced by nontrivial Jordan blocks of $S$. Only the eigenvalues that realize $k$ are relevant: if $k=\kappa_-/\alpha_-$, these are the eigenvalues $\mu$ of $A$ with $\mathrm{Re}\mu=-\kappa_-$ and the eigenvalues $\lambda$ of $S$ with $\mathrm{Re}\lambda=-\alpha_-$, and if $k=\kappa_+/\alpha_+$, these are the eigenvalues $\mu$ of $A$ with $\mathrm{Re}\mu=\kappa_+$ and the eigenvalues $\lambda$ of $S$ with $\mathrm{Re}\lambda=\alpha_+$. If $A$ and $S$ are semisimple at these eigenvalues, the corresponding polynomial factors are absent, and $\sim_k$ may hold, as Example~\ref{ex:non-unique} shows; whether $\sim_{k^-}$ can always be improved to $\sim_k$ in this case remains open.
\end{remark}

\begin{example}\label{ex:k-andk}
The following two cases illustrate the two mechanisms described in Remark~\ref{rem:k-minus}.

\noindent\textbf{(a) Nontrivial Jordan block of $A$.} Consider the scalar exosystem $\dot\omega=\omega$, $u=\omega$, and the system
\begin{equation*}
    \dot{x} = \begin{bmatrix}2&1\\[3pt]0&2\end{bmatrix}x + \begin{bmatrix}0\\[3pt]1\end{bmatrix}u,
\end{equation*}
for which $\alpha_+=1$, $\alpha_-=0$, $\kappa_+=2$, $\kappa_-=\infty$, and $k=2$, and $S$ is semisimple. The Sylvester equation \eqref{eq:sylvester} gives the solution $\pi^1$, and the function $\delta\pi$ below solves the homogeneous equation $D\delta\pi(\omega) \omega=A \delta\pi(\omega)$:
\begin{equation*}
    \pi^1(\omega)=\begin{bmatrix}\phantom{-}\omega\\[3pt]-\omega\end{bmatrix},\quad
    \delta\pi(\omega)=\begin{bmatrix}\omega^2\log|\omega|\\[3pt]\omega^2\end{bmatrix}.
\end{equation*}
Hence $\pi^2:=\pi^1+\delta\pi$ is a second $\mathcal C^1$ solution.

\noindent\textbf{(b) Nontrivial Jordan block of $S$.} Consider the exosystem
\begin{equation*}
    \dot\omega=\begin{bmatrix}1&1\\[3pt]0&1\end{bmatrix}\omega,~u=\omega_1,
\end{equation*}
and the system $\dot x=2x+u$, for which $\alpha_+=1$, $\alpha_-=0$, $\kappa_+=2$, $\kappa_-=\infty$, and $k=2$, and $A$ is semisimple. The Sylvester equation \eqref{eq:sylvester} gives $\pi^1(\omega)=-\omega_1-\omega_2$. The function $\delta\pi(\omega):=\omega_1\omega_2-\omega_2^2\log|\omega_2|$, with $\delta\pi(\omega):=0$ if $\omega_2=0$, is $\mathcal C^1$ and satisfies $D\delta\pi(\omega)S\omega=2\delta\pi(\omega)$, so that $\pi^2:=\pi^1+\delta\pi$ is a second $\mathcal C^1$ solution.

In both cases, $\pi^2-\pi^1$ is $O(\|\omega\|^b)$ for every $b<2$ but not $O(\|\omega\|^2)$; in \textbf{(b)}, for instance, $(\pi^2-\pi^1)(0,\omega_2)=-\omega_2^2\log|\omega_2|$. Hence $\pi^1\sim_{2^-}\pi^2$ while $\pi^1\not\sim_{2}\pi^2$. By contrast, in Example~\ref{ex:non-unique} the spectral ratio is also $k=2$ but both $A$ and $S$ are semisimple, and the solutions there have contact of order exactly $2$. 
\end{example}

Corollary~\ref{cor:moment} shows that contact between solutions to the invariance equation is preserved under composition with the output map $h$. Recall that we assumed $h$ to be smooth with $h(0)=0$.

\begin{corollary}\label{cor:moment}
Consider the system \eqref{eq:nl} and the exosystem \eqref{eq:nl-exo} satisfying the strip condition. Define $k$ as in the definition \eqref{eq:Nmax} and let $\pi^{1},\pi^{2}$ be two $\mathcal{C}^{1}$ solutions to the invariance equation \eqref{eq:inv}. Then $h\circ\pi^{1}\sim_{k^-} h\circ\pi^{2}$. 
\end{corollary}

\begin{proof}
Just as in the proof of Theorem~\ref{thm:finite-order}, it suffices to prove $h \circ \pi^1 \sim_{k^\prime} h \circ \pi^2$ for all $k^\prime \in (1,k)$. Since $h$ is smooth, it is Lipschitz on a neighborhood of the origin, and since $\pi^i(0)=0$ and $\pi^i$ is continuous, there exist a neighborhood $U$ of the origin and $c_h>0$ such that $\|h(\pi^1(\omega))-h(\pi^2(\omega))\|\le c_h\|\pi^1(\omega)-\pi^2(\omega)\|$ for all $\omega\in U$. The proof is complete since $\pi^1(\omega)-\pi^2(\omega)=O(\|\omega\|^{k^\prime})$ by Theorem~\ref{thm:finite-order}.
\end{proof}

Motivated by Corollary~\ref{cor:moment}, we define the nonlinear moment as follows: 

\begin{definition}\label{def:moment}
Consider the system \eqref{eq:nl} and the exosystem \eqref{eq:nl-exo} satisfying the strip condition, and define $k$ as in the definition \eqref{eq:Nmax}. Then the nonlinear moment of the system \eqref{eq:nl} at $(l,s)$ is defined as $[h \circ \pi]_{k^-}$, where $\pi$ is any $\mathcal C^1$ solution to the invariance equation, of which at least one exists by Theorem~\ref{thm:existence}. 
\end{definition}

{
From Proposition~\ref{prop:unique-jets}, for any positive integer $N \le N_{\max}$ and for any $\mathcal C^N$ solution $\pi$ to the invariance equation, $T_N \pi = T_N q_{N_{\max}}$. Although $q_{N_{\max}}$ is not a solution to the invariance equation in general, we show in Section~\ref{sec:consequences} that, if $k<\infty$, the class $[\pi]_{k^-}$ is determined by the polynomial $q_{N_{\max}}$, and we investigate the consequences of this fact.


This raises the question of how Definition~\ref{def:moment} relates to the classical definition of the nonlinear moment \cite{astolfi2010model}. Definition~\ref{def:moment} defines the nonlinear moment beyond the classical setting. In the classical setting, however, it may appear to weaken the classical definition: the classical moment is a function, whereas the contact class $[h\circ\pi]_{\infty^-}$ determines only the formal Taylor series of $h\circ\pi$ at the origin. In the following, we will see that this is not the case, i.e., the classical notion of moment can be recovered from Definition~\ref{def:moment} in the classical setting.
}

\subsection{Relation to the classical nonlinear moment}\label{subsec:on-uniqueness}
Recent works on nonlinear moments either assume that the invariance equation has a unique solution \cite[Assumption~1]{ionescu2016nonlinear}, \cite[Assumption~1]{moreschini2025closed}, or define the moment through any solution \cite[Definition~2.13]{scarciotti2017nonlinear}, \cite[Definition~2]{faedo2021approximation}. In this subsection, we identify two settings in which the $\mathcal C^1$ solution is unique on a neighborhood of the origin, so that these definitions are unambiguous and the classical moment is recovered from the contact class of Definition~\ref{def:moment}: Proposition~\ref{prop:steady-state-response} for an exponentially stable system, and Proposition~\ref{prop:unique} for a system that is only hyperbolic. Both results are consequences of classical facts on center manifolds; their role here is to relate Definition~\ref{def:moment} to the classical definition of the nonlinear moment.

\begin{proposition}\label{prop:steady-state-response}
Consider the system \eqref{eq:nl} and the exosystem \eqref{eq:nl-exo}, let $\varphi_t$ denote the flow of $s$, and suppose that $\sigma(A)\subset\C_{<0}$ and that the exosystem is neutrally stable, i.e., for every $R>0$ there exists $\delta>0$ such that, for every $\omega\in B_\delta$, $\varphi_t(\omega)$ is defined and satisfies $\|\varphi_t(\omega)\|<R$ for all $t\in\R$. Then the following hold.

\noindent \textbf{(a)} $\sigma(S)\subset i\R$, $\alpha_+=\alpha_-=0$, $\kappa_+=\infty$, so the strip condition holds with $k=\infty$.

\noindent \textbf{(b)} There exist an open neighborhood $W$ of the origin that is invariant under $\varphi_t$ and a $\mathcal C^1$ solution $\pi:W\to\R^n$ to the invariance equation \eqref{eq:inv} with the following property: for every $\mathcal C^1$ solution $\hat\pi$ to the invariance equation \eqref{eq:inv}, there exists an open neighborhood $W'\subset W$ of the origin that is invariant under $\varphi_t$ and on which $\hat\pi=\pi$.

\noindent \textbf{(c)} For every $N\in\Z_{>0}$ and every $\omega\in W$, $e^{At}(\pi-q_N)(\omega_{-t})$ converges to zero as $t$ goes to infinity, where $\omega_{-t}:=\varphi_{-t}(\omega)$; that is, $\pi$ satisfies the limit conditions \eqref{eq:LP-limit} on $W$.

\noindent \textbf{(d)} There exist $\delta_x>0$, $c>0$, and $\kappa_0>0$ such that, for every initial condition $(\omega(0),x(0))$ of the cascade with $\omega(0)\in W$ and $\|x(0)-\pi(\omega(0))\|<\delta_x$, the solution exists for all $t\ge0$ and satisfies
\begin{equation*}
    \|x(t)-\pi(\omega(t))\|\le c\|x(0)-\pi(\omega(0))\|e^{-\kappa_0 t},\quad\forall t\ge0.
\end{equation*}
Moreover, $t\mapsto(\omega(t),\pi(\omega(t)))$ is a solution of the cascade for every $\omega(0)\in W$.
\end{proposition}

Proposition~\ref{prop:steady-state-response} covers the setting in which the moment is related to the steady-state response \cite[Assumption~2.1 and Lemma~2.1]{scarciotti2017nonlinear}, \cite[Theorem~3]{moreschini2025closed}: the system is exponentially stable and the exosystem is neutrally stable. The neutral stability assumed in Proposition~\ref{prop:steady-state-response}, namely the stability of the origin in forward and backward time, is weaker than the neutral stability of \cite[Definition~2.9]{scarciotti2017nonlinear} and \cite[Definition~8]{moreschini2025closed}, namely Lyapunov stability together with Poisson stability of every point near the origin; see Appendix~\ref{app:steady-state}. Proposition~\ref{prop:steady-state-response} collects facts that are classical in this setting. Part~(b) follows from the center manifold theorem and the fact that every $\mathcal C^1$ center manifold contains all solutions that remain in a sufficiently small neighborhood of the origin for all $t\in\R$ \cite[Proposition~1]{palmer1987stability}, and part~(d) is the local attractivity of the center manifold \cite[Lemma and Proposition~2]{palmer1987stability}. The proof is provided in Appendix~\ref{app:steady-state}. By part~(b), the steady-state response $\pi(\omega(t))$ used to define the nonlinear moment in this setting is given by the unique $\mathcal C^1$ solution to the invariance equation on invariant neighborhoods of the origin; this makes precise the uniqueness assumed, for instance, in \cite[Assumption~1]{ionescu2016nonlinear}. Note that, in this setting, the limit conditions \eqref{eq:LP-limit} hold for every bounded solution, so that what singles out the classical moment is the invariance of the domain rather than the limit conditions \eqref{eq:LP-limit} of Remark~\ref{rem:boundary-condition}. 

\begin{remark}\label{rem:shakib}
The function $\pi$ in \cite[Definition~2]{shakib2023model} is only continuous, and \cite[Remark~6]{shakib2023model} stresses that the definition does not require $\pi$ to solve the invariance equation. Nevertheless, the invariance of the graph of a continuous $\pi$ is equivalent to the invariance equation along the flow of the exosystem, $\frac{d}{dt}\pi(\varphi_t(\omega))|_{t=0}=f(\pi(\omega),l(\omega))$, since $t\mapsto\pi(\varphi_t(\omega))$ is then the $x$-component of a solution of the cascade; at every point at which $\pi$ is differentiable, this is the invariance equation \eqref{eq:inv}. Since the proof of Lemma~\ref{lem:invariant-uniqueness} uses only the invariance of the graphs, it applies to continuous functions with invariant graphs; hence, if the vector fields in \cite{shakib2023model} are smooth, as assumed in Section~\ref{sec:prelim}, and if $\sigma(A)\subset\C_{<0}$ and the exosystem is neutrally stable, the function $\pi$ of \cite{shakib2023model} coincides, near the origin, with the unique $\mathcal C^1$ solution of Proposition~\ref{prop:steady-state-response}. In this sense, the approach of \cite{shakib2023model} interprets the invariance equation in a weak sense; its advantages are that it is global and that the existence and uniqueness of $\pi$ follow from convergence rather than from spectral conditions.
\end{remark}

Proposition~\ref{prop:steady-state-response} requires $A$ to be Hurwitz. To see what is needed when $A$ is only hyperbolic, consider the mechanism behind the non-uniqueness in the example of the Introduction. In polar coordinates, the exosystem \eqref{eq:sig-locally-flat} is written as $\frac{d}{dt}\|\omega\| = -\|\omega\|^3,~\dot{\theta} = 1,~u = \|\omega\| \cos \theta,$ where $\theta$ is the polar angle, so that every nonzero trajectory of the exosystem approaches the origin, much more slowly than the system state $x$; this is the source of the nontrivial flat difference $\phi$ given by the formula \eqref{eq:locally-flat}. The same mechanism occurs when $A$ is not Hurwitz. Reversing time, consider the system $\dot x=x+u$ and the exosystem $\dot\omega=-s(\omega)$, $u=\omega_1$, with $s$ as in the exosystem \eqref{eq:sig-locally-flat}, for which $A=1$ and $k=\infty$. Since $D\phi(\omega)(-s(\omega))=\phi(\omega)$, if $\pi$ is a $\mathcal C^1$ solution to the corresponding invariance equation, so is $\pi+c\phi$ for every $c\in\R$; here, every nonzero trajectory of the exosystem approaches the origin in backward time. It is therefore natural to expect uniqueness if the flow of $s$ is prevented from approaching the origin in both directions of time, for instance, by a first integral.

\begin{proposition}\label{prop:unique}
    Consider the system \eqref{eq:nl} and the exosystem \eqref{eq:nl-exo} satisfying the strip condition. Assume that there exist a neighborhood $U$ of the origin and $V\in\mathcal C^2(U,\R)$ such that
    \begin{equation}
        \label{eq:conserved}
        DV(\omega) s(\omega) = 0,~\forall\omega\in U;~V(0)=0,~DV(0)=0,~D^{2}V(0)\succ 0.
    \end{equation}
    Then any two local $\mathcal C^1$ solutions to the invariance equation \eqref{eq:inv} coincide on a neighborhood of the origin. 
\end{proposition}

The proof of Proposition~\ref{prop:unique} is provided in Appendix~\ref{app:unique}. We give three remarks. First, the condition \eqref{eq:conserved} implies $\sigma(S)\subset i\mathbb{R}$ and that the flow generated by $s$ is complete on a neighborhood of the origin: differentiating $DV(\omega)s(\omega)=0$ twice at the origin and using $s(0)=0$ and $DV(0)=0$ yields $S^{\top}D^{2}V(0) + D^{2}V(0) S = 0$, so $S$ is skew-symmetric with respect to the inner product induced by $D^{2}V(0)$ and hence has purely imaginary eigenvalues; completeness of the flow is proved in Appendix~\ref{app:unique}. Therefore, Proposition~\ref{prop:unique} only covers the case $k=\infty$, or more precisely, $\alpha_+ = \alpha_- = 0$, in which the strip condition reduces to the hyperbolicity of $A$. Second, the sublevel sets of $V$ near the origin are invariant, so that the exosystem is neutrally stable; hence, if $\sigma(A)\subset\C_{<0}$, Proposition~\ref{prop:unique} is also covered by Proposition~\ref{prop:steady-state-response}, and its new content is the case in which $A$ has eigenvalues with positive real parts. In that case, the graph of every $\mathcal C^1$ solution is still a local center manifold of the cascade and every orbit on it remains in an arbitrarily small neighborhood of the origin for all $t\in\R$, but \cite[Proposition~1]{palmer1987stability} does not apply, since it requires the remaining eigenvalues to have negative real parts; the proof in Appendix~\ref{app:unique} extends the argument of \cite{palmer1987stability} by means of the exponential dichotomy (Lemma~\ref{lem:invariant-uniqueness}). Third, the condition \eqref{eq:conserved} is satisfied, in particular, whenever the exosystem is a simple harmonic oscillator, with $V(\omega)=\|\omega\|^{2}$. For instance, Proposition~\ref{prop:unique} applies to the controlled Van der Pol oscillator driven by a simple harmonic oscillator in \cite[Example~2]{moreschini2025closed}: the linearization of the oscillator at the origin has two eigenvalues with positive real part, so that the steady-state interpretation is not available, but the invariance equation has a unique local $\mathcal C^1$ solution.

We now return to the question raised before Section~\ref{subsec:on-uniqueness}. First, Definition~\ref{def:moment} is consistent with the classical definition: under the assumptions of Proposition~\ref{prop:steady-state-response} or of Proposition~\ref{prop:unique}, all $\mathcal C^1$ solutions to the invariance equation coincide with $\pi$ on a neighborhood of the origin, so that the classical moment $h\circ\pi$ is a representative of the moment of Definition~\ref{def:moment}, namely the flat-equivalence class $[h\circ\pi]_{\infty^-}$. Second, although passing to the class discards information, in these settings this information is recovered, up to restriction of the domain, by singling out the solution $\pi$ of Proposition~\ref{prop:steady-state-response} or of Proposition~\ref{prop:unique}. Third, outside these settings, a canonical representative is not available in general. The fixed point of the truncated L--P operator used in the proof of Theorem~\ref{thm:existence} depends on the choice of the cutoff function and of $\rho$, and the example in the Introduction shows that distinct $\mathcal C^\infty$ solutions may coexist even when $\sigma(S)\subset i\R$. In such cases, a definition of the moment through any solution \cite{scarciotti2017nonlinear,faedo2021approximation} depends on the choice of the solution, and the class is the information on the moment that is determined by the system and the exosystem alone.

\section{Consequences of the equivalence class definition}\label{sec:consequences}
In this section, we derive properties of the contact classes $[\pi]_{k^-}$ and $[h\circ \pi]_{k^-}$. In Section~\ref{subsec:representatives}, we show that the polynomial $q_{N_{\max}}$ belongs to the contact class $[\pi]_{k^-}$ if $k<\infty$, and that, if $k=\infty$, the truncations of the formal power series $\hat q$ defined by the recursion \eqref{eq:qN-recursion} have contact of every order with $\pi$. In Section~\ref{subsec:transient}, we show that, when the system is exponentially stable, $[h\circ\pi]_{k^-}$ describes the output shared by all trajectories starting sufficiently close to the graph of $\pi$, up to the order $k^-$; this gives the moment a time-domain interpretation when the exosystem is not neutrally stable and the steady-state interpretation is not available. Finally, in Section~\ref{subsec:nlmor}, we investigate how the concept of moment matching can be extended according to the refined definition of nonlinear moment, in which the polynomial or formal power series representative of Section~\ref{subsec:representatives} plays a crucial role.

\subsection{Polynomial and formal power series representatives}\label{subsec:representatives}

\begin{proposition}\label{prop:inv-eq-power-series}
Assume the hypotheses of Theorem~\ref{thm:finite-order}. Let $N \in \Z_{>0}$ be such that $N \le N_{\max}$, and let $q_N$ be as in Theorem~\ref{thm:approximate-jet}. Then every $\mathcal C^1$ solution $\pi$ to the invariance equation satisfies $\pi \sim_{k^\prime} q_N$ for every $k^\prime \in (1, N+1] \cap (1, k)$. 
\end{proposition}

\begin{proof}
Let $N^\prime \in \Z_{>0}$ be such that $k^\prime \in (N^\prime, N^\prime +1]$ and let $z:= \pi - q_N$. From Proposition~\ref{prop:unique-jets} with $N=1$, $z(0) = 0$ and $Dz(0) = 0$ so that $z(\omega) = O(\|\omega\|)$. From the invariance equation \eqref{eq:inv} and the definition \eqref{eq:residual-operator}, $z$ satisfies $Dz(\omega) s(\omega) = (A + \Delta(\omega))z(\omega) - \mathcal R(q_N)(\omega)$. Here, $\Delta(\omega)$ is defined analogously to the definition \eqref{eq:B-def} with $\pi^1$ replaced by $q_N$ and $\delta \pi$ by $z$, and $\| \Delta(\omega)\| \le C_\Delta \| \omega\|$. From the estimate \eqref{eq:qN-residual}, we have $\mathcal R(q_N)(\omega) = O(\| \omega\|^{N+1})$. Since $k^\prime \le N+1$, $\mathcal R(q_N)(\omega) = O(\|\omega\|^{k^\prime})$.

Starting from $z(\omega) = O(\|\omega\|)$, apply Lemma~\ref{lem:order-raising} successively. Specifically, if $N^\prime \ge 2$, apply Lemma~\ref{lem:order-raising} from $(a,b) = (1,2)$ to $(N^\prime -1, N^\prime)$ and $(N^\prime , k^\prime)$. If $N^\prime = 1$, apply Lemma~\ref{lem:order-raising} with $(a,b) = (1,k^\prime)$. In both cases, $-\mathcal R(q_N)$ takes the role of $g$ of Lemma~\ref{lem:order-raising}. Successive application is possible because for all pairs $(a,b)$, $0<a<b\le a+1,$ $b\le k^\prime<k$, and $\mathcal R(q_N)(\omega) = O(\|\omega\|^b)$. From this successive application, we get $z(\omega) = O(\|\omega\|^{k^\prime})$ and the proof is complete.
\end{proof}

Corollary~\ref{cor:inv-eq-power-series} is a direct consequence of Proposition~\ref{prop:inv-eq-power-series}, and its proof is omitted.

\begin{corollary}\label{cor:inv-eq-power-series}
Assume the hypotheses of Theorem~\ref{thm:finite-order}.

\noindent\textbf{(a)} If $k<\infty$ and $q_{N_{\max}}$ is as in Theorem~\ref{thm:approximate-jet}, then every $\mathcal C^1$ solution $\pi$ to the invariance equation satisfies $\pi \sim_{k^-} q_{N_{\max}}$. In particular, $q_{N_{\max}}$ is the unique element of $\mathcal P_1 \oplus \cdots \oplus \mathcal P_{N_{\max}}$ that belongs to $[\pi]_{k^-}$.

\noindent\textbf{(b)} If $k = \infty$ and $\hat q$ is formally defined by $\hat q := \sum_{j=1}^\infty p_j$, where $p_j$ is defined by the recursion \eqref{eq:qN-recursion}, then every $\mathcal C^1$ solution $\pi$ to the invariance equation satisfies $T_N \hat q \sim_{N+1} \pi$ for every $N \in \Z_{>0}$. Additionally, if $\hat q$ converges, then $\hat q$ is the unique analytic function that is flat-equivalent to all $\mathcal C^1$ solutions to the invariance equation. 
\end{corollary}

For a given control law $u=c(\omega)$, Corollary~\ref{cor:inv-eq-power-series} justifies the power-series approximation of the solution $\pi$ to the invariance part of the FBI equation in~\cite{huang1992approximation} even beyond the stable case, at least up to the degree $N_{\max}$. Since, in the FBI equation, $c$ is itself an unknown determined together with $\pi$ by the zero-output condition, the corollary does not address the solvability of the FBI equation as a whole.

Although $q_{N_{\max}} \in [\pi]_{k^-}$, it is possible that $q_{N_{\max}}$ itself is not a solution to the invariance equation. Similarly, $\hat q$ need not be a solution to the invariance equation even when it converges, unless $f$, $l$, and $s$ are analytic. If $f$, $l$, and $s$ are analytic and $\hat q$ converges, then the residual $D \hat q (\omega) s(\omega) - f(\hat q(\omega), l(\omega))$ is analytic and all of its derivatives vanish at the origin, so that it vanishes identically near the origin. 

\begin{corollary}\label{cor:moment-power-series}
Assume the hypotheses of Theorem~\ref{thm:finite-order}.

\noindent\textbf{(a)} If $k<\infty$, let $q_{N_{\max}}$ be as in Corollary~\ref{cor:inv-eq-power-series}\textbf{(a)}. Define
\begin{equation}\label{eq:moment-polynomial-representative}
    m:=T_{N_{\max}} ((T_{N_{\max}}h)\circ q_{N_{\max}} ).
\end{equation}
Then, $m\sim_{k^-}h\circ\pi$ for every $\mathcal C^1$ solution $\pi$ to the invariance equation.

\noindent\textbf{(b)} If $k=\infty$, let $\hat q$ be as in Corollary~\ref{cor:inv-eq-power-series}\textbf{(b)}, and let $\hat h$ be the (formal) Taylor series of $h$ at the origin. Define the formal power series
\begin{equation}\label{eq:moment-formal-power-series-rep}
    \hat m:=\hat h\circ\hat q.
\end{equation}
Then, $T_{N}\hat m\sim_{N+1} h\circ\pi$ for every $N\in\Z_{>0}$ and every $\mathcal C^1$ solution $\pi$. In particular, if $\hat m$ converges, then it is flat-equivalent to every moment representative $h\circ\pi$. 
\end{corollary}

\begin{proof}
    For the case \textbf{(a)}, since $h$ is locally Lipschitz and $q_{N_{\max}}\sim_{k^-}\pi$, \begin{equation}\label{eq:equivalence-1}
        h\circ q_{N_{\max}}\sim_{k^-}h\circ\pi.
    \end{equation}
    Moreover, $h(x)=T_{N_{\max}}h(x)+O(\|x\|^{N_{\max}+1}),$ and $q_{N_{\max}}(\omega)=O(\|\omega\|)$. Therefore $h(q_{N_{\max}}(\omega))=(T_{N_{\max}}h)(q_{N_{\max}}(\omega))+O(\|\omega\|^{N_{\max}+1}).$ Moreover, $(T_{N_{\max}}h)\circ q_{N_{\max}}$ is a polynomial whose terms of degree at most $N_{\max}$ are exactly $m$, so that $(T_{N_{\max}}h)\circ q_{N_{\max}}-m=O(\|\omega\|^{N_{\max}+1})$ as well. Since $k\le N_{\max}+1$, the remainder $O(\|\omega\|^{N_{\max}+1})$ belongs to $O(\|\omega\|^{k^\prime})$ for every $k^\prime \in (1,k)$, and therefore
    \begin{equation}\label{eq:equivalence-2}
        h\circ q_{N_{\max}}\sim_{k^-}m.
    \end{equation}
    Combining the relations \eqref{eq:equivalence-1} and \eqref{eq:equivalence-2} proves \textbf{(a)}.

    For \textbf{(b)}, fix $N$. Corollary~\ref{cor:inv-eq-power-series}\textbf{(b)} gives $T_{N}\hat q\sim_{N+1}\pi$. Repeating the preceding finite-order argument with $N$ in place of $N_{\max}$ shows that the truncation of the formal composition satisfies $T_{N}\hat m\sim_{N+1} h\circ\pi$. Since $N$ is arbitrary, the result follows.
\end{proof}

Corollary~\ref{cor:moment-power-series} gives a direct computational procedure. In the finite-order case, compute the polynomial $q_{N_{\max}}$, substitute it into $T_{N_{\max}}h$, and discard all terms of degree higher than $N_{\max}$. In the infinite-order case, the same coefficient matching determines the formal power series $\hat m$ order by order.

Corollaries~\ref{cor:inv-eq-power-series} and \ref{cor:moment-power-series} also provide another interpretation of $\Pi$ and $C\Pi$ of the linear moment theory. If the system and the exosystem are linear and we assume the strip condition, $\Pi\omega$ corresponds to $q_{N_{\max}}(\omega)$ (or $\hat q(\omega)$) and $C\Pi \omega$ corresponds to $m(\omega)$ (or $\hat m(\omega)$).

\begin{remark}
Let $h$ be analytic. If $\hat q$ converges, then $\hat m$ also converges since $h \circ \hat q$ is a well-defined analytic function. The converse does not hold, e.g., when $h \equiv 0$.  
\end{remark}

\begin{remark}
As mentioned right after Corollary~\ref{cor:inv-eq-power-series}, it could happen that $m \neq h \circ \pi$ for every $\mathcal{C}^1$ solution $\pi$ to the invariance equation if $N_{\max} < \infty$. However, if $N_{\max}=\infty$, $\hat q$ converges, and $h$, $f$, $l$, and $s$ are analytic so that $\hat q$ is an analytic solution to the invariance equation, then $\hat m=h\circ \hat q$ around the origin. 
\end{remark}


\subsection{The time-domain interpretation of the moment}\label{subsec:transient}
Proposition~\ref{prop:steady-state-response} interprets the moment through the steady-state response, which requires the exosystem to be neutrally stable. When $k<\infty$, the exosystem is not neutrally stable, since $\alpha_+>0$ or $\alpha_->0$. In this case, the moment can still be defined through a solution to the invariance equation, but its relation to the steady-state response is lost \cite[pp.~248--249]{scarciotti2017nonlinear}; in particular, if the trajectories of the exosystem converge to the origin, the steady-state response is trivial, and such inputs are excluded in \cite[Assumption~1]{scarciotti2015model} to guarantee a well-defined steady-state response. In this subsection, we show that, if $A$ is Hurwitz, the moment retains a time-domain interpretation in terms of the polynomial representative $m$ of the definition \eqref{eq:moment-polynomial-representative}, which does not require the initial condition to lie on an invariant manifold.

\begin{proposition}\label{prop:transient}
    Consider the system \eqref{eq:nl} and the exosystem \eqref{eq:nl-exo} satisfying the strip condition with $\sigma(A)\subset\C_{<0}$, define $k$ as in the definition \eqref{eq:Nmax}, and assume $k<\infty$. Let $m$ be as in the definition \eqref{eq:moment-polynomial-representative}, and let $\pi$ be a $\mathcal C^1$ solution to the invariance equation \eqref{eq:inv} defined on a neighborhood $U$ of the origin. Then, for every $b\in(0,k)$, there exist $\rho>0$, $\delta_x>0$, and $c>0$ with the following property. Let $(\omega(t),x(t))$ be a solution of the cascade, let $y(t):=h(x(t))$, and let $T>0$. If $\omega(0)\neq0$, $\|x(0)-\pi(\omega(0))\|<\delta_x$, and $\omega(\tau)\in B_\rho$ for all $\tau\in[0,T]$, then $x(t)$ is defined on $[0,T]$ and
    \begin{equation}\label{eq:transient-bound}
        \|y(t)-m(\omega(t))\|\le c\left(1+\frac{\|x(0)-\pi(\omega(0))\|}{\|\omega(0)\|^b}\right)\|\omega(t)\|^b,\quad\forall t\in[0,T].
    \end{equation}
    In particular, if $\omega(t)\in B_\rho$ for all $t\ge0$, then $x(t)$ is defined for all $t\ge0$ and the estimate \eqref{eq:transient-bound} holds for all $t\ge0$, with the constant $c\bigl(1+\|x(0)-\pi(\omega(0))\|/\|\omega(0)\|^b\bigr)$ independent of $t$.
\end{proposition}

\begin{proof}
    Fix $b\in(0,k)$. Since $\sigma(A)\subset\C_{<0}$, we have $E_+=\{0\}$, $\kappa_+=\infty$, and $P_-$ is the identity. Since $k<\infty$, the definition \eqref{eq:Nmax} gives $\alpha_->0$ and $k=\kappa_-/\alpha_-$, so that $b\alpha_-<\kappa_-$. Choose $\epsilon>0$ and $\kappa_-'\in(0,\kappa_-)$ such that $b(\alpha_-+\epsilon)<\kappa_-'$, and let $K=K(\kappa_-')$ be as in the inequality \eqref{eq:dichotomy}, so that $\|e^{At}\|\le Ke^{-\kappa_-'t}$ for $t\ge0$. Choose $c_\Delta>0$ such that
    \begin{equation*}
        \kappa_0:=\kappa_-'-Kc_\Delta\ge b(\alpha_-+\epsilon).
    \end{equation*}
    Since $f$ is smooth, $l(0)=0$, and $D_xf(0,0)=A$, there exist $\rho_1>0$ and $\eta>0$ such that $\|D_xf(x,l(\xi))-A\|\le c_\Delta$ whenever $\|x\|\le\eta$ and $\|\xi\|\le\rho_1$. By Corollary~\ref{cor:moment-power-series}\textbf{(a)}, $m\sim_{k^-}h\circ\pi$, so there exist $\rho_2>0$ and $c_2>0$ such that $\overline B_{\rho_2}\subset U$ and $\|h(\pi(\xi))-m(\xi)\|\le c_2\|\xi\|^b$ for $\xi\in B_{\rho_2}$. Let $\rho_\epsilon$ and $C_\epsilon$ be as in Lemma~\ref{lem:cutoff}, and choose $\rho\in(0,\min\{\rho_\epsilon,\rho_1,\rho_2\})$ such that $\|\pi(\xi)\|\le\eta/2$ for $\xi\in B_\rho$. Let $c_h$ be a Lipschitz constant of $h$ on $\overline B_\eta$, and set $\delta_x:=\eta/(2K)$ and $c:=\max\{c_hKC_\epsilon^b,c_2\}$.

    Let $(\omega(t),x(t))$ and $T$ be as in the statement, and define $z(t):=x(t)-\pi(\omega(t))$. Since $\pi$ satisfies the invariance equation \eqref{eq:inv} on $U\supset B_\rho$,
    \begin{equation*}
        \dot z(t)=f(\pi(\omega(t))+z(t),l(\omega(t)))-f(\pi(\omega(t)),l(\omega(t)))=\left(A+\Delta(t)\right)z(t),
    \end{equation*}
    where $\Delta(t):=\int_0^1\left[D_xf(\pi(\omega(t))+\vartheta z(t),l(\omega(t)))-A\right]d\vartheta$. As long as $\|z(t)\|\le\eta/2$, we have $\|\pi(\omega(t))+\vartheta z(t)\|\le\eta$, so that $\|\Delta(t)\|\le c_\Delta$. On such an interval, the variation-of-constants formula and Gronwall's inequality yield $\|z(t)\|\le K\|z(0)\|e^{-\kappa_0 t}<K\delta_x=\eta/2$. Hence, by a standard continuation argument, $x(t)$ is defined on $[0,T]$ and
    \begin{equation}\label{eq:transient-z}
        \|z(t)\|\le K\|z(0)\|e^{-\kappa_0 t},\quad\forall t\in[0,T].
    \end{equation}

    Fix $t\in[0,T]$. Since $\omega(\tau)\in B_\rho$ for all $\tau\in[0,t]$ and the vector field $s_\rho$ of Lemma~\ref{lem:cutoff} coincides with $s$ on $B_\rho$, we have $\omega(0)=\varphi_{-t}^\rho(\omega(t))$. The backward-time bound in the estimate \eqref{eq:flow-bound} gives $\|\omega(0)\|\le C_\epsilon\|\omega(t)\|e^{(\alpha_-+\epsilon)t}$, that is,
    \begin{equation*}
        e^{-(\alpha_-+\epsilon)t}\le\frac{C_\epsilon\|\omega(t)\|}{\|\omega(0)\|}.
    \end{equation*}
    Since $\kappa_0\ge b(\alpha_-+\epsilon)$, it follows that
    \begin{equation*}
        e^{-\kappa_0 t}\le e^{-b(\alpha_-+\epsilon)t}\le \frac{C_\epsilon^b\|\omega(t)\|^b}{\|\omega(0)\|^b},
    \end{equation*}
    and combining this with the estimate \eqref{eq:transient-z} gives $\|z(t)\|\le KC_\epsilon^b\|z(0)\|\|\omega(0)\|^{-b}\|\omega(t)\|^b$.
    Finally, since $x(t)$ and $\pi(\omega(t))$ belong to $\overline B_\eta$ and $\omega(t)\in B_{\rho}\subset B_{\rho_2}$,
    \begin{align*}
        \|y(t)-m(\omega(t))\|&\le\|h(x(t))-h(\pi(\omega(t)))\|+\|h(\pi(\omega(t)))-m(\omega(t))\|\\
        &\le c_h\|z(t)\|+c_2\|\omega(t)\|^b,
    \end{align*}
    which yields the estimate \eqref{eq:transient-bound}.
\end{proof}

\begin{remark}\label{rem:transient}
    The proof of Proposition~\ref{prop:transient} combines a Gronwall argument for the local exponential attractivity of the invariant graph, as in \cite{palmer1987stability}, with the backward-time estimate of the exosystem flow in Lemma~\ref{lem:cutoff} and Corollary~\ref{cor:moment-power-series}. The point of the proposition is that the threshold $b<k$, below which the transient $e^{-\kappa_-'t}$ is of higher order than $\|\omega(t)\|^b$, is the same spectral ratio $k$ that limits the contact between solutions in Theorem~\ref{thm:finite-order}: the information on the output that does not depend on the initial condition coincides with the information determined by the moment. Proposition~\ref{prop:transient} shows that, when $A$ is Hurwitz, the moment $[h\circ\pi]_{k^-}$ describes the part of the output that is common to all trajectories starting near the invariant manifold: every such output satisfies $\|y(t)-m(\omega(t))\|\le c_0\|\omega(t)\|^b$ for all $t\ge0$, with $c_0$ independent of $t$, for every $b<k$, regardless of the initial condition of the system. In this sense, the moment admits a time-domain interpretation that replaces the steady-state interpretation of Proposition~\ref{prop:steady-state-response} when the exosystem is not neutrally stable. The order $k$ cannot be exceeded in general, because the difference between two trajectories decays as $e^{-\kappa_- t}$, which is of order $\|\omega(t)\|^{k}$ when $\|\omega(t)\|$ decays as $e^{-\alpha_- t}$; see Example~\ref{ex:transient-example}. If $k=\infty$, the same argument shows that the estimate \eqref{eq:transient-bound} holds with $T_N\hat m$ in place of $m$ for every $N\in\Z_{>0}$ and every $b\in(0,N+1]$, with $\hat m$ as in the definition \eqref{eq:moment-formal-power-series-rep}. If $E_+\neq\{0\}$, the invariant manifold is not attractive in the directions of $E_+$, and no such interpretation is available for initial conditions off the manifold.
\end{remark}

\subsection{Extending moment-matching model-order reduction}\label{subsec:nlmor}
One of the most common applications that use the concept of nonlinear moment is model-order reduction based on moment matching. We now discuss how the concept of nonlinear moment-matching model-order reduction can be extended by introducing contact classes. We first recall the concept of nonlinear moment-matching model-order reduction \cite{astolfi2010model,ionescu2016nonlinear,scarciotti2017nonlinear}. We highlight that these notions are based on nonlinear moments in terms of functions, not contact classes.

An $\overline n$-dimensional model $(\overline h , \overline f)$ such that $\overline f$ and $\overline h$ are smooth, $\overline f(0,0) = 0$, $\overline h(0) = 0$, and
\begin{equation}\label{eq:reduced-model}
    \dot{\overline x} = \overline f(\overline x, u), ~ \overline y = \overline h(\overline x),~ \overline x \in \mathbb{R}^{\overline n},~\overline y \in \R^{n_y}
\end{equation}
is said to match the moment of the system \eqref{eq:nl} at $(l,s)$ if there exists a pair of local solutions $(\pi, \overline \pi)$ to the invariance equation for the system $(h,f)$ and the reduced-order model $(\overline h,\overline f)$ that satisfies
\begin{equation}\label{eq:moment-matching}
    h \circ \pi(\omega) = \overline h \circ \overline \pi(\omega).
\end{equation}
If $\overline n <n$, $(\overline h , \overline f )$ is referred to as a reduced-order model of the system \eqref{eq:nl} obtained by moment matching at $(l,s)$.

The condition \eqref{eq:moment-matching} compares the moments as functions. By Corollary~\ref{cor:moment}, however, the contact class of $h\circ\pi$ does not depend on the choice of the solution $\pi$, which leads to Definition~\ref{def:refined-moment-matching} in terms of contact classes.

\begin{definition}\label{def:refined-moment-matching}
Consider the system \eqref{eq:nl}, the model \eqref{eq:reduced-model}, and the exosystem \eqref{eq:nl-exo}, where both the system and the model satisfy the strip condition. Let $k$ be defined as in the definition \eqref{eq:Nmax}, define $\overline k$ analogously for the model, and let $\tilde k:=\min\{k,\overline k\}$. The system $(h,f)$ and the model $(\overline h,\overline f)$ \emph{match the moment at $(l,s)$} if
\begin{equation}\label{eq:equiv-moment-matching}
    [h\circ\pi]_{\tilde k^-}=[\overline h\circ\overline\pi]_{\tilde k^-},
\end{equation}
where $\pi$ and $\overline\pi$ are $\mathcal C^1$ solutions to the invariance equations for the system and for the model, respectively. By Corollary~\ref{cor:moment} and the implications \eqref{eq:basic-prop-for-equiv-1}, the condition \eqref{eq:equiv-moment-matching} does not depend on the choice of $\pi$ and $\overline\pi$. If, in addition, $\overline n<n$, then the model $(\overline h,\overline f)$ is a \emph{moment-matching reduced-order model} of $(h,f)$ at $(l,s)$.
\end{definition}

\begin{remark}\label{rem:moment-matching-consistency}
If there exists a pair of solutions $(\pi,\overline\pi)$ that satisfies the condition \eqref{eq:moment-matching}, then the condition \eqref{eq:equiv-moment-matching} holds, so that Definition~\ref{def:refined-moment-matching} contains the classical definition of moment matching \cite{astolfi2010model,ionescu2016nonlinear,scarciotti2017nonlinear}. The converse does not hold, since the condition \eqref{eq:moment-matching} also constrains $h\circ\pi$ beyond the order $\tilde k$. Note that the classical definition is too restrictive from the viewpoint of contact classes since if the system admits solutions $\pi^1$ and $\pi^2$ with $h\circ\pi^1\neq h\circ\pi^2$, then the condition \eqref{eq:moment-matching} cannot hold for every pair of solutions even when $(\overline h,\overline f)=(h,f)$. In Definition~\ref{def:refined-moment-matching}, the classes are compared at the order $\tilde k^-$ because the moment $[h\circ\pi]_{k^-}$ of the system and the moment $[\overline h\circ\overline\pi]_{\overline k^-}$ of the model cannot be compared beyond the order $\tilde k$. If $\overline k\ge k$ (which can always be achieved by Proposition~\ref{prop:parametrized-family-ROM} below), then $\tilde k=k$ and the whole moment of the system is matched.
\end{remark}

Let $m$ and $\hat m$ be the representatives of the moment of the system given by Corollary~\ref{cor:moment-power-series}, and let $\overline m$ and $\overline{\hat m}$ be the corresponding representatives for the model, obtained by applying Corollary~\ref{cor:moment-power-series} to $(\overline h,\overline f)$. Proposition~\ref{prop:refined-moment-matching} shows that, in order to check the contact of order $\tilde k^-$ between the moments of the system and the model, it suffices to compare the coefficients of these representatives up to order $\tilde N$, with $\tilde N$ such that $\tilde k \in (\tilde N, \tilde N+1]$.

\begin{proposition}\label{prop:refined-moment-matching}
    Assume the hypotheses of Definition \ref{def:refined-moment-matching}. For $M \in \Z_{>0}$, define $m_M:=T_M m$ if $k<\infty$ and $m_M:=T_M \hat m$ if $k = \infty$. Analogously define $\overline m_M$. If $\tilde k < \infty$, let $\tilde N \in \Z_{>0}$ be the unique positive integer such that $\tilde k \in (\tilde N, \tilde N+1]$. Then, the system and the model match the moment at $(l,s)$ if and only if 
    \begin{equation}\label{eq:moment-matching-by-representatives}
    m_{\tilde N} = \overline m_{\tilde N}.
    \end{equation}
    If $\tilde k = \infty$, then they match the moment at $(l,s)$ if and only if $m_N = \overline m_N$ for all $N\in \Z_{>0}$. 
\end{proposition}

\begin{proof}
Assume $\tilde k <\infty$. By Corollary~\ref{cor:moment-power-series} and the implications \eqref{eq:basic-prop-for-equiv-1}, $m\sim_{\tilde k^-}h\circ\pi$ and $\overline m\sim_{\tilde k^-}\overline h\circ\overline\pi$, so that the condition \eqref{eq:equiv-moment-matching} holds if and only if $m\sim_{\tilde k^-}\overline m$. Since $m-\overline m$ is a polynomial and $\tilde k\in(\tilde N,\tilde N+1]$, the relation $m\sim_{\tilde k^-}\overline m$ holds precisely when every term of $m-\overline m$ has degree at least $\tilde N+1$, that is, when the condition \eqref{eq:moment-matching-by-representatives} holds. The case $\tilde k=\infty$ is identical, with $\hat m$ and $\overline{\hat m}$ in place of $m$ and $\overline m$.
\end{proof}

\begin{remark}\label{rem:information-content}
By Proposition~\ref{prop:refined-moment-matching}, moment matching in the sense of Definition~\ref{def:refined-moment-matching} amounts to matching finitely many Taylor coefficients of the moments if $\tilde k<\infty$, which is close in spirit to the power-series approach of \cite[Section 4.2]{huang2004nonlinear}, \cite{huang1992approximation}. In particular, if $k\le2$, then $N_{\max}=1$ and the moment $[h\circ\pi]_{k^-}$ is determined by the linear moment $C\Pi$, so that Definition~\ref{def:refined-moment-matching} reduces to linear moment matching. This loss of nonlinear information is not an artifact of Definition~\ref{def:moment}: Examples~\ref{ex:non-unique} and~\ref{ex:k-andk} show that solutions to the invariance equation are, in general, not unique beyond the order $k$, so that the Taylor coefficients of $h\circ\pi$ of order higher than $k$ are not determined by the system and the exosystem. Moreover, the case $k<\infty$ lies outside the classical setting of \cite{astolfi2010model}, in which $k=\infty$ by Proposition~\ref{prop:steady-state-response}\textbf{(a)}, so that Definition~\ref{def:moment} extends, rather than weakens, the classical definition in this case.
\end{remark}

There exists a parametrized family of models matching the moment in the sense of Definition~\ref{def:refined-moment-matching}, similar to the families in \cite[Section 3.C]{astolfi2010model} or \cite[Proposition 2.1]{scarciotti2017nonlinear}. We mention that given a formal power series $\hat m$, there exists a smooth function $m$ such that the Taylor series of $m$ equals $\hat m$ (Borel--Ritt theorem) \cite[Theorem A.3.2]{murdock2003normal}.

\begin{proposition}\label{prop:parametrized-family-ROM}
Assume the hypotheses of Theorem~\ref{thm:finite-order}. If $k<\infty$, let $m$ be as in the definition \eqref{eq:moment-polynomial-representative}, and if $k = \infty$, let $\hat m$ be as in the definition \eqref{eq:moment-formal-power-series-rep} {and let $m$ be a smooth function} that has $\hat m$ as its Taylor series. Finally, let $(L,S)$ be observable and let $\overline k \in (1,\infty]$ be given, such that
\begin{equation*}
\begin{cases}
    \overline k= \infty, & \text{if}~\alpha_+=\alpha_-=0, \\ 
    \overline k\in(1,\infty], & \text{if exactly one of}~\alpha_+,\alpha_-~\text{is zero},\\ 
    \overline k\in(1,\infty), & \text{if}~\alpha_+>0~\text{and}~\alpha_->0.
\end{cases}
\end{equation*}
Then, there exists a smooth function $G:\R^\nu \rightarrow \R^{\nu \times n_u}$ such that the model
\begin{equation}\label{eq:mm-canonical-form}
\dot{\overline x} = \overline f(\overline x, u):= s(\overline x) + G(\overline x) (u - l(\overline x)),~\overline y = \overline h(\overline x):=m(\overline x),
\end{equation}
satisfies the strip condition with $\overline k$. Moreover, for such $G$, the model \eqref{eq:mm-canonical-form} and the system match the moment at $(l,s)$ with $\tilde{k} = \min\{\overline{k}, k\}$; in particular $\tilde{k} = k$ whenever $\overline{k} \ge k$, which is always possible; indeed, if $k=\infty$, then $\alpha_-=0$ or $\alpha_+=0$ because $\R^n=E_-\oplus E_+$ is nontrivial, so that $\overline k=\infty$ is admissible.
\end{proposition}

Note that the model \eqref{eq:mm-canonical-form} has dimension $\overline n=\nu$, so that it is a reduced-order model of the system if and only if $\nu<n$.

\begin{proof}
    The linearized system dynamics around the origin of the model becomes $\dot{\overline x} = (S - G(0)L) \overline x + G(0) u$. Since $(L,S)$ is observable, the eigenvalues of $S-G(0)L$ can be assigned arbitrarily, provided that complex eigenvalues appear in conjugate pairs, by choosing $G(0)$. Since the model and the system share the exosystem, $\alpha_+$ and $\alpha_-$ are the same for both. If $\overline k<\infty$ and $\alpha_->0$, we assign all eigenvalues of $S-G(0)L$ to $-\overline k\,\alpha_-$, so that the corresponding quantities in the definition \eqref{eq:spectral-params} for the model are $\kappa_-=\overline k\,\alpha_-$ and $\kappa_+=\infty$; if $\overline k<\infty$ and $\alpha_-=0$, then $\alpha_+>0$ and we assign all eigenvalues to $\overline k\,\alpha_+$. If $\overline k=\infty$, then $\alpha_-=0$ or $\alpha_+=0$; we assign all eigenvalues to $-1$ in the former case and to $1$ in the latter case. In every case, the model satisfies the strip condition and its spectral ratio defined as in the definition \eqref{eq:Nmax} equals the preassigned $\overline k$. Now let $G$ be any smooth function with this value of $G(0)$, e.g., $G(x) \equiv G(0)$. The invariance equation for the interconnection of the model and the exosystem becomes $D \overline \pi (\omega) s(\omega)= s(\overline \pi(\omega)) + G(\overline \pi(\omega))(l(\omega) - l(\overline \pi (\omega)))$, and the identity map is a solution to the invariance equation. Therefore, the Taylor series of $m$ or its truncation is the unique polynomial or formal power series representative of the moment of the interconnection of the model and the exosystem. By definition of $m$, the Taylor series of $m$ is the unique polynomial or formal power series representative of the moment of the interconnection of the system and the exosystem. The proof is complete by Proposition~\ref{prop:refined-moment-matching}.
\end{proof}

\section{Illustrative example}\label{sec:example}
We close with an example in which the exosystem is not neutrally stable, so that the steady-state interpretation of the nonlinear moment is not available, and $k$ is finite. The example illustrates the non-uniqueness of solutions, the polynomial representative of the moment, its time-domain interpretation, and the construction of a reduced-order model.

\begin{figure}[t]
    \centering
    \includegraphics[width=\textwidth]{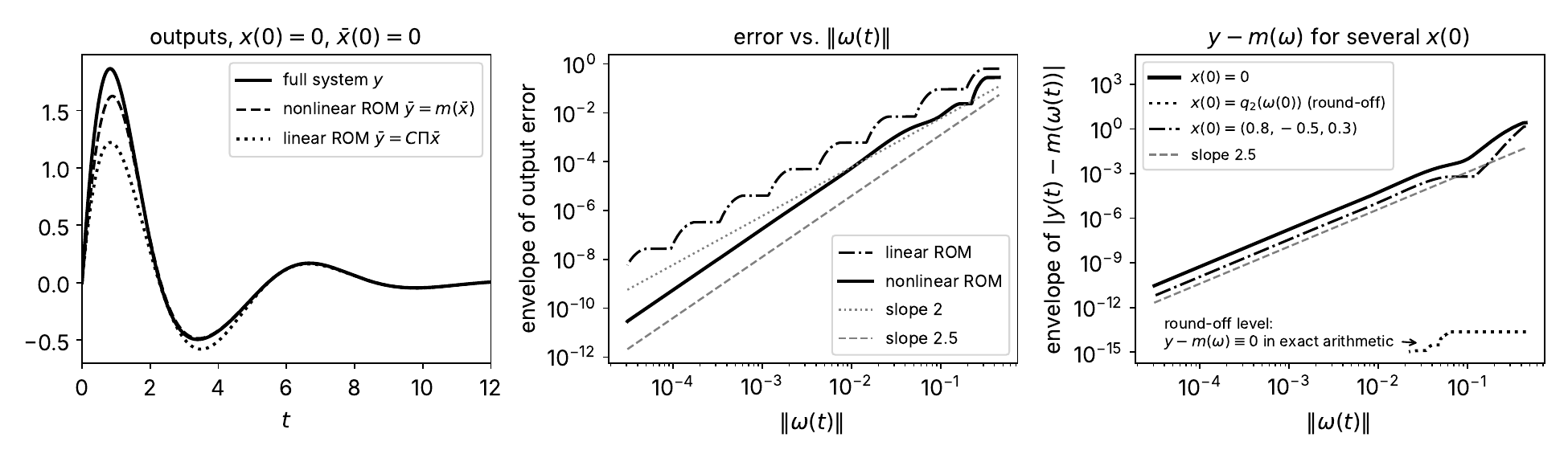}
    \caption{Simulation of Example~\ref{ex:transient-example} with $\omega(0)=(1.5,0)$. Left: outputs of the system, of the reduced-order model $\overline y=m(\overline x)$, and of the linear model $\overline y=C\Pi\overline x$, with $x(0)=0$ and $\overline x(0)=0$. Middle: envelopes of the output errors of the two models versus $\|\omega(t)\|$. Right: envelopes of $|y(t)-m(\omega(t))|$ for three initial conditions $x(0)$. For $x(0)=q_2(\omega(0))$, the plotted error is at the round-off level, since $q_2$ is an exact solution and $h\circ q_2=m$, so that $y(t)-m(\omega(t))$ vanishes identically in exact arithmetic.}
    \label{fig:transient-example}
\end{figure}

\begin{example}\label{ex:transient-example}
Consider the exosystem and the system
\begin{gather*}
    \dot\omega=S\omega:=\begin{bmatrix}-\tfrac25&-1\\[3pt]1&-\tfrac25\end{bmatrix}\omega,~u=\omega_1=L\omega;\\
    \dot x=Ax+Bu+\begin{bmatrix}\phantom{-}0\\[3pt]\phantom{-}1\\[3pt]-1\end{bmatrix}u^2,~y=x_1+x_2+x_3+x_1^2;\\
    A=\begin{bmatrix}-1&\phantom{-}0&\phantom{-}0\\[3pt]\phantom{-}0&-2&\phantom{-}0\\[3pt]\phantom{-}0&\phantom{-}0&-3\end{bmatrix},~B=\begin{bmatrix}1\\[3pt]1\\[3pt]1\end{bmatrix}.
\end{gather*} The exosystem generates exponentially decaying oscillations, since $\sigma(S)=\{-\tfrac25\pm i\}$; in particular, it is not neutrally stable, and Proposition~\ref{prop:steady-state-response} does not apply. We have $\alpha_+=0$, $\alpha_-=\tfrac25$, $\kappa_-=1$, and $\kappa_+=\infty$, so that the strip condition holds with $k=\kappa_-/\alpha_-=2.5$ and $N_{\max}=2$.

\noindent\textbf{Solutions and their non-uniqueness.} A direct calculation shows that the polynomial $q_2$ of Theorem~\ref{thm:approximate-jet} is a solution to the invariance equation:
\begin{equation*}
    q_2(\omega)=\Pi\omega+\begin{bmatrix}0\\ \tfrac{215}{408}\omega_1^2+\tfrac{25}{68}\omega_1\omega_2+\tfrac{125}{408}\omega_2^2\\ -\tfrac{855}{2431}\omega_1^2-\tfrac{50}{221}\omega_1\omega_2-\tfrac{250}{2431}\omega_2^2\end{bmatrix},\quad
    \Pi=\begin{bmatrix}\tfrac{15}{34}&\tfrac{25}{34}\\[2pt] \tfrac{40}{89}&\tfrac{25}{89}\\[2pt] \tfrac{65}{194}&\tfrac{25}{194}\end{bmatrix},
\end{equation*}
where $\Pi$ is the solution to the Sylvester equation \eqref{eq:sylvester}. In polar coordinates, the exosystem is written as $\frac{d}{dt}\|\omega\|=-\tfrac25\|\omega\|$, $\dot\theta=1$, where $\theta$ is the polar angle. Since $\theta+\frac52\log\|\omega\|$ is constant along each trajectory of the exosystem and $\|\omega(t)\|^{5j/2}=e^{-jt}\|\omega(0)\|^{5j/2}$, the function $\|\omega\|^{5j/2}\Psi_j(\theta+\frac52\log\|\omega\|)$ is multiplied by $e^{-jt}$ along the flow for $j=1,2,3$. Here, $\Psi_j$ is required to be $2\uppi$-periodic because $\theta$ is defined only modulo $2\uppi$. Hence, for every triple $\Psi=(\Psi_1,\Psi_2,\Psi_3)$ of $2\uppi$-periodic $\mathcal C^2$ functions,
\begin{equation*}
    \pi_\Psi(\omega):=q_2(\omega)+\begin{bmatrix}\|\omega\|^{5/2}\,\Psi_1\!\left(\theta+\tfrac52\log\|\omega\|\right)\\[2pt] \|\omega\|^{5}\,\Psi_2\!\left(\theta+\tfrac52\log\|\omega\|\right)\\[2pt] \|\omega\|^{15/2}\,\Psi_3\!\left(\theta+\tfrac52\log\|\omega\|\right)\end{bmatrix}
\end{equation*}
is a $\mathcal C^2$ solution to the invariance equation. If $\Psi_1$ is a nonzero constant, $\pi_\Psi$ is not $\mathcal C^3$, which is consistent with Theorem~\ref{thm:existence}, which guarantees $\mathcal C^N$ solutions only for $N<k=2.5$. The difference $\pi_\Psi-q_2$ is $O(\|\omega\|^{5/2})$ but not $O(\|\omega\|^{b})$ for $b>5/2$ unless $\Psi_1\equiv0$, so that the solutions have contact of order exactly $k$, in accordance with Theorem~\ref{thm:finite-order} and Remark~\ref{rem:k-minus}, since $A$ and $S$ are semisimple. The differences between these solutions illustrate the two orders $k$ and $N_*$ discussed after Theorem~\ref{thm:finite-order}. Depending on which of $\Psi_1$, $\Psi_2$, and $\Psi_3$ vanish identically, the difference $\pi_\Psi-q_2$ first appears at the order $5/2$, $5$, or $15/2$. On the other hand, since $A$ is Hurwitz, $A_-=A$ and $\frac{\max_{\mu\in\sigma(A_-)}|\mathrm{Re}\mu|}{\min_{\lambda\in\sigma(S)}|\mathrm{Re}\lambda|}=7.5$, and $\mathcal L_1,\ldots,\mathcal L_7$ are invertible, so that the conditions of \cite[Theorem~1.2]{cabre2003parameterization1} hold for $N_*=8$. Hence, $q_2$ is the unique Lipschitz solution to the invariance equation of the form $q_2(\w)+O(\|\w\|^8)$. Note that Theorem~\ref{thm:finite-order} says that every $\mathcal C^1$ solution and $q_2$ have contact of order $k^-=(5/2)^-$. There may exist solutions that have contact of order $k^-$ with $q_2$ but are not of the form $q_2(\w) + O(\| \w\|^j)$ for some $j \in (2,8)$: if $\Psi_1\equiv0$ and $\Psi_2\not\equiv0$, then $q_2(\w) - \pi_\Psi(\w) = O(\|\w\|^5)$, but not $O(\|\w\|^b)$ for $b>5$.

\noindent\textbf{Moment.} By Corollary~\ref{cor:moment-power-series}\textbf{(a)}, the moment is $[m]_{2.5^-}$, where $m=T_2(h\circ q_2)=h\circ q_2$ is given by
\begin{equation*}
    m(\omega)\approx1.2257\,\omega_1+1.1451\,\omega_2+0.3699\,\omega_1^2+0.7902\,\omega_1\omega_2+0.7442\,\omega_2^2.
\end{equation*}
The linear part of $m$ is the linear moment $C\Pi\omega$, and the quadratic part is the nonlinear information carried by the moment; since $k>2$, this information is determined by the system and the exosystem (see Remark~\ref{rem:information-content}).

\noindent\textbf{Reduced-order model.} Since $(L,S)$ is observable, Proposition~\ref{prop:parametrized-family-ROM} yields the two-dimensional model
\begin{equation*}
    \dot{\overline x}=S\overline x+G(u-L\overline x),\quad\overline y=m(\overline x),\quad G=\begin{bmatrix}\phantom{-}\tfrac{16}{5}\\[3pt]-\tfrac{39}{25}\end{bmatrix},
\end{equation*}
for which $\sigma(S-GL)=\{-2\}$, so that the model satisfies the strip condition with $\overline k=2/\alpha_-=5\ge k$. Hence the model matches the moment with $\tilde k=k=2.5$, that is, it reproduces the coefficients of $m$ up to order $\tilde N=2$ by Proposition~\ref{prop:refined-moment-matching}.

\noindent\textbf{Simulation.} Figure~\ref{fig:transient-example} shows the simulation results for $\omega(0)=(1.5,0)$. In the left panel, the system and the model are initialized at $x(0)=0$ and $\overline x(0)=0$, which are not on the invariant manifolds, and the output of the model is compared with that of the linear model obtained by replacing $m$ by its linear part $C\Pi\overline x$. The middle panel shows that the output error of the model decays as $\|\omega(t)\|^{2.5}$, whereas that of the linear model decays only as $\|\omega(t)\|^{2}$. The right panel shows that $|y(t)-m(\omega(t))|$ decays as $\|\omega(t)\|^{2.5}$ for several initial conditions $x(0)$, as predicted by Proposition~\ref{prop:transient}.
\end{example}

We make two remarks on Example~\ref{ex:transient-example}. First, every trajectory of the cascade with $\omega(0)\neq0$ lies on the graph of $\pi_\Psi$ for some $\Psi$, since $x(t)-q_2(\omega(t))=e^{At}(x(0)-q_2(\omega(0)))$ and the value of each $\Psi_j$ can be prescribed at the value that $\theta+\frac52\log\|\omega\|$ takes on the trajectory: the non-uniqueness of the solutions to the invariance equation reflects the freedom in the initial condition of the system, and the moment $[h\circ\pi]_{k^-}$ collects exactly the part of the output that is shared by all of them. Second, the reduced-order model reproduces this shared part without requiring $x(0)=\pi(\omega(0))$ or $\overline x(0)=\omega(0)$, since both the system and the model are exponentially stable.

\section{Conclusion}\label{sec:conclusion}
Using the L--P approach under the strip condition, this paper has addressed the existence of solutions to the invariance equation, which is central in nonlinear moment theory, and has established an equivalence relation among all its $\mathcal C^1$ solutions. Although the existence of local invariant manifolds of the cascade can be derived within the classical theory of invariant manifolds \cite{kelley1967stable,hirsch1977invariant,irwin1980pseudostable,vanderbauwhede1989centre}, the contact of all $\mathcal C^1$ solutions beyond the case of purely imaginary spectrum, its interpretation as the nonlinear moment, and its consequences for moment matching were, to the best of our knowledge, not available. Since Theorem~\ref{thm:finite-order} concerns the invariance equation itself, it also applies to other contexts in which this equation arises for a system driven by an autonomous exosystem, such as the invariance part of the FBI equation in output regulation for a given control law $u=c(\omega)$; see the discussion after Corollary~\ref{cor:inv-eq-power-series}. Based on the contact of solutions, we refined the definition of nonlinear moments as a contact class of functions, rather than a single function. Example~\ref{ex:k-andk} showed that the order $k^-$ cannot be improved in general. In the settings of the classical nonlinear moment, namely when the system is exponentially stable and the exosystem is neutrally stable, or when $A$ is hyperbolic and the exosystem admits a first integral, the $\mathcal C^1$ solution is unique on a neighborhood of the origin, so that the contact class recovers the classical moment. Our results rely on the polynomial $q_N$ and on the spectral ratio $k$, and hence on the smoothness of $f$, $l$, and $s$ at the origin assumed in Section~\ref{sec:prelim}; if $f$ is not differentiable at the origin, neither the linearization nor $k$ is defined, and the invariant graph of \cite{shakib2023model}, obtained from convergence, plays the role of the center manifold. The two approaches are thus complementary: ours is local and spectral, whereas that of \cite{shakib2023model} is global and based on convergence.

The equivalence class definition of the moment naturally extends the concept of moment matching. The equivalence class admits a unique polynomial or formal power series representative, which plays a crucial role in the extended concept of moment matching. The concept of parametrized family of moment-matching reduced-order models remains valid. Example~\ref{ex:transient-example} illustrated these results for a decaying exosystem, for which the steady-state interpretation of the nonlinear moment is not available: the moment admits a time-domain interpretation shared by all trajectories, and a two-dimensional reduced-order model with a quadratic output reproduces the output with an error of order $k^-$, whereas a linear model achieves only order $2$.

Several directions remain open for future investigation. First, although the present formulation substantially reduces the asymmetry between the linear and the nonlinear moment theories, the strip condition is still much stronger than its linear counterpart $\sigma(S) \cap \sigma(A) = \varnothing$; it compares the stable part of $A$ only with the growth of the exosystem in backward time and the unstable part of $A$ only with the growth in forward time, but it still requires $\sigma(S)\cup i\R$ to be separated from $\sigma(A)$ by a vertical strip. Relaxing it is the most important open problem left by this work. Second, it would be of interest to investigate the system-theoretic properties of the nonlinear system $(h,f)$ through a nonlinear moment analogue of the frequency response, and whether contact of order $k^-$ can be improved to contact of order $k$ when $A$ and $S$ are semisimple at the eigenvalues that realize $k$, as mentioned in Remark~\ref{rem:k-minus}. 

\appendix

\section{Derivatives of compositions}\label{app:derivatives}
This appendix collects two elementary facts on higher-order derivatives of compositions, which are used in Appendices~\ref{app:cutoff} and~\ref{app:existence}. Throughout, for $c \in \R^n$ and a row vector $v^\top \in \R^{1 \times \nu}$, $c\, v^\top \in \R^{n \times \nu}$ denotes the usual matrix product. The first fact is a direct consequence of the chain rule (or the multivariable Fa\`a di Bruno formula \cite{constantine1996multivariate}).

\begin{lemma}\label{lem:FaaDiBruno}
Let $j \in \Z_{>0}$ and let $g, \psi \in \mathcal C^j$. Then
\begin{equation}\label{eq:FaaDiBruno}
D^j (g \circ \psi) = D g(\psi)D^j \psi + \sum_{i=2}^j \sum_{\substack{j_1+\cdots+j_i = j\\ j_1,\cdots,j_i \in \Z_{>0}}} c_{j_1\cdots j_i} D^i g(\psi) [D^{j_1} \psi,\cdots,D^{j_i}\psi],
\end{equation}
where $c_{j_1\cdots j_i} \in \Z_{\ge0}$ is a coefficient that depends only on $(j_1,\cdots,j_i)$, and $c_{1\cdots1}=1$.
\end{lemma}

The second fact concerns a composition of the form $\eta\mapsto F(\eta,r(\eta))$ with a change of variables $\psi$, and identifies the unique term that contains the highest derivative of $r$. For $F \in \mathcal C^j(\R^\nu\times\R^n,\R^n)$, $r\in\mathcal C^{j-1}(\R^\nu,\R^n)$, integers $a,b\ge0$, and $d_1,\ldots,d_b\in\{1,\ldots,j-1\}$ with $i:=a+d_1+\cdots+d_b\le j$, define $\mathcal G_{a,d}[r](\eta)\in\mathrm{Sym}^i(\R^\nu,\R^n)$, $d:=(d_1,\ldots,d_b)$, as the symmetrization of the $i$-linear map
\begin{multline}\label{eq:app-G-def}
    (v_1,\ldots,v_i)\mapsto D_\eta^{a}D_z^{b}F (\eta,r(\eta) )  \big[ v_1,\ldots,v_a;\\ D^{d_1}r(\eta)[v_{a+1},\ldots,v_{a+d_1}],\ldots,D^{d_b}r(\eta)[v_{i-d_b+1},\ldots,v_i] \big],
\end{multline}
that is, the first $a$ arguments are inserted into the $\eta$-slots of $D_\eta^aD_z^bF$, and the next $d_\iota$ arguments are inserted into $D^{d_\iota}r$ for $\iota=1,\ldots,b$.

\begin{lemma}\label{lem:top-order}
Let $j\in\Z_{>0}$, $F:\R^\nu\times\R^n\to\R^n$ and $\psi: \R^\nu\to \R^\nu$ be $\mathcal C^j$, $r:\R^\nu\to\R^n$ be $\mathcal C^{j-1}$, and $\Xi_r(\eta):=F(\eta,r(\eta))$. Define $\mathcal E_j[r]:\R^\nu\to\mathrm{Sym}^j(\R^\nu,\R^n)$ by
\begin{equation}\label{eq:app-Rj-def}
    \mathcal E_j[r](\omega):=\sum_{i=1}^{j}\ \sum_{\substack{j_1+\cdots+j_i=j\\ j_1,\ldots,j_i\in\Z_{>0}}}\ \sum_{(a,d)} c_{a,d}\,c_{j_1\cdots j_i}\, \mathcal G_{a,d}[r](\psi(\omega))\big[D^{j_1}\psi(\omega) ,\cdots,D^{j_i}\psi(\omega)\big],
\end{equation}
where $c_{j_1\cdots j_i}$ is as in Lemma~\ref{lem:FaaDiBruno}, the innermost sum runs over all $(a,d)$, $d=(d_1,\ldots,d_b)$, with $b \ge 0$, $a+d_1+\cdots+d_b=i$ and $1 \le d_\iota\le j-1$, and $c_{a,d}\in\Z_{\ge0}$ are the coefficients produced by Lemma~\ref{lem:FaaDiBruno} applied to $g=F$ and $\eta\mapsto(\eta,r(\eta))$, which depend only on $(a,d)$. If, in addition, $r\in\mathcal C^j$, then
\begin{equation}\label{eq:app-top-order}
    D^j(\Xi_r\circ\psi)(\omega)=D_zF\big(\psi(\omega),r(\psi(\omega))\big)\,D^jr(\psi(\omega))\big[D\psi(\omega)\cdot,\ldots,D\psi(\omega)\cdot\big]+\mathcal E_j[r](\omega).
\end{equation}
\end{lemma}
\begin{proof}
Lemma~\ref{lem:FaaDiBruno} with $g=\Xi_r$ gives
\begin{equation*}
D^j(\Xi_r\circ\psi)=\sum_{i=1}^j\sum_{j_1+\cdots+j_i=j}c_{j_1\cdots j_i}D^i\Xi_r(\psi)[D^{j_1}\psi,\ldots,D^{j_i}\psi].
\end{equation*}
Applying Lemma~\ref{lem:FaaDiBruno} once more with $g=F$ and the map $\eta\mapsto(\eta,r(\eta))$, whose first derivative is $(I,Dr)$ and whose $j'$th derivative is $(0,D^{j'}r)$ for $j'\ge2$, shows that $D^i\Xi_r$ is a sum, with nonnegative integer coefficients, of terms of the form \eqref{eq:app-G-def} with $a+d_1+\cdots+d_b=i$ and $d_\iota\ge1$. Such a term contains $D^jr$ only if $d_\iota=j$ for some $\iota$; since $a+\sum_\iota d_\iota=i\le j$, this forces $i=j$, $a=0$, $b=1$ and $d_1=j$, and the corresponding term of $D^j\Xi_r$ is $D_zF(\eta,r(\eta))D^jr(\eta)$ with coefficient one. Moreover, $i=j$ together with $j_1+\cdots+j_i=j$ and $j_\iota\ge1$ forces $j_1=\cdots=j_i=1$, and $c_{1\cdots1}=1$. This yields the first term of the identity \eqref{eq:app-top-order}, and all the remaining terms involve $D^{j'}r$ only with $j'\le j-1$; they constitute $\mathcal E_j[r]$.
\end{proof}

We refer to the first term on the right-hand side of the identity \eqref{eq:app-top-order} as the \emph{top-order term}; it is the only term that contains the highest derivative $D^jr$, and it is linear in $D^jr$. Note that $\mathcal E_j[r]$ is well defined for $r\in\mathcal C^{j-1}$. For instance, for $j=1$ and $j=2$, with all derivatives of $F$ evaluated at $(\psi(\omega),r(\psi(\omega)))$ and all derivatives of $r$ at $\psi(\omega)$,
\begin{align*}
    D(\Xi_r\circ\psi)(\cdot)
    &=\boxed{D_zF\,Dr[D\psi(\cdot)]}+D_\eta F[D\psi(\cdot)],\\
    D^{2}(\Xi_r\circ\psi)(\cdot,\cdot)
    &=\boxed{D_zF\,D^{2}r[D\psi(\cdot),D\psi(\cdot)]}+D^{2}_\eta F[D\psi(\cdot),D\psi(\cdot)]\\
    &\qquad +2D_\eta D_zF[D\psi(\cdot);Dr[D\psi(\cdot)]]+D^{2}_zF[Dr[D\psi(\cdot)],Dr[D\psi(\cdot)]]\\
    &\qquad+D_\eta F[D^2\psi(\cdot,\cdot)]+D_zF\,Dr[D^2\psi(\cdot,\cdot)],
\end{align*}
where the boxed terms are the top-order terms and the remaining terms constitute $\mathcal E_1[r]$ and $\mathcal E_2[r]$, respectively.

\section{Proof of Lemma~\ref{lem:cutoff}}\label{app:cutoff}
Write $s_{\mathrm{nl}}(\omega):=s(\omega)-S\omega.$ 
By the estimate \eqref{eq:qN-residual}, setting $z=0$ in the definition \eqref{eq:FN-def} gives $\mathcal F_N(\omega,0)=O(\|\w\|^{N+1})$; Taylor's theorem therefore gives
\begin{equation}\label{eq:FN-omega-derivatives}
    D_\omega^j\mathcal F_N(\omega,0)=O(\|\omega\|^{N+1-j}),~\text{for}~j=0,\ldots,N,
\end{equation}
while $D_xf(0,0)=A$ together with $q_N(0)=0$ and $l(0)=0$ gives 
\begin{equation}\label{eq:FN-z-derivative}
    \|D_z\mathcal F_N(\omega,z)\|\le C(\|\omega\|+\|z\|),
\end{equation}
around the origin.

Since $s_{\mathrm{nl}}(0)=0, Ds_{\mathrm{nl}}(0)=0,$ and the estimates \eqref{eq:FN-omega-derivatives}--\eqref{eq:FN-z-derivative} hold in a neighborhood of the origin, one can choose $0<\rho_{\mathrm{loc}}\le1$ and constants $c_s,c_{Ds},c_z,c_0,\ldots,c_N>0$ such that
\begin{align}
    &\|s_{\mathrm{nl}}(\omega)\| \le c_s\|\omega\|^2,~\|Ds_{\mathrm{nl}}(\omega)\|\le c_{Ds}\|\omega\|, \label{eq:app-snl-bounds}\\
    &\|D_\omega^j\mathcal F_N(\omega,0)\| \le c_j\|\omega\|^{N+1-j}~\text{for}~j=0,\ldots,N, \label{eq:app-FN-jet-local}\\
    &\|D_z\mathcal F_N(\omega,z)\|\le c_z(\|\omega\|+\|z\|) \label{eq:app-FN-z-local}
\end{align}
whenever \(\|\omega\|\le2\rho_{\mathrm{loc}}\), and, in the estimate \eqref{eq:app-FN-z-local}, also \(\|z\|\le2\rho_{\mathrm{loc}}\).

\medskip
\noindent\textbf{Proof of \textup{(a)}.}
Temporarily, fix \(\rho\in(0,\rho_{\mathrm{loc}})\). If \(\|\omega\|\le2\rho\), then the estimate \eqref{eq:app-snl-bounds} gives $\|s_\rho(\omega)-S\omega\|\le \|s_{\mathrm{nl}}(\omega)\|\le c_s\|\omega\|^2\le 2c_s\rho\|\omega\|.$
If \(\|\omega\|>2\rho\), then the cutoff vanishes and
\(s_\rho(\omega)=S\omega\). Hence, we have the uniform estimate
\begin{equation}\label{eq:app-srho-linear-bound}
    \|s_\rho(\omega)-S\omega\|
    \le 2c_s\rho\|\omega\|,~\forall\omega\in\mathbb R^\nu,~\forall \rho\in (0,\rho_{\mathrm{loc}}).
\end{equation}
In particular, $\|s_\rho(\omega)\|\le(\|S\|+2c_s\rho_{\mathrm{loc}})\|\omega\|,$ so every \(s_\rho\), \(0<\rho<\rho_{\mathrm{loc}}\), has at most linear growth. Therefore its maximal flow is complete.

Let \(\epsilon>0\). By the Jordan normal form, there exists a constant
\(M_\epsilon>0\), depending on \(\epsilon\) and \(S\), but not on
\(\rho\), such that $\|e^{St}\| \le M_\epsilon e^{(\alpha_++\epsilon/2)t}$ and $\|e^{-St}\| \le M_\epsilon e^{(\alpha_-+\epsilon/2)t}$ for all $t \ge 0$; the second bound follows from the first one applied to $-S$, because $\max_{\lambda\in\sigma(-S)}\mathrm{Re}\lambda\le\alpha_-$ by the definition \eqref{eq:spectral-params}. For \(t\ge0\), the variation-of-constants formula and the estimate \eqref{eq:app-srho-linear-bound} yield $\|\varphi_t^\rho(\omega)\|
    \le M_\epsilon e^{(\alpha_++\epsilon/2)t}\|\omega\|+2M_\epsilon c_s\rho
      \int_0^t e^{(\alpha_++\epsilon/2)(t-\tau)}
      \|\varphi_\tau^\rho(\omega)\| d\tau.$
Gronwall's inequality gives
\begin{equation}\label{eq:app-flow-before-choice}
    \|\varphi_t^\rho(\omega)\|
    \le M_\epsilon\|\omega\|
       e^{(\alpha_++\epsilon/2+2M_\epsilon c_s \rho)t}.
\end{equation}
The flow of the vector field $-s_\rho$ is $t\mapsto\varphi_{-t}^\rho$, and the estimate \eqref{eq:app-srho-linear-bound} also holds with $s_\rho$ and $S$ replaced by $-s_\rho$ and $-S$. Hence the same argument with the bound on $\|e^{-St}\|$ gives
\begin{equation}\label{eq:app-flow-before-choice-backward}
    \|\varphi_{-t}^\rho(\omega)\|
    \le M_\epsilon\|\omega\|
       e^{(\alpha_-+\epsilon/2+2M_\epsilon c_s \rho)t},~\forall t\ge0.
\end{equation}
Define
\begin{equation}\label{eq:app-rho-flow}
    \rho_{\epsilon}^{\mathrm{flow}}
    :=\min\left\{
        \rho_{\mathrm{loc}},
        \frac{\epsilon}{4M_\epsilon c_s}
    \right\}.
\end{equation}
Then, for every \(\rho\in(0,\rho_{\epsilon}^{\mathrm{flow}})\), the exponents in the estimates \eqref{eq:app-flow-before-choice} and \eqref{eq:app-flow-before-choice-backward} are smaller than $(\alpha_++\epsilon)t$ and $(\alpha_-+\epsilon)t$, respectively, for $t>0$. Thus the estimate \eqref{eq:flow-bound} holds with \(C_\epsilon:=M_\epsilon\),
uniformly for \(\rho\in(0,\rho_{\epsilon}^{\mathrm{flow}})\).

Differentiation gives $Ds_\rho(\omega)-S=\frac1\rho { s_{\mathrm{nl}}(\omega) D\chi\!\left(\frac\omega\rho\right)}+\chi\!\left(\frac\omega\rho\right) Ds_{\mathrm{nl}}(\omega).$ On \(B_{2\rho}\), the two terms are bounded, respectively, by $4c_s\|D\chi\|_\infty\rho~\text{and}~2c_{Ds}\rho,$ and outside \(B_{2\rho}\) both terms vanish. Consequently, with $C_D^*:=4c_s\|D\chi\|_\infty+2c_{Ds},$ we obtain
\begin{equation}\label{eq:Dsrho-close}
    \sup_{\omega\in\mathbb R^\nu}
    \|Ds_\rho(\omega)-S\|
    \le C_D^*\rho,
    ~\forall \rho\in(0,\rho_{\mathrm{loc}}).
\end{equation}
In particular, the constant \(C_D^*\) is independent of \(\rho\).

Let $\Gamma_\rho(t,\omega):=D\varphi_t^\rho(\omega).$ It satisfies
\begin{equation}\label{eq:first-variational}
    \dot\Gamma_\rho(t,\omega)
    =Ds_\rho(\varphi_t^\rho(\omega))\Gamma_\rho(t,\omega),~\Gamma_\rho(0,\omega)=I.
\end{equation}
Using the estimate \eqref{eq:Dsrho-close}, the variation-of-constants formula for
the equation \eqref{eq:first-variational}, and the same Gronwall argument as
above, we obtain
\begin{equation*}
\|\Gamma_\rho(t,\omega)\|\le M_\epsilon e^{(\alpha_++\epsilon/2+M_\epsilon C_D^*\rho)t},~\forall t\ge0.
\end{equation*} {Since $\Gamma_\rho(-t,\omega)$, $t\ge0$, satisfies the equation \eqref{eq:first-variational} with $s_\rho$ replaced by $-s_\rho$, and the estimate \eqref{eq:Dsrho-close} also holds with $s_\rho$ and $S$ replaced by $-s_\rho$ and $-S$, the same argument with the bound on $\|e^{-St}\|$ gives $\|\Gamma_\rho(-t,\omega)\|\le M_\epsilon e^{(\alpha_-+\epsilon/2+M_\epsilon C_D^*\rho)t}$, for all $t\ge0$.} Set $\rho_{\epsilon}^{\mathrm{var}}:=\min\left\{\rho_{\mathrm{loc}},\frac{\epsilon}{2M_\epsilon C_D^*} \right\},$ and finally define
\begin{equation}\label{eq:app-rhoepsilon-final}
    \rho_\epsilon
    :=\min\left\{
        \rho_{\epsilon}^{\mathrm{flow}},
        \rho_{\epsilon}^{\mathrm{var}}
    \right\}.
\end{equation}
Then, for every \(\rho\in(0,\rho_\epsilon)\), both the estimate \eqref{eq:flow-bound} and 
\begin{equation*}
\|D\varphi_t^\rho(\omega)\| \le M_\epsilon e^{(\alpha_++\epsilon)t},~~\|D\varphi_{-t}^\rho(\omega)\| \le M_\epsilon e^{(\alpha_-+\epsilon)t}
\end{equation*}
hold for all $t\ge0$ and \(\omega\in\mathbb R^\nu\). Thus, the estimate \eqref{eq:flow-deriv-bound} holds with \(C_\epsilon=M_\epsilon\), as in the estimate \eqref{eq:flow-bound}, independent of \(\rho\in(0,\rho_\epsilon)\).

It remains to estimate the higher variations, for which we use Lemma~\ref{lem:FaaDiBruno} in Appendix~\ref{app:derivatives}.
Fix \(\rho\in(0,\rho_\epsilon)\). Since \(\omega\mapsto s_\rho(\omega)-S\omega\) is smooth and compactly supported, $\Lambda_{a,\rho}:= \sup_{\omega\in\mathbb R^\nu}\|D^as_\rho(\omega)\|<\infty$ for every \(a\ge2\). Thus, $\frac{d}{dt}D^j \varphi_t^\rho = D^j \frac{d}{dt}\varphi_t^\rho = D^j (s_\rho \circ \varphi_t^\rho)$. From Lemma~\ref{lem:FaaDiBruno} with $g = s_\rho$ and $\psi = \varphi_t^\rho$, we obtain
\begin{equation}\label{eq:app-jth-variation}
    \frac{d}{dt}D^j\varphi_t^\rho(\omega)
    =Ds_\rho(\varphi_t^\rho(\omega))
       D^j\varphi_t^\rho(\omega)
     +Q_{j,\rho}(t,\omega),
\end{equation}
where $Q_{j,\rho}$ corresponds to the double sum of the right-hand side of the formula \eqref{eq:FaaDiBruno}, namely $Q_{j,\rho}(t,\omega) =\sum_{i}\ \sum_{{ j_1,\ldots,j_i}} c_{{ j_1\ldots j_i}}  D^i s_\rho(\varphi_t^\rho(\omega)) [D^{{j_1}}\varphi_t^\rho(\omega),\ldots, D^{{ j_i}}\varphi_t^\rho(\omega)].$ Since $i\ge2$ and every ${ j_a}\ge1$, each index satisfies ${ j_a}\le j-1$ for $a=1,\ldots,i$. Thus, the flow derivatives $D^{{ j_1}}\varphi_t^\rho,\ldots,D^{{ j_i}}\varphi_t^\rho$ can be bounded by the induction hypotheses, and their orders sum to $j$. Let $\Phi_\rho(t,\tau;\omega) :=D\varphi_{t-\tau}^\rho(\varphi_\tau^\rho(\omega)).$ Since $\varphi_{t-\tau}^\rho$ is a forward-time flow map if $t\ge\tau$ and a backward-time flow map if $t\le\tau$, the first-variation estimate gives

\begin{equation}\label{eq:app-transition-bound}
    \|\Phi_\rho(t,\tau;\omega)\|
    \le\begin{cases}
    C_\epsilon e^{(\alpha_++\epsilon)(t-\tau)}, & \text{if}~t\ge\tau,\\
    C_\epsilon e^{(\alpha_-+\epsilon)(\tau-t)}, & \text{if}~t\le\tau.
    \end{cases}
\end{equation}
Bounding each term $D^i s_\rho(\varphi_t^\rho(\omega))$ by $\Lambda_{i,\rho}$, we obtain
\begin{equation*}
\|Q_{j,\rho}(t,\omega)\| \le \tilde C_{j,\rho,\epsilon} e^{j(\alpha_++\epsilon)t},~~\|Q_{j,\rho}(-t,\omega)\| \le \tilde C_{j,\rho,\epsilon} e^{j(\alpha_-+\epsilon)t},~~\forall t\ge0,
\end{equation*} uniformly in $\omega$. Since \(D^j\varphi_0^\rho=0\) for \(j\ge2\), the variation-of-constants formula for the equation \eqref{eq:app-jth-variation} and the estimate \eqref{eq:app-transition-bound} yield, for \(t\ge0\), 
\begin{align*}
\|D^j\varphi_t^\rho(\omega)\| &\le C_\epsilon\tilde C_{j,\rho,\epsilon} \int_0^t e^{(\alpha_++\epsilon)(t-\tau)} e^{j(\alpha_++\epsilon)\tau} d\tau\\
&\le \frac{C_\epsilon\tilde C_{j,\rho,\epsilon}} {(j-1)(\alpha_++\epsilon)} e^{j(\alpha_++\epsilon)t},
\end{align*}
where the last inequality uses $j\ge2$ and $\alpha_++\epsilon>0$. The same argument on the interval $[-t,0]$, with $\alpha_+$ replaced by $\alpha_-$, gives the bound on $\|D^j\varphi_{-t}^\rho(\omega)\|$; here $\alpha_-+\epsilon>0$ is used. Since $\alpha_+\ge0$ and $\alpha_-\ge0$ by the definition \eqref{eq:spectral-params}, these exponents are positive. This proves
the estimate \eqref{eq:higher-flow-deriv-bound} for \(j=2,\ldots,N\), with constants
\(C_{j,\rho,\epsilon}\) that may depend on the fixed \(\rho\).

\medskip
\noindent\textbf{Proof of \textup{(b)}.}
By the fundamental theorem of calculus in the \(z\)-variable, $\mathcal F_N(\omega,z)=\mathcal F_N(\omega,0)+\int_0^1 D_z \mathcal F_N(\omega,\vartheta z) z  d \vartheta$, so that
the estimate \eqref{eq:app-FN-jet-local} with \(j=0\) and
the estimate \eqref{eq:app-FN-z-local} give, after increasing a constant if
necessary,
\begin{equation}\label{eq:app-FN-basic}
    \|\mathcal F_N(\omega,z)\|
    \le C_F (
       \|\omega\|^{N+1}
       +\|\omega\|\|z\|
       +\|z\|^2
     )
\end{equation}
for \(\|\omega\|,\|z\|\le2\rho_{\mathrm{loc}}\), where \(C_F\) depends on
\(\rho_{\mathrm{loc}}\), but not on \(\rho\).

Next, $D_z\mathcal F_{N,\rho}(\omega,z)
=\chi\!\left(\frac\omega\rho\right)
\Bigg[
\frac1\rho { \mathcal F_N(\omega,z) D\chi\!\left(\frac z\rho\right)}
+\chi\!\left(\frac z\rho\right)
       D_z\mathcal F_N(\omega,z)
\Bigg].$ Whenever the right-hand side is nonzero, \(\|\omega\|,\|z\|\le2\rho\). Therefore, the estimates \eqref{eq:app-FN-basic}, \eqref{eq:app-FN-z-local}, \(\rho_{\mathrm{loc}}\le1\), and \(N\ge1\) imply $\|D_z\mathcal F_{N,\rho}(\omega,z)\| \le \frac{\|D\chi\|_\infty}{\rho} C_F ((2\rho)^{N+1}+8\rho^2 ) +4c_z\rho \le  [\|D\chi\|_\infty C_F(2^{N+1}+8)+4c_z ]\rho,$ in which $\rho^{N}\le\rho$ has been used. Thus the estimate \eqref{eq:FN-rho-lipschitz} holds for every
\(\rho\in(0,\rho_{\mathrm{loc}})\), with the uniform constant
\begin{equation}\label{eq:app-Cz-choice}
    C_z:=\|D\chi\|_\infty C_F(2^{N+1}+8)+4c_z.
\end{equation}

Finally, since \(\chi(0)=1\), one has $\mathcal F_{N,\rho}(\omega,0)=\chi(\omega/\rho)\mathcal F_N(\omega,0)$, and, for \(0\le j\le N\), Leibniz's rule gives, with the natural
symmetrization of multilinear products,
\begin{equation}\label{eq:app-Leibniz-jet}
\begin{aligned}
D_\omega^j\mathcal F_{N,\rho}(\omega,0)
={}&\sum_{a=0}^j\binom ja\rho^{-a}
D^a\chi\!\left(\frac\omega\rho\right)
D_\omega^{j-a}\mathcal F_N(\omega,0).
\end{aligned}
\end{equation}
The term with \(a=0\) is bounded by
\(c_j\|\omega\|^{N+1-j}\); for \(j=0\) this is the only term, so that the estimate \eqref{eq:FN-rho-jet-bound} holds with \(C_0:=c_0\). If \(a\ge1\), then
\(D^a\chi(\omega/\rho)\ne0\) only when $\rho\le\|\omega\|\le2\rho.$ On this annulus, $\rho^{-a}\le2^a\|\omega\|^{-a},$ and the estimate \eqref{eq:app-FN-jet-local} yields $\rho^{-a} \left\|D^a\chi\!\left(\frac\omega\rho\right)\right\| \left\|D_\omega^{j-a}\mathcal F_N(\omega,0)\right\|\le 2^a\|D^a\chi\|_\infty c_{j-a} \|\omega\|^{N+1-j}.$ That is, the factor $\rho^{-a}$ produced by differentiating the cutoff is absorbed by the $a$ additional powers of $\|\omega\|$ available on the annulus, which is what keeps the constant independent of $\rho$. Summing the finitely many terms in the formula \eqref{eq:app-Leibniz-jet} proves
the estimate \eqref{eq:FN-rho-jet-bound}, with a constant
\(C_j>0\) independent of
\(\rho\in(0,\rho_{\mathrm{loc}})\).

For each fixed \(\rho>0\), the map \(\mathcal F_{N,\rho}\) is smooth and compactly supported in \(B_{2\rho}\times B_{2\rho}\). Therefore all its derivatives of order at most \(N+1\) are bounded; these latter bounds are allowed to depend on the fixed \(\rho\). {Finally, the estimate \eqref{eq:FN-z-local-main} is the estimate \eqref{eq:app-FN-z-local} with \(C_{\mathrm{loc}}:=c_z\).}

To summarize, \(\rho_{\mathrm{loc}}\) and the local constants in the estimates \eqref{eq:app-snl-bounds}--\eqref{eq:app-FN-z-local} are chosen first. Given \(\epsilon>0\), the constants \(C_\epsilon=M_\epsilon\) and the threshold \(\rho_\epsilon\le\rho_{\mathrm{loc}}\) are then fixed by the definitions \eqref{eq:app-rho-flow}--\eqref{eq:app-rhoepsilon-final}. Estimates \eqref{eq:flow-bound} and \eqref{eq:flow-deriv-bound} hold with these same constants for every actual truncation radius \(\rho\in(0,\rho_\epsilon)\), while the constants in the estimate \eqref{eq:higher-flow-deriv-bound} depend on the actual truncation radius $\rho\in(0,\rho_\epsilon)$. All estimates in \textbf{(b)} hold uniformly for every \(\rho\in(0,\rho_{\mathrm{loc}})\). \hfill\proofbox

\section{Fixed point of the truncated L--P operator}\label{app:existence}
Throughout this appendix, $N\in\Z_{>0}$ satisfies the condition \eqref{eq:N-cond}, $q_N$ is given by Theorem~\ref{thm:approximate-jet}, and $\rho_{\mathrm{loc}}$, $\rho_\epsilon$, $C_\epsilon$, $C_{j,\rho,\epsilon}$, $C_z$, $C_j$, and $C_{\mathrm{loc}}$ denote the constants of Lemma~\ref{lem:cutoff}. Choose $\kappa_-'\in(N\alpha_-,\kappa_-)$ and $\kappa_+'\in(N\alpha_+,\kappa_+)$, which is possible by the condition \eqref{eq:N-cond}, and then $\epsilon>0$ so that
\begin{equation}\label{eq:weighted-gap-preparation}
    N(\alpha_-+\epsilon)<\kappa_-',~N(\alpha_++\epsilon)<\kappa_+',
\end{equation}
and set $\beta_-:=\alpha_-+\epsilon$ and $\beta_+:=\alpha_++\epsilon$. Choose $\gamma$ that satisfies
\begin{equation}\label{eq:q-choice}
    N<\gamma<\min\left\{N+1,\frac{\kappa_-'}{\beta_-},\frac{\kappa_+'}{\beta_+}\right\}.
\end{equation}
Set $\zeta_-:=\kappa_-'-\gamma\beta_-$, $\zeta_+:=\kappa_+'-\gamma\beta_+$, and $\zeta:=\min\{\zeta_-,\zeta_+\}$, which is positive by the condition \eqref{eq:q-choice}, and let $K=K(\kappa_-',\kappa_+')$ be as in the inequality \eqref{eq:dichotomy}. By the estimate \eqref{eq:flow-bound}, for every $\rho\in(0,\rho_\epsilon)$, $t\ge0$, and $\omega\in\R^\nu$,
\begin{equation}\label{eq:app-onesided-weights}
    \|\omega_t^\rho\|^{\gamma}\le C_\epsilon^{\gamma}\|\omega\|^{\gamma}e^{\gamma\beta_+t},~
    \|\omega_{-t}^\rho\|^{\gamma}\le C_\epsilon^{\gamma}\|\omega\|^{\gamma}e^{\gamma\beta_-t},
\end{equation}
and the inequality \eqref{eq:dichotomy} gives
\begin{equation}\label{eq:app-onesided-integrals}
    \int_0^\infty Ke^{-\kappa_-'t}e^{\gamma\beta_-t}dt=\frac{K}{\zeta_-}\le\frac{K}{\zeta},~
    \int_0^\infty Ke^{-\kappa_+'t}e^{\gamma\beta_+t}dt=\frac{K}{\zeta_+}\le\frac{K}{\zeta}.
\end{equation}
In the first integral of the definition \eqref{eq:T-def}, which involves $e^{A_-t}P_-$, only the backward-time flow $\omega_{-t}^\rho$ appears, and in the second integral, which involves $e^{-A_+t}P_+$, only the forward-time flow $\omega_t^\rho$ appears. Therefore, every estimate of the first integral below combines $\kappa_-'$ with $\beta_-$, and every estimate of the second integral combines $\kappa_+'$ with $\beta_+$. For $j=0,\ldots,N$, define the weighted spaces
\begin{equation}\label{eq:weighted-spaces}
\mathcal X_{\gamma-j}^{(j)}:= \{\Phi\in
    \mathcal C^0 (\mathbb R^\nu,\mathrm{Sym}^j(\mathbb R^\nu,\mathbb R^n) ): \Phi(0)=0,~\|\Phi\|_{\gamma-j}<\infty \},
\end{equation}
where $\mathrm{Sym}^0(\mathbb R^\nu,\mathbb R^n):=\mathbb R^n$ and $\|\Phi\|_{\gamma-j}:=\sup_{\omega\ne0}\|\Phi(\omega)\|/\|\omega\|^{\gamma-j}$. Because $\gamma>N$, every exponent $\gamma-j$ is positive. Each space in the definition \eqref{eq:weighted-spaces} is a Banach space by the standard pointwise-limit argument. We abbreviate $\mathcal X_\gamma^{(0)}$ as $\mathcal X_\gamma$.

\begin{lemma}\label{lem:weighted-LP}
    Let $\kappa_-'$, $\kappa_+'$, $\epsilon$, and $\gamma$ be as in the conditions \eqref{eq:weighted-gap-preparation}--\eqref{eq:q-choice}. Then there exists $\rho\in(0,\rho_\epsilon)$ such that the operator $\mathcal I_{\rho,N}$ defined in the definition \eqref{eq:T-def} has a unique fixed point $r_\rho$ in a closed ball of $\mathcal X_\gamma$ centered at the origin. Moreover, $r_\rho\in\mathcal C^N(\R^\nu,\R^n)$ and $r_\rho(0)=0$.
\end{lemma}

\begin{proof}
Fix $\rho \in \left(0, \min\left\{\rho_\epsilon,~\frac{\zeta}{4KC_zC_\epsilon^{\gamma}},~\left(\frac{\zeta}{2^{N+3}KC_0C_\epsilon^{\gamma}}\right)^{1/N}\right\} \right)$. Then,
\begin{equation}\label{eq:app-graph-inside-cutoff}
    \Theta_\rho:=\frac{2KC_zC_\epsilon^{\gamma}\rho}{\zeta}<\frac12,~M_0:=\frac{4KC_0C_\epsilon^{\gamma}(2\rho)^{N+1-\gamma}}{\zeta} <\frac{\rho}{(2\rho)^\gamma}.
\end{equation}
Let $\mathcal B:=\{r\in\mathcal X_\gamma:\|r\|_\gamma\le M_0\}$. Note that $\mathcal F_{N,\rho}(\eta,z)=0$ whenever $\|\eta\|\ge2\rho$; hence the integrands in the definition \eqref{eq:T-def} vanish unless $\|\omega_{\mp t}^\rho\|<2\rho$.

Let $r\in\mathcal B$ and $\|\eta\|<2\rho$. Then $\|r(\eta)\|\le M_0\|\eta\|^\gamma<M_0(2\rho)^\gamma<\rho$ by the estimate \eqref{eq:app-graph-inside-cutoff}. Since $\chi(z/\rho)=1$ for $\|z\|\le\rho$, $\mathcal F_{N,\rho}(\eta,z)=\chi(\eta/\rho)\mathcal F_N(\eta,z)$ for all $z$ in the open ball $B_\rho$, which contains $r(\eta)$. Hence $D_z\mathcal F_{N,\rho}(\eta,r(\eta))=\chi(\eta/\rho)D_z\mathcal F_N(\eta,r(\eta))$, and the estimate \eqref{eq:FN-z-local-main} together with $\|r(\eta)\|\le M_0\|\eta\|^{\gamma}\le M_0(2\rho)^{\gamma-1}\|\eta\|$ yields
\begin{equation}\label{eq:app-Dz-sharp}
     \|D_z\mathcal F_{N,\rho} (\eta,r(\eta) ) \|
    \le C_*\|\eta\|,~C_*:=C_{\mathrm{loc}} (1+M_0(2\rho)^{\gamma-1} ),
\end{equation}
which holds trivially also for $\|\eta\|\ge2\rho$. Thus, along graphs of functions in $\mathcal B$, the $z$-derivative of $\mathcal F_{N,\rho}$ vanishes at the origin. 

Let $r\in\mathcal B$ and $\eta\in\R^\nu$. By the fundamental theorem of calculus, $\mathcal F_{N,\rho}(\eta,r(\eta))=\mathcal F_{N,\rho}(\eta,0)+\int_0^1D_z\mathcal F_{N,\rho}(\eta,\vartheta r(\eta))r(\eta)d\vartheta$. Hence the estimate \eqref{eq:FN-rho-jet-bound} with $j=0$, the estimate \eqref{eq:FN-rho-lipschitz}, and $\|r(\eta)\|\le M_0\|\eta\|^\gamma$ give $\|\mathcal F_{N,\rho}(\eta,r(\eta))\|\le C_0\|\eta\|^{N+1}+C_z\rho M_0\|\eta\|^\gamma$. Since the left-hand side vanishes for $\|\eta\|\ge2\rho$, and $\|\eta\|^{N+1}\le(2\rho)^{N+1-\gamma}\|\eta\|^\gamma$ for $\|\eta\|<2\rho$, we obtain
\begin{equation}\label{eq:app-F-along-graph}
    \|\mathcal F_{N,\rho}(\eta,r(\eta))\|
    \le \left(C_0(2\rho)^{N+1-\gamma}+C_z\rho M_0 \right)\|\eta\|^\gamma,~\forall\eta\in\R^\nu.
\end{equation}
Applying the inequality \eqref{eq:dichotomy} and the estimate \eqref{eq:app-F-along-graph} with $\eta=\omega^\rho_{-t}$ and $\eta=\omega^\rho_{t}$ to the first and the second integrals in the definition \eqref{eq:T-def}, respectively, and using the estimates \eqref{eq:app-onesided-weights}--\eqref{eq:app-onesided-integrals},
\begin{align*}
    \|(\mathcal I_{\rho,N}r)(\omega)\|
    &\le \left(C_0(2\rho)^{N+1-\gamma}+C_z\rho M_0\right)C_\epsilon^\gamma\|\omega\|^\gamma\\
    &\qquad\times\left(\int_0^\infty Ke^{-\kappa_-' t}e^{\gamma\beta_- t}dt+\int_0^\infty Ke^{-\kappa_+' t}e^{\gamma\beta_+ t}dt\right)\\
    &\le\frac{2KC_\epsilon^\gamma}{\zeta}\left(C_0(2\rho)^{N+1-\gamma}+C_z\rho M_0\right)\|\omega\|^\gamma\\
    &=\left(\frac{M_0}{2}+\Theta_\rho M_0\right)\|\omega\|^\gamma\le M_0\|\omega\|^\gamma,
\end{align*}
where the last inequality follows from $\Theta_\rho<1/2$. The integrands are continuous in $\omega$ and, on bounded sets of $\omega$, dominated by an integrable function of $t$; hence $\mathcal I_{\rho,N}r$ is continuous by dominated convergence, and $(\mathcal I_{\rho,N}r)(0)=0$ by the bound above. Thus $\mathcal I_{\rho,N}$ maps $\mathcal B$ into itself.

Next, let $r_1,r_2\in\mathcal B$. By the fundamental theorem of calculus, $\mathcal F_{N,\rho}(\eta,r_1(\eta))-\mathcal F_{N,\rho}(\eta,r_2(\eta))=\int_0^1D_z\mathcal F_{N,\rho}\big(\eta,r_2(\eta)+\vartheta(r_1(\eta)-r_2(\eta))\big)d\vartheta\,(r_1(\eta)-r_2(\eta)),$ so that the estimate \eqref{eq:FN-rho-lipschitz} gives $\|\mathcal F_{N,\rho}(\eta,r_1(\eta))-\mathcal F_{N,\rho}(\eta,r_2(\eta))\|\le C_z\rho\|r_1-r_2\|_\gamma\|\eta\|^\gamma$. The same computation as above yields
\begin{equation}\label{eq:app-T-contraction}
    \|\mathcal I_{\rho,N}r_1-\mathcal I_{\rho,N}r_2\|_\gamma
    \le\frac{2KC_zC_\epsilon^\gamma\rho}{\zeta}\|r_1-r_2\|_\gamma=\Theta_\rho\|r_1-r_2\|_\gamma .
\end{equation}
Since $\mathcal B$ is a closed subset of the Banach space $\mathcal X_\gamma$ and $\Theta_\rho<1/2$, the Banach fixed-point theorem gives a unique fixed point $r_\rho\in\mathcal B$, and $r_\rho(0)=0$ by the definition of $\mathcal X_\gamma$. It remains to prove $r_\rho\in\mathcal C^N$.

Let $r_0:=0$ and $r_{\ell+1}:=\mathcal I_{\rho,N}r_\ell$. Then $r_\ell\in\mathcal B$ for every $\ell$, and the estimate \eqref{eq:app-T-contraction} {implies that $r_\ell$ converges to $r_\rho$ in $\mathcal X_\gamma$.} We prove, by induction on $j=0,1,\ldots,N$, the statement
\medskip

\noindent $(\mathrm S_j)$: \textit{$r_\ell\in\mathcal C^{j}$ with $D^{j}r_\ell\in\mathcal X^{(j)}_{\gamma-j}$ for every $\ell$; the quantity $M_j:=\sup_\ell\|D^{j}r_\ell\|_{\gamma-j}$ is finite; and $D^{j}r_\ell$ converges in $\mathcal X^{(j)}_{\gamma-j}$ to $D^{j}r_\rho\in\mathcal X^{(j)}_{\gamma-j}$.}
\medskip

\noindent We already proved $(\mathrm S_0)$, and $(\mathrm S_N)$ implies $r_\rho\in\mathcal C^N$. Let $j\in\{1,\ldots,N\}$, assume $(\mathrm S_{j'})$ for all $j'<j$, and set
\begin{equation}\label{eq:app-admissible}
    \mathcal Q^{(j)} := \{r\in\mathcal B\cap\mathcal C^{j-1}: \|D^{j'} r\|_{\gamma-j'}\le M_{j'}\ \text{ for } j'=1,\ldots,j-1 \},
\end{equation}
so that $r_\rho\in\mathcal Q^{(j)}$ and $r_\ell\in\mathcal Q^{(j)}$ for every $\ell$; for $j=1$, $\mathcal Q^{(1)}=\mathcal B$.

Write $\Xi_r(\eta):=\mathcal F_{N,\rho}(\eta,r(\eta))$, so that the integrands in the definition \eqref{eq:T-def} are $e^{A_-t}P_-\Xi_r\circ\varphi^\rho_{-t}$ and $e^{-A_+t}P_+\Xi_r\circ\varphi^\rho_{t}$, $t\in[0,\infty)$. If $r\in\mathcal C^j$ and differentiation under the integral sign is legitimate, $D^j(\mathcal I_{\rho,N}r)$ is obtained by differentiating these integrands $j$ times in $\omega$ and integrating the result over $t\in[0,\infty)$. 
Lemma~\ref{lem:top-order} with $F=\mathcal F_{N,\rho}$ and $\psi=\varphi^\rho_{t}$ or $\psi=\varphi^\rho_{-t}$ shows that the result is the sum of a term that is linear in $D^jr$ and a term that depends on $r$ only through $D^{j'}r$ with $j'\le j-1$. This suggests obtaining $D^jr_\rho$ as the fixed point of the affine operator on $\mathcal X^{(j)}_{\gamma-j}$ obtained by replacing $D^jr$ with an unknown $\Phi\in\mathcal X^{(j)}_{\gamma-j}$. Precisely, for $r\in\mathcal Q^{(j)}$ and $\Phi\in\mathcal X^{(j)}_{\gamma-j}$, we define the \emph{linear part}
\begin{align}
\begin{split}\label{eq:app-Ak}
(\mathcal A_r^{(j)}\Phi)(\omega)
:=-\int_0^\infty e^{-A_+t}P_+ D_z\mathcal F_{N,\rho}(\omega^\rho_{t},r(\omega^\rho_{t}))\, \Phi(\omega_t^\rho) [D\omega_t^\rho,\cdots,D\omega_t^\rho] dt\\
+\int_0^\infty e^{A_-t}P_- D_z\mathcal F_{N,\rho}(\omega^\rho_{-t},r(\omega^\rho_{-t}))\, \Phi(\omega_{-t}^\rho) [D\omega_{-t}^\rho,\cdots,D\omega_{-t}^\rho] dt,
\end{split}
\end{align}
and the \emph{inhomogeneous part}
\begin{equation}\label{eq:app-Hk}
\mathcal H^{(j)}[r](\omega):=\int_0^\infty e^{A_-t}P_-\mathcal E_j^{-t}[r](\omega)dt-\int_0^\infty e^{-A_+t}P_+\mathcal E_j^{t}[r](\omega)dt,
\end{equation}
where $\mathcal E_j^{t}[r]$ denotes $\mathcal E_j[r]$ of Lemma~\ref{lem:top-order} with $F=\mathcal F_{N,\rho}$ and $\psi=\varphi^\rho_{t}$, and $\mathcal E_j^{-t}[r]$ is defined analogously with $\psi=\varphi^\rho_{-t}$. Set $\mathcal S_r^{(j)}\Phi:=\mathcal A_r^{(j)}\Phi+\mathcal H^{(j)}[r]$. By the identity \eqref{eq:app-top-order}, if $r\in\mathcal C^j$ and differentiation under the integral sign is legitimate, then $D^j(\mathcal I_{\rho,N}r)=\mathcal S_r^{(j)}D^jr$. We abbreviate $\mathcal A_\rho^{(j)}:=\mathcal A_{r_\rho}^{(j)}$, $\mathcal H_\rho^{(j)}:=\mathcal H^{(j)}[r_\rho]$, $\mathcal A_{\rho,\ell}^{(j)}:=\mathcal A_{r_\ell}^{(j)}$ and $\mathcal H_{\rho,\ell}^{(j)}:=\mathcal H^{(j)}[r_\ell]$. Now we prove that $\mathcal A^{(j)}_r$ is a contraction, that $\mathcal H^{(j)}[r]\in\mathcal X^{(j)}_{\gamma-j}$, and that the fixed point of $\mathcal S^{(j)}_{r_\rho}$ is actually $D^jr_\rho$.

Let $\Phi\in\mathcal X^{(j)}_{\gamma-j}$ and $t\ge0$. By the estimates \eqref{eq:flow-bound} and \eqref{eq:flow-deriv-bound},
\begin{align*}
\|\Phi(\omega_{t}^\rho) [D\omega_{t}^\rho,\ldots,D\omega_{t}^\rho]\|
&\le\|\Phi\|_{\gamma-j}\|\omega^\rho_{t}\|^{\gamma-j}\|D\omega^\rho_{t}\|^j\le\|\Phi\|_{\gamma-j} C_\epsilon^{\gamma} \|\omega\|^{\gamma-j}e^{\gamma\beta_+ t}{,}\\
\|\Phi(\omega_{-t}^\rho) [D\omega_{-t}^\rho,\ldots,D\omega_{-t}^\rho]\|
&\le\|\Phi\|_{\gamma-j}\|\omega^\rho_{-t}\|^{\gamma-j}\|D\omega^\rho_{-t}\|^j\le\|\Phi\|_{\gamma-j} C_\epsilon^{\gamma} \|\omega\|^{\gamma-j}e^{\gamma\beta_- t}.
\end{align*}
Combining these bounds with the inequality \eqref{eq:dichotomy}{, the estimate \eqref{eq:app-onesided-integrals},} and the estimate \eqref{eq:FN-rho-lipschitz}{, we obtain}
\begin{equation*}
\|(\mathcal A_r^{(j)}\Phi)(\omega)\|\le 2KC_z\rho C_\epsilon^\gamma\zeta^{-1}\|\Phi\|_{\gamma-j}\|\omega\|^{\gamma-j}{;}
\end{equation*}
continuity of $\mathcal A_r^{(j)}\Phi$ follows by dominated convergence. Since the estimate \eqref{eq:FN-rho-lipschitz} is uniform in $(\omega,z)$,
\begin{equation}\label{eq:app-Ak-bound}
    \|\mathcal A_r^{(j)}\|
    \le\Theta_\rho<\frac{1}{2}~\text{for every}~r\in\mathcal Q^{(j)},
\end{equation}
in particular for $r=r_\rho$ and $r=r_\ell$.

By the definition \eqref{eq:app-Rj-def}, $\mathcal E_j^{t}[r]$ is a finite sum of terms
\begin{equation}\label{eq:app-R-term}
    c_{a,d}c_{j_1\cdots j_i}\,\mathcal G_{a,d}[r](\omega^\rho_{t})\big[D^{j_1}\omega^\rho_{t} ,\cdots,D^{j_i}\omega^\rho_{t} \big],
\end{equation}
where $\mathcal G_{a,d}[r]$ is defined by the definition \eqref{eq:app-G-def} with $F=\mathcal F_{N,\rho}$, $1\le i\le j$, $a+d_1+\cdots+d_b=i$, $1\le d_\iota\le j-1$, and $j_1+\cdots+j_i=j$. We first claim that there is a constant $C_{\rho,j}>0$ such that
\begin{equation}\label{eq:app-remainder-vanishing}
     \|\mathcal G_{a,d}[r](\eta) \|\le C_{\rho,j}\|\eta\|^{N+1-i}
    ~\text{for every }\eta\in\R^\nu,~ r\in\mathcal Q^{(j)},
\end{equation}
and every $(a,d)$ appearing in the definition \eqref{eq:app-Rj-def}.

Since $\rho$ is fixed, Lemma~\ref{lem:cutoff}\textbf{(b)} makes all derivatives of $\mathcal F_{N,\rho}$ of order at most $N+1$ bounded, so that
\begin{equation}\label{eq:app-Lambda}
    \Lambda_\rho :=\max_{a+b\le N+1}\ \sup_{(\eta,z)\in\R^\nu\times\R^n}  \|D_\eta^{a}D_z^{b}\mathcal F_{N,\rho}(\eta,z) \|<\infty .
\end{equation}
Every term {$\mathcal G_{a,d}[r]$} satisfies $a+b\le i\le j\le N$, so
the estimate \eqref{eq:app-Lambda} applies to all of them. {Moreover, the estimate \eqref{eq:app-Dz-sharp} holds, and} $r\in\mathcal Q^{(j)}$ says precisely that
\begin{equation}\label{eq:app-r-weights}
    \|D^{j'}r(\eta)\|\le M_{j'}\|\eta\|^{\gamma-j'},~\text{for}~j'=0,1,\ldots,j-1 .
\end{equation}
{Since every $d_\iota$ in $\mathcal G_{a,d}[r]$ satisfies $d_\iota\le j-1$, the estimate \eqref{eq:app-r-weights} covers every factor $D^{d_\iota}r$. Since $\mathcal G_{a,d}[r](\eta)=0$ for $\|\eta\|\ge2\rho$, it suffices to consider $\|\eta\|<2\rho$}{, on which a higher power of $\|\eta\|$} is dominated by a lower one,
\begin{equation}\label{eq:app-power-trade}
    \|\eta\|^{e}\le(2\rho)^{ e-(N+1-i)} \|\eta\|^{N+1-i}
    \qquad\text{whenever } e\ge N+1-i,
\end{equation}
so it suffices to bound each term by $K'\|\eta\|^{e}$ with $e\ge N+1-i$ and with
$K'$ built only from the estimates \eqref{eq:FN-rho-jet-bound}, \eqref{eq:app-Dz-sharp}, \eqref{eq:app-Lambda}, and \eqref{eq:app-r-weights}. We do this according to the values of $a$ and $b$.

\noindent\textbf{Case 0: $b=0$, $a=i$.} No factor $D^{j'}r$ occurs, and the
fundamental theorem of calculus in the second argument gives, for
$\|z\|\le M_0\|\eta\|^{\gamma}$,
\begin{align*}
     \|D_\eta^{i}\mathcal F_{N,\rho}(\eta,z) \|
    & \le\underbrace{ \|D_\eta^{i}\mathcal F_{N,\rho}(\eta,0) \|}
        _{\le C_i\|\eta\|^{N+1-i}\ \text{by }\eqref{eq:FN-rho-jet-bound}}
      +\underbrace{\sup \|D_zD_\eta^{i}\mathcal F_{N,\rho} \|}
        _{\le \Lambda_\rho\ \text{by }\eqref{eq:app-Lambda}}
       \cdot\underbrace{\|z\|}_{\le M_0\|\eta\|^{\gamma}} \\
    &\le C_i\|\eta\|^{N+1-i}+\Lambda_\rho M_0\|\eta\|^{\gamma},
\end{align*}
and $\gamma>N\ge N+1-i$ because $i\ge1$.

\noindent\textbf{Case 1: $b=1$, $a=0$.} Here $d_1=i\le j-1$. Then
\begin{equation*}
     \|\mathcal G_{a,d}[r](\eta) \|
    \le\underbrace{ \|D_z\mathcal F_{N,\rho}(\eta,r(\eta)) \|}
        _{\le C_*\|\eta\|\ \text{by }\eqref{eq:app-Dz-sharp}}
      \cdot\underbrace{\|D^{i}r(\eta)\|}
        _{\le M_i\|\eta\|^{\gamma-i}\ \text{by }\eqref{eq:app-r-weights}}
    \le C_*M_i \|\eta\|^{ \gamma-i+1},
\end{equation*}
and $\gamma-i+1\ge N+1-i$ because $\gamma>N$.

\noindent\textbf{Case 2: $b=1$, $a\ge1$.} Here $d_1=i-a$, and
\begin{equation*}
     \|\mathcal G_{a,d}[r](\eta) \|
    \le\underbrace{ \|D_\eta^{a}D_z\mathcal F_{N,\rho}(\eta,r(\eta)) \|}
        _{\le \Lambda_\rho}
      \cdot\underbrace{\|D^{i-a}r(\eta)\|}
        _{\le M_{i-a}\|\eta\|^{\gamma-i+a}}
    \le\Lambda_\rho M_{i-a} \|\eta\|^{ \gamma-i+a},
\end{equation*}
and $\gamma-i+a\ge N+1-i$ because $\gamma+a>N+1$.

\noindent\textbf{Case 3: $b\ge2$.} Using $\sum_{\iota}d_\iota=i-a$,
\begin{equation*}
     \|\mathcal G_{a,d}[r](\eta) \|
    \le\underbrace{ \|D_\eta^{a}D_z^{b}\mathcal F_{N,\rho}(\eta,r(\eta)) \|}
        _{\le \Lambda_\rho}
      \cdot\prod_{\iota=1}^{b}
       \underbrace{\|D^{d_\iota}r(\eta)\|}_{\le M_{d_\iota}\|\eta\|^{\gamma-d_\iota}}
    \le\Lambda_\rho \Big(\prod_{\iota=1}^{b}M_{d_\iota} \Big)\|\eta\|^{ b\gamma-i+a},
\end{equation*}
and $b\gamma-i+a\ge N+1-i$ because $b\gamma+a\ge2\gamma>2N\ge N+1$.

In each case the constant $K'$ is a product of $C_i$, $C_*$, or
$\Lambda_\rho$ with at most $b+1\le N+1$ of the numbers $M_0,\ldots,M_{j-1}$, and the
exponent $e$ is determined by $a$, $b$, and $i$ alone; after
the estimate \eqref{eq:app-power-trade} the resulting constant is
$K'(2\rho)^{ e-(N+1-i)}$. None of these quantities depends on the particular
$r\in\mathcal Q^{(j)}$. Letting $C_{\rho,j}$ be the largest of them over the
finitely many {$(a,d)$ in the definition \eqref{eq:app-Rj-def}} with $1\le i\le j$ proves
the estimate \eqref{eq:app-remainder-vanishing}.

We now {bound each term \eqref{eq:app-R-term}} for $t\ge0$. {By the estimates \eqref{eq:flow-deriv-bound} and \eqref{eq:higher-flow-deriv-bound}, with $C_{1,\rho,\epsilon}:=C_\epsilon$, we have} $\|D^{j'}\omega^\rho_{t}\|\le C_{j',\rho,\epsilon} e^{j'\beta_+ t}$ for every $j'\ge1$; since $\sum_\iota j_\iota=j$, {the product of the flow derivatives in the term \eqref{eq:app-R-term}} is bounded by $ (\prod_{\iota}C_{j_\iota,\rho,\epsilon} )e^{j\beta_+ t}$, and this exponential factor is the same for every $j$ differentiations. The inner factor is bounded by $C_{\rho,j}\|\omega^\rho_{t}\|^{N+1-i}$ through the estimate \eqref{eq:app-remainder-vanishing}. The exponent $N+1-i$ does not match the weight $\gamma-j$ of $\mathcal X^{(j)}_{\gamma-j}$, and the excess is traded for a constant on the effective support $\|\omega^\rho_{t}\|<2\rho$, on which 
\begin{equation}\label{eq:app-weight-trade}
    \|\omega_{t}^\rho\|^{N+1-i}
    \le(2\rho)^{N+1-\gamma+j-i} \|\omega_{t}^\rho\|^{\gamma-j},
\end{equation}
the exponent $N+1-\gamma+j-i$ being positive because $\gamma<N+1$ and $i\le j$. Finally the estimate \eqref{eq:flow-bound} gives $\|\omega^\rho_{t}\|^{\gamma-j}\le C_\epsilon^{\gamma-j}\|\omega\|^{\gamma-j}e^{(\gamma-j)\beta_+ t}$. Multiplying the three estimates, the exponentials combine as $e^{(\gamma-j)\beta_+ t}e^{j\beta_+ t}=e^{\gamma{\beta_+} t}$ so that {$\|\mathcal E_j^{t}[r](\omega)\|\le\tilde C_{\rho,j}\|\omega\|^{\gamma-j}e^{\gamma{\beta_+} t}$} with
\begin{equation*}
    \tilde C_{\rho,j}
    :=\sum {c_{a,d}c_{j_1\cdots j_i}} C_{\rho,j}
       \Big(\prod_{\iota}C_{j_\iota,\rho,\epsilon} \Big)
      (2\rho)^{N+1-\gamma+j-i}C_\epsilon^{\gamma-j},
\end{equation*}
the sum running over the finitely many terms {\eqref{eq:app-R-term}}. The constant $\tilde C_{\rho,j}$ depends on the fixed $\rho$, through the constants of the estimate \eqref{eq:higher-flow-deriv-bound}, and is uniform over $r\in\mathcal Q^{(j)}$. The same argument with $\omega_{-t}^\rho$ in place of $\omega_t^\rho$ and $\beta_-$ in place of $\beta_+$, which uses the bounds on $\|\omega_{-t}^\rho\|$ and $\|D^{j'}\omega_{-t}^\rho\|$ in the estimates \eqref{eq:flow-bound}--\eqref{eq:higher-flow-deriv-bound}, gives $\|\mathcal E_j^{-t}[r](\omega)\|\le\tilde C_{\rho,j}\|\omega\|^{\gamma-j}e^{\gamma\beta_- t}$, after enlarging $\tilde C_{\rho,j}$ if necessary. Applying the inequality \eqref{eq:dichotomy} and the estimate \eqref{eq:app-onesided-integrals} to the definition \eqref{eq:app-Hk}, we get
\begin{equation*}
\begin{split}
     \|\mathcal H^{(j)}{[r]}(\omega) \|
    &\le\int_0^\infty Ke^{-\kappa_-' t}
        \tilde C_{\rho,j}\|\omega\|^{\gamma-j}e^{\gamma\beta_- t} dt+\int_0^\infty Ke^{-\kappa_+' t}
        \tilde C_{\rho,j}\|\omega\|^{\gamma-j}e^{\gamma\beta_+ t} dt\\
    &\le\frac{2K\tilde C_{\rho,j}}{{\zeta}}\|\omega\|^{\gamma-j}
    =:\hat C_{\rho,j}\|\omega\|^{\gamma-j}.
\end{split}
\end{equation*}
Continuity of $\mathcal H^{(j)}{[r]}$ follows by dominated convergence and $\mathcal H^{(j)}{[r]}(0)=0$. Hence $\mathcal H^{(j)}{[r]}\in\mathcal X^{(j)}_{\gamma-j}$ with $\|\mathcal H^{(j)}{[r]}\|_{\gamma-j}\le \hat C_{\rho,j}$ {for every $r\in\mathcal Q^{(j)}$, in particular, $\hat C_{\rho,j}$ bounds both $\|\mathcal H^{(j)}_\rho\|_{\gamma-j}$ and $\|\mathcal H^{(j)}_{\rho,\ell}\|_{\gamma-j}$, independently of $\ell$}. Since $\mathcal X^{(j)}_{\gamma-j}$ is a Banach space and $\|\mathcal A^{(j)}_\rho\|\le\Theta_\rho<1/2$, the affine operator {$\mathcal S^{(j)}_{r_\rho}$} has a unique fixed point $\Phi^{(j)}_\rho\in\mathcal X^{(j)}_{\gamma-j}$.

We first check that $D^j r_\ell$ exists for $\ell \in \Z_{\ge0}$. It is clear for $\ell=0$ since $r_0=0$. If $D^j r_\ell$ exists and lies in $\mathcal X_{\gamma-j}^{(j)}$, then the previous discussion (the estimates leading to the estimate \eqref{eq:app-Ak-bound} and to $\|\mathcal H^{(j')}[r]\|_{\gamma-j'}\le \hat C_{\rho,j'}$, applied with $j'=1,\dots,j$) shows that all the terms produced by differentiating the Picard relation $r_{\ell+1}=\mathcal I_{\rho,N}r_\ell$ $j$ times {under the integral sign} are dominated by a multiple of $\|\omega\|^{\gamma-j'}e^{-\zeta t}$, which is integrable over $t\in[0,\infty)$ uniformly for $\omega$ in bounded sets. Differentiation under the integral sign is therefore legitimate and gives
\begin{equation}\label{eq:app-picard-diff}
    D^j r_{\ell+1}=\mathcal A_{\rho,\ell}^{(j)}D^j r_\ell+\mathcal H_{\rho,\ell}^{(j)}
    \ \in\ \mathcal X_{\gamma-j}^{(j)} .
\end{equation}
Taking norms in the identity \eqref{eq:app-picard-diff} and using the estimate \eqref{eq:app-Ak-bound} together with $D^j r_0=0$ gives $\|D^j r_\ell\|_{\gamma-j}\le \hat C_{\rho,j}/(1-\Theta_\rho)<2\hat C_{\rho,j}$ for every $\ell$; hence $M_j:=\sup_\ell\|D^jr_\ell\|_{\gamma-j}<\infty$.

Define $\xi_\ell:= \|(\mathcal A_{\rho,\ell}^{(j)}-\mathcal A_{\rho}^{(j)}) \Phi_\rho^{(j)} \|_{\gamma-j}
+ \|\mathcal H_{\rho,\ell}^{(j)}-\mathcal H_{\rho}^{(j)} \|_{\gamma-j}$. Subtracting $\Phi_\rho^{(j)}=\mathcal A_{\rho}^{(j)}\Phi_\rho^{(j)}+\mathcal H_{\rho}^{(j)}$ from the identity \eqref{eq:app-picard-diff} yields $\|D^j r_{\ell+1}-\Phi_\rho^{(j)}\|_{\gamma-j} \le\Theta_\rho\|D^j r_\ell-\Phi_\rho^{(j)}\|_{\gamma-j}+\xi_\ell$ by the estimate \eqref{eq:app-Ak-bound}, and $\xi_\ell$ converges to zero. Indeed, set $e_\ell:=\|r_\ell-r_\rho\|_\gamma+\sum_{j'=1}^{j-1}\|D^{j'}r_\ell-D^{j'}r_\rho\|_{\gamma-j'}$, which tends to zero by $(\mathrm S_{j'})$, $j'<j$. Since all derivatives of $\mathcal F_{N,\rho}$ of order at most $N+1$ are bounded by $\Lambda_\rho$ and $a+b+1\le N+1$ for every factor $D^a_\eta D^b_z\mathcal F_{N,\rho}$ appearing in the definition \eqref{eq:app-Ak} and the term \eqref{eq:app-R-term}, the mean value theorem gives $\|D^a_\eta D^b_z\mathcal F_{N,\rho}(\eta,r_\ell(\eta))-D^a_\eta D^b_z\mathcal F_{N,\rho}(\eta,r_\rho(\eta))\|\le\Lambda_\rho e_\ell\|\eta\|^\gamma$, while $\|D^{j'}r_\ell(\eta)-D^{j'}r_\rho(\eta)\|\le e_\ell\|\eta\|^{\gamma-j'}$ for $j'\le j-1$. Writing each difference of integrands as a telescoping sum in which exactly one factor is replaced by its difference, and repeating the computations of the estimate \eqref{eq:app-Ak-bound} and of Cases 0–3 with that factor (the exponent of $\|\eta\|$ does not decrease, excess powers of $\|\eta\|$ are absorbed into constants on the support $\|\eta\|<2\rho$ as in the estimates \eqref{eq:app-power-trade}--\eqref{eq:app-weight-trade}, and the replaced factor contributes $e_\ell$), we obtain $\xi_\ell\le c_{\rho,j}e_\ell$ with a constant $c_{\rho,j}$ independent of $\ell$; hence $\xi_\ell$ converges to zero. Hence $D^j r_\ell$ converges to $\Phi_\rho^{(j)}$ in $\mathcal X_{\gamma-j}^{(j)}$.

Weighted convergence implies local uniform convergence because $\gamma-j>0$: on $B_R$ one has $\|\Phi\|_{\infty}\le\|\Phi\|_{\gamma-j}R^{\gamma-j}$. By the induction hypothesis, $D^{j-1}r_\ell$ converges to $D^{j-1}r_\rho$ locally uniformly, and $D^j r_\ell$ converges locally uniformly to $\Phi_\rho^{(j)}$; the standard derivative-convergence theorem \cite[Chapter 7]{rudin2021principles} therefore gives $r_\rho\in\mathcal C^{j}$ and $D^j r_\rho=\Phi_\rho^{(j)}\in\mathcal X_{\gamma-j}^{(j)}$. This establishes $(\mathrm S_j)$ and, for $j=N$, the proof is complete.
\end{proof}

\section{Proof of Lemma~\ref{lem:order-raising}}\label{app:order-raising}
Since $b<k$, we have $b\alpha_-<\kappa_-$ and $b\alpha_+<\kappa_+$. Choose $\epsilon>0$, $\kappa_-'$, and $\kappa_+'$ such that
\begin{equation}\label{eq:order-raising-gap}
    b(\alpha_-+\epsilon)<\kappa_-'<\kappa_-,~b(\alpha_++\epsilon)<\kappa_+'<\kappa_+,
\end{equation}
set $\beta_-:=\alpha_-+\epsilon$ and $\beta_+:=\alpha_++\epsilon$, and let $K=K(\kappa_-',\kappa_+')$ be as in the inequality \eqref{eq:dichotomy}. Let $\rho_\epsilon$ be as in Lemma~\ref{lem:cutoff} and choose $\rho\in(0,\rho_\epsilon)$ small enough so that $B_\rho \subset U$ and that there exists $C>0$ such that 
\begin{equation}\label{eq:Lem4.2eq1}
    \|\Delta(\omega)z(\omega)+g(\omega)\| \le C\|\omega\|^b,~\|z(\omega)\| \le C\| \omega\|^a,
\end{equation} 
for all $\omega \in B_\rho$ (it is possible since $b \le a+1$ and $z(\omega) = O(\|\omega\|^a)$). Recall that $\varphi_t^\rho$ is the flow of the vector field $s_\rho$ defined in the definition \eqref{eq:truncated}.
    
For $\omega \in B_\rho$, define the backward and forward exit times
\begin{align*}
T_-(\omega)&:=\sup\{T>0:\varphi_{-t}^\rho(\omega)\in B_\rho \text{ for every }t\in[0,T]\},\\
T_+(\omega)&:=\sup\{T>0:\varphi_{t}^\rho(\omega)\in B_\rho \text{ for every }t\in[0,T]\}.
\end{align*}
Since $\|\varphi_{-t}^\rho(\omega)\|\le C_\epsilon\|\omega\|e^{\beta_-t}$ and $\|\varphi_{t}^\rho(\omega)\|\le C_\epsilon\|\omega\|e^{\beta_+t}$ for $t\ge0$ by the estimate \eqref{eq:flow-bound}, we have
\begin{equation}\label{eq:exit-time-lower-bound}
    T_-(\omega)\ge\frac1{\beta_-} \log\frac{\rho}{C_\epsilon\|\omega\|},~
    T_+(\omega)\ge\frac1{\beta_+} \log\frac{\rho}{C_\epsilon\|\omega\|},
\end{equation}
whenever the left-hand sides are finite.
Fix $\delta \in \left(0, \frac{\rho}{C_\epsilon}\right)$; since $C_\epsilon \ge 1$ by the estimate \eqref{eq:flow-bound} with $t=0$, $B_\delta \subset B_\rho$ and if $\omega \in B_\delta\setminus\{0\}$, then the right-hand sides of the estimate \eqref{eq:exit-time-lower-bound} {are} positive.

Set $\eta(t):=z(\omega_t^\rho)$. Then, for $\omega \in B_\delta$ and $-T_-(\omega)<t<T_+(\omega)$, we have $\dot \eta(t) = (A+\Delta(\omega_t^\rho)) \eta(t) + g(\omega_t^\rho)$, since $s_\rho = s$ on $B_\rho$. Let $\tau \in (0,T_-(\omega))$ if $T_-(\omega)<\infty$ and let $\tau$ go to infinity if $T_-(\omega) = \infty$. From the variation-of-constants formula applied to $\eta$ with initial time $-\tau$ and final time $0$ we have $\eta(0) = e^{A\tau}\eta(-\tau) + \int_{-\tau}^0e^{-At}(\Delta(\omega_t^\rho)z(\omega_t^\rho)+g(\omega_t^\rho))dt$. Applying $P_-$ to both sides and using $\eta(0) = z(\omega)$, we get
\begin{equation}\label{eq:Lem4.2eq2}
    P_- z(\omega) = e^{A_-\tau} P_-\eta(-\tau) + \int_0^{\tau}e^{A_- t} P_- (\Delta(\omega_{-t}^\rho)z(\omega_{-t}^\rho)+g(\omega_{-t}^\rho))dt.
\end{equation}
From the estimate \eqref{eq:flow-bound}, $\| \omega_{-t}^\rho \| \le C_\epsilon \| \omega\| e^{\beta_- t}$ for $t \in [0,\tau]$. Combining this bound with the estimate \eqref{eq:Lem4.2eq1}, we obtain
\begin{align}
\begin{split}\label{eq:Lem4.2eq3}
\|\eta(-\tau)\|&\le C \|\omega^\rho_{-\tau}\|^{a}\le C C_\epsilon^{a}\|\omega\|^{a}e^{a\beta_-\tau},\\ \|\Delta(\omega_{-t}^\rho)z(\omega_{-t}^\rho)+g(\omega_{-t}^\rho)\|&\le C\|\omega^\rho_{-t}\|^{b}\le C C_\epsilon^{b}\|\omega\|^{b}e^{b\beta_- t}.    
\end{split}
\end{align}
Taking norms of both sides of the equation \eqref{eq:Lem4.2eq2} and using the triangle inequality, the inequality \eqref{eq:dichotomy}, and the estimate \eqref{eq:Lem4.2eq3}, we get $\|P_-z(\omega)\| \le CC_\epsilon ^a K\|\omega\|^a e^{-\zeta_a\tau} +C C_\epsilon^b K\|\omega\|^b\int_0^\tau e^{-\zeta_bt} dt$. Here, $\zeta_a:=\kappa_-'-a\beta_- =(\kappa_-'-b\beta_-)+(b-a)\beta_->0$ and $\zeta_b:=\kappa_-'-b\beta_->0$ (by $a<b$ and the condition \eqref{eq:order-raising-gap}). Thus, the integral term is bounded by $CC_\epsilon^bK\zeta_b^{-1}\|\omega\|^b$. If $T_-(\omega)<\infty$, the estimate \eqref{eq:exit-time-lower-bound} and $\zeta_a>0$ give $\|\omega\|^a e^{-\zeta_aT_-(\omega)} \le \|\omega\|^a \exp\left(-\frac{\zeta_a}{\beta_-} \log\frac{\rho}{C_\epsilon\|\omega\|}\right)=\left(\frac{C_\epsilon}{\rho}\right)^{\frac{\zeta_a}{\beta_-}} \|\omega\|^{\frac{\kappa_-'}{\beta_-}}$. If $T_-(\omega)=\infty$, the corresponding boundary term converges to zero. In either case it is $O(\|\omega\|^b)$ because $\kappa_-'/\beta_->b$, and we obtain $\| P_- z(\omega)\| = O(\| \omega\|^b)$. The $E_+$ projection is treated analogously by using the terminal time $\tau\in(0,T_+(\omega))$ if $T_+(\omega)<\infty$ and letting $\tau$ go to infinity if $T_+(\omega)=\infty$, the forward-time bound $\|\omega_t^\rho\|\le C_\epsilon\|\omega\|e^{\beta_+t}$ in the estimate \eqref{eq:flow-bound}, and $\kappa_+'$ and $\beta_+$ in place of $\kappa_-'$ and $\beta_-$. Hence $z(\omega)=O(\|\omega\|^b)$.\hfill\proofbox

{\section{The uniqueness lemma}
The following lemma is used in the proofs of Propositions~\ref{prop:steady-state-response} and~\ref{prop:unique}. It is a version of the classical fact that a local center manifold contains all solutions that remain in a sufficiently small neighborhood of the origin for all $t\in\R$ \cite[Corollary~4.7]{vanderbauwhede1989centre}, \cite[Proposition~1]{palmer1987stability}, adapted to the invariance equation. Since \cite[Proposition~1]{palmer1987stability} assumes that the eigenvalues of $A$ have negative real parts and \cite[Corollary~4.7]{vanderbauwhede1989centre} assumes sufficiently smooth center manifolds, we replace the Gronwall argument of \cite{palmer1987stability} by a variation-of-constants estimate based on the exponential dichotomy, which applies to hyperbolic $A$ and $\mathcal C^1$ solutions.

\begin{lemma}\label{lem:invariant-uniqueness}
    Consider the system \eqref{eq:nl} and the exosystem \eqref{eq:nl-exo} satisfying the strip condition, and let $\varphi_t$ denote the flow of $s$. There exist $\rho_0>0$ and $\eta_0>0$ with the following property: If $W\subset B_{\rho_0}$ is invariant under $\varphi_t$, i.e., for every $\omega\in W$, $\varphi_t(\omega)$ is defined and belongs to $W$ for all $t\in\R$, and if $\pi^1$ and $\pi^2$ are $\mathcal C^1$ solutions to the invariance equation \eqref{eq:inv} defined on an open set containing $W$ that satisfy $\sup_{\omega\in W}\|\pi^i(\omega)\|\le\eta_0$ for $i=1,2$, then $\pi^1=\pi^2$ on $W$.
\end{lemma}
 
\begin{proof}
    Fix $\kappa_-'\in(0,\kappa_-)$ and $\kappa_+'\in(0,\kappa_+)$, let $K=K(\kappa_-',\kappa_+')$ satisfy the inequality \eqref{eq:dichotomy}, and set
    \begin{equation*}
        c_\Delta:=\left(2K\left(\frac1{\kappa_-'}+\frac1{\kappa_+'}\right)\right)^{-1}.
    \end{equation*}
    Since $f$ is smooth, $l(0)=0$, and $D_xf(0,0)=A$, there exist $\rho_0>0$ and $\eta_0>0$ such that
    \begin{equation}\label{eq:Dxf-close}
        \|D_xf(x,l(\omega))-A\|\le c_\Delta,\quad \forall\, \|x\|\le\eta_0,~\|\omega\|\le\rho_0.
    \end{equation}
    Let $W$, $\pi^1$, and $\pi^2$ be as in the statement, define $\delta\pi:=\pi^2-\pi^1$, and define $\Delta(\omega)$ by the definition \eqref{eq:B-def}. For $\omega\in W$ and $\vartheta\in[0,1]$, $\pi^1(\omega)+\vartheta\delta\pi(\omega)=(1-\vartheta)\pi^1(\omega)+\vartheta\pi^2(\omega)$ belongs to the closed ball of radius $\eta_0$, so that the inequality \eqref{eq:Dxf-close} gives $\|\Delta(\omega)\|\le c_\Delta$ for every $\omega\in W$.
 
    Fix $\omega\in W$ and define $\eta(t):=\delta\pi(\omega_t)$ for $t\in\R$, where $\omega_t:=\varphi_t(\omega)\in W$. Since $\pi^1$ and $\pi^2$ satisfy the invariance equation \eqref{eq:inv} on an open set containing $W$,
    \begin{equation*}
        \dot\eta(t)=D\delta\pi(\omega_t)s(\omega_t)=f(\pi^2(\omega_t),l(\omega_t))-f(\pi^1(\omega_t),l(\omega_t))=\left(A+\Delta(\omega_t)\right)\eta(t),
    \end{equation*}
    and $M_\eta:=\sup_{t\in\R}\|\eta(t)\|\le2\eta_0<\infty$. Fix $t\in\R$ and $T>0$. The variation-of-constants formula on $[t-T,t]$, projected onto $E_-$, gives
    \begin{equation*}
        P_-\eta(t)=e^{A_-T}P_-\eta(t-T)+\int_{t-T}^te^{A_-(t-\tau)}P_-\Delta(\omega_\tau)\eta(\tau)d\tau,
    \end{equation*}
    so that the inequality \eqref{eq:dichotomy} yields $\|P_-\eta(t)\|\le Ke^{-\kappa_-'T}M_\eta+Kc_\Delta M_\eta/\kappa_-'$. Letting $T$ go to infinity, we obtain $\|P_-\eta(t)\|\le Kc_\Delta M_\eta/\kappa_-'$. Similarly, the variation-of-constants formula on $[t,t+T]$, projected onto $E_+$,
    \begin{equation*}
        P_+\eta(t)=e^{-A_+T}P_+\eta(t+T)-\int_t^{t+T}e^{-A_+(\tau-t)}P_+\Delta(\omega_\tau)\eta(\tau)d\tau,
    \end{equation*}
    yields $\|P_+\eta(t)\|\le Kc_\Delta M_\eta/\kappa_+'$. Hence
    \begin{equation*}
        \|\eta(t)\|\le\|P_-\eta(t)\|+\|P_+\eta(t)\|\le Kc_\Delta\left(\frac1{\kappa_-'}+\frac1{\kappa_+'}\right)M_\eta=\frac12M_\eta.
    \end{equation*}
    Taking the supremum over $t\in\R$ gives $M_\eta\le\frac12M_\eta$, so $M_\eta=0$. In particular, $\delta\pi(\omega)=\eta(0)=0$. Since $\omega\in W$ is arbitrary, $\pi^1=\pi^2$ on $W$.
\end{proof}
}

{\section{Proof of Proposition~\ref{prop:steady-state-response}}\label{app:steady-state}
We first show that the assumption of Proposition~\ref{prop:steady-state-response} is implied by Lyapunov stability together with Poisson stability of every point in a neighborhood $W_0$ of the origin. Choose $\varepsilon>0$ with $\overline B_\varepsilon\subset W_0$, and let $\delta\in(0,\varepsilon)$ be such that $\|\varphi_t(\omega)\|<\varepsilon/2$ for all $t\ge0$ whenever $\|\omega\|<\delta$. If, for some $\omega\in B_\delta$, the backward trajectory reached the sphere of radius $\varepsilon$ at $p:=\varphi_{-t_1}(\omega)$, $t_1>0$, then, by Poisson stability of $p$, there would exist $t_2>t_1$ with $\|\varphi_{t_2}(p)-p\|<\varepsilon/2$, hence $\|\varphi_{t_2}(p)\|>\varepsilon/2$, whereas $\varphi_{t_2}(p)=\varphi_{t_2-t_1}(\omega)$ satisfies $\|\varphi_{t_2}(p)\|<\varepsilon/2$. Hence $\varphi_{-t}(\omega)$ is defined and satisfies $\|\varphi_{-t}(\omega)\|<\varepsilon$ for all $t\ge0$, and, since $\varepsilon$ can be taken arbitrarily small, the assumption of Proposition~\ref{prop:steady-state-response} holds. The converse does not hold: for $s(\omega)=\omega_2^2(-\omega_2,\omega_1)$, $\|\omega\|$ is conserved, so that the origin is stable in forward and backward time, whereas $\dot\theta=\omega_2^2\ge0$ in polar coordinates, so that every point with $\omega_2\neq0$ converges to an equilibrium on the $\omega_1$-axis and is not Poisson stable.

\textbf{(a)} If $S$ had an eigenvalue with positive real part, the origin of $\dot\omega=s(\omega)$ would be unstable \cite[Theorem~4.7]{khalil2002nonlinear}, which contradicts neutral stability. Applying the same argument to $\dot\omega=-s(\omega)$, whose flow is $\varphi_{-t}$ and whose linearization is $-S$, excludes eigenvalues with negative real part. Hence $\sigma(S)\subset i\R$ and $\alpha_+=\alpha_-=0$ by the definition \eqref{eq:spectral-params}. Since $\sigma(A)\subset\C_{<0}$, we have $\kappa_->0$ and $\kappa_+=\infty$, so that the strip condition holds and $k=\infty$ by the definition \eqref{eq:Nmax}.

\textbf{(b)} Since $\sigma(A)\subset\C_{<0}$, this part also follows from \cite[Proposition~1]{palmer1987stability}; we give a proof based on Lemma~\ref{lem:invariant-uniqueness}. Let $\rho_0$ and $\eta_0$ be as in Lemma~\ref{lem:invariant-uniqueness}. By \textbf{(a)} and Theorem~\ref{thm:existence} with $N=1$, there exists a $\mathcal C^1$ solution $\tilde\pi$ to the invariance equation \eqref{eq:inv} defined on a neighborhood $U$ of the origin. Since $\tilde\pi(0)=0$, we can choose $R\in(0,\rho_0]$ such that $\overline B_R\subset U$ and $\|\tilde\pi(\xi)\|\le\eta_0/2$ for all $\xi\in B_R$. Let $\delta>0$ correspond to $R$ in the definition of neutral stability, and define
\begin{equation*}
    W:=\{\varphi_t(\omega):\omega\in B_\delta,~t\in\R\}.
\end{equation*}
Then $W$ is invariant under $\varphi_t$, $B_\delta\subset W\subset B_R$, and $W$ is open because it is the union of the open sets $\varphi_t(B_\delta)$, $t\in\R$. Define $\pi:=\tilde\pi|_W$, which is a $\mathcal C^1$ solution to the invariance equation \eqref{eq:inv} on $W$ satisfying $\sup_{\omega\in W}\|\pi(\omega)\|\le\eta_0/2$. Now let $\hat\pi$ be a $\mathcal C^1$ solution to the invariance equation \eqref{eq:inv} defined on a neighborhood $\hat U$ of the origin. Choose $\hat R\in(0,R]$ such that $\overline B_{\hat R}\subset\hat U$ and $\|\hat\pi(\xi)\|\le\eta_0$ for all $\xi\in B_{\hat R}$, let $\hat\delta\in(0,\delta]$ correspond to $\hat R$ in the definition of neutral stability, and define $W':=\{\varphi_t(\omega):\omega\in B_{\hat\delta},~t\in\R\}$. As above, $W'$ is an open neighborhood of the origin that is invariant under $\varphi_t$, and $W'\subset W\cap B_{\hat R}\subset B_{\rho_0}$. Since $\sup_{\omega\in W'}\|\pi(\omega)\|\le\eta_0$ and $\sup_{\omega\in W'}\|\hat\pi(\omega)\|\le\eta_0$, Lemma~\ref{lem:invariant-uniqueness} yields $\hat\pi=\pi$ on $W'$.

\textbf{(c)} By \textbf{(a)}, $E_+=\{0\}$, $P_-$ is the identity, and every $N\in\Z_{>0}$ satisfies the condition \eqref{eq:N-cond}, so that $q_N$ is defined for every $N\in\Z_{>0}$. Since $W\subset B_R$ is bounded and $q_N$ is a polynomial, $C_N:=\sup_{\xi\in W}\|\pi(\xi)-q_N(\xi)\|<\infty$. For $\omega\in W$, $\omega_{-t}\in W$ for all $t\ge0$, so the inequality \eqref{eq:dichotomy} gives $\|e^{At}(\pi-q_N)(\omega_{-t})\|\le KC_Ne^{-\kappa_-'t}$, which converges to $0$ as $t$ goes to infinity.

\textbf{(d)} This part is the local attractivity of the center manifold \cite[Lemma and Proposition~2]{palmer1987stability}; we give a short proof for completeness. Let $c_\Delta$, $\kappa_-'$, $\kappa_+'$, and $K$ be as in the proof of Lemma~\ref{lem:invariant-uniqueness}, so that $Kc_\Delta\le\kappa_-'/2$ and the inequality \eqref{eq:Dxf-close} holds; since $E_+=\{0\}$, the inequality \eqref{eq:dichotomy} becomes $\|e^{At}\|\le Ke^{-\kappa_-'t}$ for $t\ge0$. Set $\delta_x:=\eta_0/(2K)$, $c:=K$, and $\kappa_0:=\kappa_-'/2$. Let $\omega(0)\in W$, so that $\omega(t)\in W$ for all $t\in\R$, and let $\|x(0)-\pi(\omega(0))\|<\delta_x$. Define $z(t):=x(t)-\pi(\omega(t))$. Since $\pi$ satisfies the invariance equation \eqref{eq:inv},
\begin{equation*}
    \dot z(t)=f(\pi(\omega(t))+z(t),l(\omega(t)))-f(\pi(\omega(t)),l(\omega(t)))=\left(A+\Delta(t)\right)z(t),
\end{equation*}
where $\Delta(t):=\int_0^1\left[D_xf(\pi(\omega(t))+\vartheta z(t),l(\omega(t)))-A\right]d\vartheta$. As long as $\|z\|\le\eta_0/2$, we have $\|\pi(\omega)+\vartheta z\|\le\eta_0$, so that $\|\Delta(t)\|\le c_\Delta$ by the inequality \eqref{eq:Dxf-close}. On such an interval $[0,t]$, the variation-of-constants formula $z(t)=e^{At}z(0)+\int_0^te^{A(t-\tau)}\Delta(\tau)z(\tau)d\tau$ and Gronwall's inequality yield
\begin{equation*}
    \|z(t)\|\le K\|z(0)\|e^{-(\kappa_-'-Kc_\Delta)t}\le c\|z(0)\|e^{-\kappa_0 t}<K\delta_x=\eta_0/2.
\end{equation*}
Hence, by a standard continuation argument, $\|z(t)\|<\eta_0/2$ for all $t\ge0$ in the maximal interval of existence. Since $\omega(t)\in W\subset B_R$ and $\|x(t)\|\le\|\pi(\omega(t))\|+\|z(t)\|<\eta_0$, the solution remains in a compact set and therefore exists for all $t\ge0$, and the estimate holds for all $t\ge0$. Finally, since $W$ is invariant and $\pi$ satisfies the invariance equation \eqref{eq:inv} on $W$, $\frac{d}{dt}\pi(\omega(t))=D\pi(\omega(t))s(\omega(t))=f(\pi(\omega(t)),l(\omega(t)))$, so $t\mapsto(\omega(t),\pi(\omega(t)))$ is a solution of the cascade.\hfill\proofbox
}

\section{Proof of Proposition~\ref{prop:unique}}\label{app:unique}
By Taylor's theorem and $D^{2}V(0)\succ0$, there exist $R>0$ and constants $0<c_{1}\le c_{2}$ such that $\overline{B}_R\subset U$ and
\begin{equation}\label{eq:quadratic-V-bound}
    c_{1}\|\xi\|^{2}\le V(\xi)\le c_{2}\|\xi\|^{2},
    \qquad \xi\in B_R.
\end{equation}
Choose $R_0\in(0,R)$ so that $c_{2}R_0^{2}<c_{1}R^{2}$. If $\omega\in B_{R_0}$, conservation of $V$ guarantees $\|\omega_t\| < R$ for all $t$. Hence, the trajectory remains in the compact set $\overline{B}_R$ throughout its maximal interval of existence, and the standard continuation theorem implies that it is defined for all $t\in\mathbb R$. Write $\omega_{t}:=\varphi_{t}(\omega)$ and $c_3:= \sqrt{c_2/c_1}\ge1$. Applying the estimate \eqref{eq:quadratic-V-bound} at both $\omega$ and $\omega_t$, and using $V(\omega_t)=V(\omega)$, gives
\begin{equation} \label{eq:two-sided}
    \frac{\|\omega\|}{c_3} \le\|\omega_{t}\|\le c_3\|\omega\|,
\end{equation}
for every $\omega\in B_{R_0}$ and every $t\in\mathbb R$.
 
Let $\pi^{1},\pi^{2}$ be two $\mathcal C^1$ solutions to the invariance equation \eqref{eq:inv}, and let $\rho_0$ and $\eta_0$ be as in Lemma~\ref{lem:invariant-uniqueness}. Since $\pi^i(0)=0$ and $\pi^i$ is continuous, we can choose $\rho\in(0,R_0)$ so small that $c_3\rho\le\rho_0$, that $\overline{B}_{c_3\rho}$ is contained in the common domain of $\pi^1$ and $\pi^2$, and that $\|\pi^i(\xi)\|\le\eta_0$ for all $\xi\in\overline B_{c_3\rho}$ and $i=1,2$. Define $W:=\{\omega_t:\omega\in B_\rho,~t\in\R\}$. Then $W$ is invariant under $\varphi_t$, $B_\rho\subset W$, and $W\subset B_{c_3\rho}\subset B_{\rho_0}$ by the estimate \eqref{eq:two-sided}. Therefore, Lemma~\ref{lem:invariant-uniqueness} yields $\pi^1=\pi^2$ on $W$, and in particular on the neighborhood $B_\rho$ of the origin.\hfill\proofbox

\section*{Acknowledgments}
Heein Park would like to thank Seokho Jeong for helpful advice on this research.

\bibliographystyle{siamplain}
\bibliography{references}
\end{document}

%% file: ex_shared.tex
\usepackage{lipsum}
\usepackage{amsfonts}
\usepackage{graphicx}
\usepackage{epstopdf}
\usepackage{algorithmic}
\ifpdf
  \DeclareGraphicsExtensions{.eps,.pdf,.png,.jpg}
\else
  \DeclareGraphicsExtensions{.eps}
\fi

\newsiamremark{remark}{Remark}
\newsiamremark{hypothesis}{Hypothesis}
\crefname{hypothesis}{Hypothesis}{Hypotheses}
\newsiamthm{claim}{Claim}
\newsiamremark{fact}{Fact}
\crefname{fact}{Fact}{Facts}

\headers{Nonlinear Moments as Equivalence Classes}{H. Park, H. Jang, A. Astolfi, H. Shim, and J. G. Lee}

\title{Nonlinear Moments as Equivalence Classes:
A Lyapunov--Perron Approach to the Invariance Equation
    \thanks{Submitted to the editors, \today.
        \funding{This work was supported by the National Research Foundation of Korea (NRF) grant funded by the Korea government (MSIT) (RS-2026-25504174) and the Strategic Hub for International Research Collaboration project of Seoul National University (SNU).}
    }
}

\author{Heein Park\thanks{Automation Systems and Research Institute, Department of Electrical and Computer Engineering, Seoul National University, Seoul, Republic of Korea 
  (\email{phoinee@cdsl.kr}, \email{hyjang@cdsl.kr}, \email{hshim@snu.ac.kr}, \email{jingyu.lee@snu.ac.kr}) }
\and Hyeonyeong Jang\footnotemark[2]
\and Alessandro Astolfi\thanks{Computer, Electrical and Mathematical Sciences and Engineering Division, King Abdullah University of Science and Technology, Thuwal, Kingdom of Saudi Arabia (\email{alessandro.astolfi@kaust.edu.sa})}
\and Hyungbo Shim\footnotemark[2]
\and Jin Gyu Lee\footnotemark[2]
}

\usepackage{amsopn}
